\documentclass[useAMS,usenatbib]{mnras}

\usepackage{lscape}
\usepackage{graphicx}
\usepackage{colortbl}
\usepackage{amssymb}
\usepackage{amsmath}
\usepackage{natbib}
\usepackage{times}
\usepackage{threeparttable}

 \newcommand{\Msun}{\mbox{${M_{\odot}}$}}

\usepackage{color}

\def\simpropto{\lower.2ex\hbox{$\; \buildrel \propto \over \sim \;$}}
\def\ltsim{\lower.5ex\hbox{$\; \buildrel < \over \sim \;$}}
\def\gtsim{\lower.5ex\hbox{$\; \buildrel > \over \sim \;$}}

\begin{document}

\title[Substructure in HSTFF clusters]{Mapping substructure in the HST Frontier Fields cluster lenses and in cosmological simulations}
\author[Natarajan, et al.]{Priyamvada Natarajan$^1$\thanks{E-mail:
priyamvada.natarajan@yale.edu}, Urmila Chadayammuri$^1$, Mathilde Jauzac$^{2,3,4}$, Johan Richard$^5$, 
\newauthor Jean-Paul Kneib$^6$, Harald Ebeling$^7$, Fangzhou Jiang$^{1,8}$, Frank van den Bosch$^1$,  
\newauthor Marceau Limousin$^9$, Eric Jullo$^9$, Hakim Atek$^{1,10}$, Annalisa Pillepich$^{11}$, Cristina Popa$^{12}$,  
\newauthor Federico Marinacci$^{13}$, Lars Hernquist$^{11}$, Massimo Meneghetti$^{14}$ and Mark Vogelsberger$^{13}$\\
\\
$^1$Department of Astronomy, 52 Hillhouse Avenue, Steinbach Hall, Yale University, New Haven, CT 06511, USA\\
$^2$ Centre for Extragalactic Astronomy, Department of Physics, Durham University, Durham DH1 3LE, U.K.\\
$^{3}$Institute for Computational Cosmology, Durham University, South Road, Durham DH1 3LE, U.K.\\
$^{4}$Astrophysics and Cosmology Research Unit, School of Mathematical Sciences, University of KwaZulu-Natal, Durban 4041, South Africa\\
$^{5}$CRAL, Observatoire de Lyon, Universit\'e Lyon 1, 9 Avenue Ch. Andr\'e, 69561 Saint Genis Laval Cedex, France\\
$^{6}$Laboratoire d'Astrophysique, Ecole Polytechnique F\'ed\'erale de Lausanne (EPFL), Observatoire de Sauverny, CH-1290 Versoix, Switzerland\\
$^{7}$Institute for Astronomy, University of Hawaii, 2680 Woodlawn Drive, Honolulu, Hawaii 96822, USA\\
$^{8}$The Hebrew University, Jerusalem 91904, Israel\\
$^{9}$Laboratoire d'Astrophysique de Marseille - LAM, Universit\'e d'Aix-Marseille $\&$ CNRS, UMR7326, 38 rue F. Joliot-Curie, 13388 Marseille Cedex 13, France\\
$^{10}$Institut d'Astrophysique de Paris, 98 bis bd Arago, F-75014 Paris, France\\
$^{11}$Harvard-Smithsonian Center for Astrophysics, 60 Garden Street, Cambridge, MA 02138, USA\\ 
$^{12}$Physics Department, Harvard University, Cambridge, MA, 02138\\
$^{13}$Department of Physics, Kavli Institute for Astrophysics and Space Research, Massachusetts Institute of Technology, Cambridge, MA 02139, USA\\
$^{14}$Osservatorio Astronomico di Bologna, INAF, via Ranzani 1, 40127, Bologna, Italy}

\date{00 Jul 2016}
\pagerange{\pageref{firstpage}--\pageref{lastpage}} \pubyear{0000}
\maketitle

\label{firstpage}

\begin{abstract}
We map the lensing-inferred substructure in the first three clusters observed by the \textit{Hubble Space Telescope Frontier Fields} Initiative (HSTFF): Abell 2744 ($z{=}0.308$), MACSJ\,0416, ($z{=}0.396$) and MACSJ\,1149 ($z{=}0.543$). Statistically resolving dark-matter subhaloes down to ${\sim}10^{9.5}\,\Msun$, we compare the derived subhalo mass functions (SHMFs) to theoretical predictions from analytical models and with numerical simulations in a Lambda Cold Dark Matter (LCDM) cosmology. Mimicking our observational cluster member selection criteria in the HSTFF, we report excellent agreement in both amplitude and shape of the SHMF over four decades in subhalo mass ($10^{9-13}\,\Msun$). Projection effects do not appear to introduce significant errors in the determination of SHMFs from simulations. We do not find evidence for a substructure crisis, analogous to the missing satellite problem in the Local Group, on cluster scales, but rather excellent agreement of the count-matched HSTFF SHMF down to $M_{\rm sub halo}/M_{\rm halo}{\sim}10^{-5}$. However, we do find discrepancies in the radial distribution of sub haloes inferred from HSTFF cluster lenses compared to determinations from simulated clusters. This suggests that although the selected simulated clusters match the HSTFF sample in mass, they do not adequately capture the dynamical properties and complex merging morphologies of these observed cluster lenses. Therefore, HSTFF clusters are likely observed in a transient evolutionary stage that is presently insufficiently sampled in cosmological simulations. The abundance and mass function of dark matter substructure in cluster lenses continues to offer an important test of the LCDM paradigm, and at present we find no tension between model predictions and observations.
\end{abstract}

\begin{keywords}
{cosmology: theory, dark matter, large scale structure of the Universe, galaxies: haloes, galaxies: clusters: general
  galaxies: substructure}
\end{keywords}

\section{Introduction}

While the bulk of the matter content of our Universe is inventoried to be dark matter -- cold, collisionless particles that drive the formation of all observed structure -- its nature remains elusive.  Fortunately, observational cosmology provides us with luminous probes that nonetheless enable us to map dark matter on a range of scales, namely galaxies that reside at the centers of dark-matter halos. The gravitational influence exerted by dark matter, as reflected  dynamically (in the motions of stars in a galaxy or galaxies in a cluster) and in the deflection of light rays from distant sources, yields insights into its spatial distribution and role in structure formation in the universe.  In particular gravitational lensing offers a unique and powerful probe of the detailed distribution of dark matter, as it is achromatic and independent of the dynamical state of the object producing the lensing. Lensing of faint, distant background galaxies by clusters of galaxies, the most recently assembled massive structures that are extremely dark-matter dominated ($\sim$90\% of their content), results in dramatic observational effects that can be studied in two regimes. Strong lensing -- which creates highly distorted, magnified and occasionally multiple images of a single source -- and weak lensing -- which results in modestly yet systematically deformed shapes of background galaxies -- provide robust constraints on the projected distribution of dark matter within lensing clusters \citep{natarajan97,bradac05,limousin07b,Merten09,umetsu16}. Lensing by clusters has many other applications, as it allows, in combination with multi-wavelength data, studies of the masses and assembly history of clusters \citep{clowe04,merten11,eckert15}, and probes faint, distant galaxy populations that would otherwise be inaccessible to observation. The luminosity function of galaxies at very high redshift derived from lensing has been instrumental for studies of the re-ionization of the universe; for a status report see the review by \cite{Finkelstein15} and references therein; as well as recent results in \citet{Bradac14,atek14a,Bouwens15,Coe15,Laporte15,McLeod16}.  In addition, cosmography -- mapping the geometry of the universe -- has been demonstrated to be another powerful application of gravitational lensing that provides constraints on dark energy complementary to those from other probes \citep{Jullo10,D'Aloisio11,Caminha15}. Exploiting strong gravitational lensing, we here present a test of the currently accepted Lambda Cold Dark Matter (LCDM) paradigm and its implementation in cosmological simulations from the abundance and properties of substructure in massive clusters.

The high-resolution of the imaging cameras aboard the Hubble Space Telescope (HST) has transformed this field in the last two decades \citep[detailed in a review by][]{KN11}, allowing   the secure identification of multiply imaged systems that provide critical constraints on the mass model, from deep imaging data. While data from the \textit{Advanced Camera for Surveys} ({\it ACS}) and the \textit{Wide-Field Planetary Camera-2} ({\it WFPC-2}) tremendously advanced early studies of gravitational lensing by clusters compared to groundbased work, HST's on-going Frontier Fields (HSTFF) programme has truly revolutionized this area of research \citep{lotz15}. As part of the HSTFF program six clusters ranging in redshift from $z{=}0.3$ to 0.55 have been selected for a total of 140 orbit observations per cluster with the ACS in the F435W, F606W, and F814W, as well as with WFC3/IR in the F105W, F125W, F140W, and F160W passbands. In addition to these multi-filter data, coordinated observational efforts in other wavelengths with dedicated Spitzer and Chandra programs coupled with ground-based spectroscopic follow-up of cluster galaxies and lensed images are in the process of compiling exquisite and comprehensive data sets for these cluster lenses.\footnote{For further details see http://www.stsci.edu/hst/campaigns/frontier-fields/}

In this paper, we study the detailed distribution of substructure derived directly from mass models constrained by more than a hundred lensed images each gleaned from the HSTFF imaging data for Abell 2744, MACSJ\,0416.1--2403 \citep[hereafter MACSJ\,0416;][]{mann12} and 65 images for MACSJ\,1149.5+2223 \citep[hereafter MACSJ\,1149;][]{ebeling10}. These three clusters, spanning a redshift range 0.308-0.554, also represent various stages of cluster mass assembly. All three clusters have complex mass distributions involving the on-going merger of several sub-components \citep{jauzac14,lam14,diego15,wang15,jauzac15b,medezinski16,jauzac16}. Merging clusters with complex interaction geometries like in these three cases turn out to be more efficient as lenses compared to relaxed clusters, as they generate a larger number of multiply lensed systems \citep{Owers11,wong12,wong13}.  While lensing is independent of the dynamical state of the cluster, the efficiency of lensing is enhanced when sub-clusters merge due to the resultant higher surface mass densities produced \citep{natarajan98,torri04}. The positions, magnitudes and multiplicities of lensed images provide strong constraints for the mass modeling of cluster lenses. In addition, to calibrate the strength of the lensing signal, the redshifts of the images need to be known either spectroscopically or photometrically. In the case of highly magnified objects the HSTFF filter set choice provides photometric redshifts with reasonable accuracy. Follow-up spectroscopy by several independent groups has been on-going for the bright, highly magnified multiple images in these clusters as well as for faint objects with GTO/MUSE observations for Abell 2744 and MACSJ\,0416. In this paper, we present the best-to-date model for the mass distribution in these three clusters from which we derive properties of the dark matter substructure content. The inferred substructure - also referred to as the subhalo mass function (SHMF thereafter) - is then compared with mimicked "measurements" from simulated clusters in the Illustris cosmological boxes \citep[][details are available at http://www.illustris-project.org/]{vogelsberger14};  as well as analytic estimates that take halo-to-halo scatter into account. This exercise offers a concrete and powerful test of the standard LCDM model, its implementation in cosmological simulations and our analytic calculational framework. 

The motivation for this entire exercise is to carefully examine if any gaps emerge between theoretical predictions in LCDM and the observationally inferred degree of substructure. Earlier work on lower mass scales - namely galaxy scales - had claimed a crisis in LCDM due to the discrepancy in abundance between predicted and observed substructure \citep{moore99, klypin99}. Convincing resolutions to this "crisis" have since been proposed, that implicate the paucity of observed substructures to their intrinsic faintness as well as our lack of understanding of the efficiency of star formation in the smallest dark-matter haloes \citep{read05,pontzen14,dicintio16,wetzel16} reflecting our ignorance of the detailed relationship between baryons and dark matter. Alternative, less persuasive explanations for the mismatch challenging the collisionless nature of dark matter have also been proposed \citep{rocha13,lovell14}. Given that LCDM is a hierarchical theory it is imperative to explore if any such discrepancy is replicated on the next higher mass scale, on that of clusters. For this purpose, the HSTFF data-set offers unique leverage due to the large range in the SHMF that it permits scrutiny of. Utilizing strong lensing to reconstruct the detailed mass distribution of clusters, here we present the detailed substructure distribution in the first three HSTFF clusters, Abell 2744, MACSJ\,0416 and MACSJ\,1149. In this work, we derive the mass function of subhalos from the lensing data and compare these results with those obtained from the high-resolution Illustris cosmological N-body simulations and analytic predictions. Here we present the detailed comparison with zoom-in simulations of two clusters that are as massive -- with a virial mass of ${\sim}10^{15}\, \Msun$ -- as the HSTFF targets considered here. These Ilustris clusters will hereafter be referred to as {\it iCluster Zooms}, specifically the one with a mass of ${\sim}10^{15.3}\, \Msun$ as {\it iCluster Zoom 1} and the one with mass  $\sim \, 10^{14.8}\, \Msun$ as {\it iCluster Zoom 2}. We also include in our study a larger sample of less massive simulated clusters with masses between $10^{14}$ and $10^{15}\, \Msun$ that form in the small box size of the Illustris suite. We will refer to these systems as the $10^{14}$ {\it Illustris Haloes}.  We chose the Illustris simulations for comparison with results from our lensing data because they represent the state of the art regarding the treatment of baryonic physics. Morever, they allow us to mimic several key aspects of the observations, thereby enabling a detailed and robust comparison. We note, however, the important caveat is that the massive, actively merging cluster lenses targeted by the HSTFF project are in a dynamical state for which there is no equal in any of the currently simulated volumes. We also compare our lensing determined SHMF to the analytical prediction for parent halos with masses of ${\sim}10^{15} \Msun$ to understand the impact of cosmic variance. The HSTFF data allow probing the SHMFs down to several orders of magnitude below previous studies and offer the best current tests of the abundance and properties of substructure in the LCDM model.     

This paper is structured as follows: in Section 2, we describe the predictions for substructure in the LCDM model; followed by a synopsis of previous work in Section 3. In Section 4, the general methods we employ to derive SHMFs are described, after which the overall mass models for Abell 2744, MACSJ\,0416 and MACSJ\,1149 inferred from the HSTFF are described in Section 5. In Section 6, we present and discuss the derived SHMFs. We then provide a brief description of the Illustris suite of simulations in Section 7, before detailing the comparison of the lensing derived subhalo properties with those from the simulations and analytic methods (in Section 8). We close the paper in Section 9 with a discussion of the implications of our results for the LCDM model.

\section{LCDM substructure predictions}

Cold dark matter predicts the existence of copious substructure within collapsed halos of all masses. As a description of the underlying world model that best describes our universe, the LCDM model has been incredibly successful, tested with several observational probes ranging from the measured properties of the fluctuations in the Cosmic Microwave Background Radiation to the observed abundance, clustering and properties of galaxy populations. Precision measurements of cosmological parameters have now determined that we appear to live in a collisionless, cold dark matter dominated, dark energy driven accelerating expanding universe \citep{hinshaw13}. However, despite this highly successful paradigm for structure formation, in the past two decades attention has been drawn to challenges on "small scales" where discrepancies between the theory and observations appeared to lurk. There are two related small scale problems that continue to be active areas of study. The first pertains to the long-standing and debated question of whether the rotation curves of low surface brightness galaxies are compatible with the LCDM model and the second concerns the abundance of substructure on galaxy scales. The work presented in this paper addresses the second issue. The substructure crisis, as it was originally referred to, noted that the amount of dark matter substructure theoretically predicted on Milky Way scale galaxies was highly discrepant with the number of observationally detected satellites - believed to be hosted by dark matter subhaloes - in the Local Group. Interestingly, both the substructure problem and the mismatch of rotation curves of low surface brightness galaxies came to the fore only when high resolution N-body simulations could be performed with sub-kpc to parsec scale spatial resolutions. These higher resolution simulations in which the inner profile of dark matter halos could also be studied revealed that these dark matter halos on galaxy, group and cluster mass scales were filled with a large number of self-bound dark matter satellites. This plethora of clumpy structure had not been seen in earlier lower resolution studies in which all halos appeared to be significantly smoother \citep{kuhlen08,helmi11}. We now know that in fact the existence of vast amounts of substructure is a generic prediction of hierarchical structure formation in LCDM models where assembly of collapsed mass structures occurs via a merging hierarchy during which a large fraction of the infalling dense clumps survive as dynamically distinct substructures inside virialized halos until late times albeit after dynamical modification via tidal stripping, tidal heating and dynamical friction. And clusters of galaxies as the most recently assembled mass structures retain copious amounts of bound substructures within them. 

Recent work comparing the abundance of simulated satellites to those observationally detected in the Milky Way (with a mass of ${\sim}10^{12}\,\Msun$), do suggest that we have detected all the substructure associated with the most massive subhaloes bound to the Milky Way halo and what we might be missing are likely only the extremely faint galaxy population - though an increasing number of these are also being found in deeper images \citep{torrealba16,deason14}. The paucity of the detection of these galaxies can be explained by a combination of factors: their faintness makes them observationally challenging to detect and the efficiency of star formation in such low mass dark matter host halos might be suppressed due to baryonic physics. The leading hypothesis is that the reason for the existence of a sea of low mass dark matter halos being largely devoid of stars has to do with the physics of feedback processes in galaxy formation. For instance, feedback processes wherein the photo-ionizing UV background or the expelling of gas via strong, powerful supernovae winds leads to highly inefficient star formation can and are largely believed to explain away the substructure problem on galaxy scales \citep{wetzel16}. Independent of prior disagreements on galaxy scales, the shape and amplitude of the SHMF within clusters offers a powerful probe of the LCDM model. Any deviations from theoretical predictions of LCDM on these scales could be used as a diagnostic of the nature of dark matter and perhaps signal new physics. In the LCDM model, structure aggregates via gravity and is essentially scale-free, and the best-fit functional form to the sub halo abundance per unit parent halo mass can be written as:
\begin{eqnarray}
\frac{dn}{dm}\,=\,{10^{-3.2}}\,(\frac{m}{\,\,\Msun} h^{-1})^{-1.9}
\end{eqnarray}
\citep{gao04,vandenbosch05}. Since abundant substructure is endemic to LCDM, if there was a real substructure problem on galaxy scales, it is expected to be replicated on cluster scales. Therefore, testing substructure predictions on cluster scales offers an extremely critical analysis of the LCDM model. The sub halo mass range over which such an inventory can be performed directly impacts the degree of accuracy to which the LCDM model can be tested. To perform this test, here we obtain the mass spectrum of clumps in these HSTFF lensing clusters. These observationally determined quantities are directly compared with results of cosmological simulations and analytic estimates from theoretical calculations. Contrary to galaxy scales, in clusters we expect many more dark matter substructures to host visible galaxies, making the comparison of the SHMF less sensitive to uncertainties in the physics of galaxy formation. All the while, we however need to keep in mind that dynamically similar counter-parts to the HSTFF clusters are not available either in the Illustris volume or in any other simulated volume at present, even if comparable mass clusters like {\it iCluster Zooms} are available. The veracity of this expectation is also tested in our analysis here. Despite this, full consistency with the abundance of optically detected member galaxies, substructures in the LCDM simulations and subhaloes detected by lensing can be expected and asked for; and finding strong concordance can be viewed as a stringent test of the LCDM paradigm. This is precisely what we attempt in this paper.  

In earlier analytic work on the calculation of the SHMFs in the context of hierarchical CDM theories \cite{vandenbosch05} have computed the substructure content on cluster mass scales. Their work and findings are relevant here for the interpretation of our results of the lensing determined SHMF and comparison with those derived from the Illustris {\it iCluster Zooms} and the {$10^{14}${\it  Illustris Clusters} samples. In a two-step process that takes into account the assembly of clusters, \citep{vandenbosch05} first derive the masses of sub haloes at the time of their initial accretion using Monte Carlo realizations of their merger histories. Subsequent to being accreted these sub haloes are subject to a variety of processes that lead to mass loss, namely dynamical friction, tidal stripping and tidal heating. While the detailed mass loss is apt to depend on the motions of individual sub haloes within the cluster, they find that an average mass loss rate can be computed by simply averaging over all possible orbital configurations. Coupled with the additional assumption that the distribution of orbits is actually independent of the host/parent halo mass, they express the average mass loss rate as a function of two key variables: the mass ratio of the subhalo to the parent halo and redshift. This result is natural as it intuitively suggests that the formation time of the massive parent halo is an important variable. Comparing the predictions of this model with high-resolution dark matter only cosmological simulations to calibrate this picture, they found that contrary to earlier claims, the SHMF does depend on the mass of the parent halo. Both the slope and the normalization of the SHMF depend on the formation time of the parent halo, and explicitly depend on the ratio of the parent halo mass to the characteristic non-linear mass scale. Therefore, in early assembling clusters, in-falling sub haloes are subject to dynamical modification for longer and since the most massive clusters form later in hierarchical CDM, their sub haloes experience less stripping. One of the advantages of this formalism is the ability to easily compute and quantify the halo-to-halo variation that can be expected in absence of an ensemble of simulated clusters to average over. Estimating cosmic  variance is challenging for simulations that are limited by the essential compromise between box size and resolution which results in the paucity of high mass clusters. The estimated halo-to-halo variance depends as expected on the detailed mass accretion history during the process of cluster assembly. During the assembly of massive clusters there are two effects that need to be understood - the mass loss suffered by individual infalling sub haloes all the while as the parent host halo itself gains mass due to cosmic accretion as part of its growth in a dark matter dominated universe. \cite{vandenbosch05} find that the recent cosmic accretion history is what is most relevant, in fact, cosmic accretion in the previous Gyr or so. As predicted by their model, this dependence is what is reflected in the observed halo-to-halo scatter. The predictions of this model for LCDM are specially salient to examine the trends with parent halo mass and we compare our lensing derived SHMFs with these analytic predictions.

\section{Deriving substructure from cluster lensing data}

In order to derive the SHMF  from lensing data, we adopt the methodology that we have developed over the last decade for analyzing cluster lensing data. We start with modeling the mass distribution in the cluster with a set of large and small scale self-similar parametric mass profiles. The cluster itself is visualized as a composite of large-scale smooth mass components with several small-scale sub clumps, which are both modeled with the analytic PIEMD (pseudo-isothermal elliptical mass distribution) profile \citep{natarajan97,limousin07b}. The small scale subhaloes in our conception of the cluster are associated with the locations of bright, early-type cluster galaxies under the explicit assumption that light traces mass.  This is entirely akin to the process by which we will derive the substructure for the {\it iCluster Zooms} as well as the Illustris $10^{14}$ {\it Illustris Haloes} as described below. The location, brightness and redshifts of the magnified, multiply imaged background sources are used in Abell 2744, MACSJ\,0416 and MACSJ\,1149 to statistically quantify the masses of sub-clumps using a Bayesian scheme. Deploying an MCMC method we are then able to derive a family of best-fit models modulo the assumed priors for the choice of parametric profiles including the self-similar scaling and the association of mass with light. In this work, we also explore a couple of distinct scaling relations to characterize the relation between mass and light for cluster galaxies to examine the dependence of these assumptions on our final results. 

\subsection{Determination of cluster members}

 Cluster membership for galaxies in these clusters was determined using methods described in detail in \cite{richard14}. Here we summarise the key steps. Galaxy catalogs were first generated using SExtractor \citep{BA96} and cluster membership was assigned using complementary colour-magnitude diagrams ($m_{\rm F606W}-m_{\rm F814W}$ versus $m_{\rm F814W}$ and $m_{\rm F435W}-m_{\rm F606W}$ versus $m_{\rm F814W}$). Spectroscopically confirmed cluster members were used to identify the red sequence; cluster membership was assigned to all galaxies that lie within 3$\sigma$ of a linear fit to the red sequence. 
 We used a fixed value for the dispersion, obtained by collapsing the red sequence along the best linear fit down to the (preset) limiting magnitude. We are thus effectively fitting a superposition of many Gaussians of ever increasing width with a single Gaussian.  We note that a moving sigma that becomes smaller as we move along the red sequence toward the BCG would not result in a very different galaxy selection: usually the gap between the red sequence and the green valley galaxies widens too. For MACS\,J0416, we used the cluster member catalogue of \citet{Grillo15} which comprises 175 galaxies, 63 of them spectroscopically identified and the remaining 112 selected using a spectro-photometric method described in detail in their paper \citep[see also][]{Rosati14,Balestra16}. This catalogue was provided to the lens-modeling community by C.\ Grillo and the CLASH collaboration in the context of the magnification map-making project in September 2015\footnote{https://archive.stsci.edu/prepds/frontier/}. For Abell 2744 and MACSJ\,1149, our selection is based on spectroscopically confirmed members from \cite{Owers11} and \citet{ebeling14}, respectively. For all three clusters, the selection technique adopted to select cluster members from our galaxy catalogues extends to a uniform limiting bolometric luminosity of $0.01\,L_{*}$. Despite this uniform cut, we find that this yields a differing number of cluster galaxies for each of the clusters due to the range that they span in redshift. For MACS\,J0416, note that we used the catalogue of mostly spectroscopically confirmed galaxies provided by Grillo and co-workers.

The F814W magnitudes of the resulting set of cluster members range from 18.49 to 26.3 in Abell 2744 (563 galaxies), from 19.04 to 23.91 in MACS\,J0416 (175 galaxies), and from 18.96 to 25.66 in MACS\,J1149 (217 galaxies). The galaxies thus selected were included as small-scale perturbers in our high-fidelity lensing models. The details of these mass models are given in \cite{jauzac15b} for Abell 2744, \cite{jauzac14} for MACS\,J0416, and \cite{jauzac16}  for MACS\,J1149. We note that the models for Abell 2744 and MACS\,J0416 have been updated since their publication as part of the 2015 mass-mapping effort using new spectroscopic redshifts for multiple-image systems, and the cluster member catalogue of \cite{Grillo15}. We describe the resulting changes in detail in the following section. In addition, as part of the data-sharing for this map-making project we have used spectroscopic redshifts for multiple-image families in these clusters that were provided by other teams including the GLASS collaboration and K.\ Sharon's team \citep{johnson14,schmidt14,treu16}. Our cluster member selection is by construction incomplete at large cluster-centric radii and at low luminosities as outlined above.

\subsection{Mass Modeling: Methodology}

In this section, we outline the modeling framework, and note that further details can be found in several earlier papers \citep{natarajan97,natarajan98} and a more recent review \citep{KN11}.  In order to extract the properties of the population of subhaloes in cluster lenses, as mentioned above, the range of mass scales is modeled using a parametric form for the surface mass density profile of the lens. Motivated by the regularity of X-ray surface brightness maps of clusters we envision the cluster as composed of a super-position of several smooth large-scale gravitational potentials and smaller scale perturber potentials that are associated with the locations of bright early-type cluster members:
\begin{equation}
\phi_{\rm tot} = \Sigma_i\,\phi_{\rm s_i} + \Sigma_n\,\phi_{\rm p_n},
\end{equation}
where $\phi_{\rm s_i}$ are the gravitational potentials of the smooth components and 
$\phi_{\rm  p_n}$ are the potentials of the $n$ subhaloes associated with the $n$
cluster galaxies treated as perturbers. The lensing amplification matrix $A^{-1}$ can 
also be decomposed into contributions from the main clump and the perturbing potentials:
\begin{eqnarray} 
A^{-1}\,=\,(1\,-\,\Sigma_i \kappa_{\rm s_i}\,-\,\Sigma_n \kappa_{\rm
p})\,I - \Sigma_i \gamma_{\rm s_i}J_{2\theta_{\rm s_i}} \\ - \Sigma_n \,\gamma_{\rm
p_n}J_{2\theta_{\rm p_n}} \nonumber;
\end{eqnarray}
where $\kappa$ is the magnification
and $\gamma$ the shear. The quantity relevant to lensing is the projected surface mass density. The distortion induced by the overall potential 
with the smooth and individual galaxy-scale halos modeled self-similarly as linear superposition of two PIEMD distributions, has the following form:
\begin{eqnarray}
{\frac{\Sigma(R)}{\Sigma_0}}\,=\,{r_0  \over {1 - r_0/r_t}}
\left({1 \over \sqrt{r_0^2+R^2}}\,-\,{1 \over \sqrt{r_t^2+R^2}}\right),
\end{eqnarray}
with a model core-radius $r_0$ and a truncation radius $r_t\,\gg\,
r_0$. The projected coordinate $R$ is a function of $x$, $y$ and the ellipticity, 
\begin{eqnarray}
R^2\,=\,{x^2 \over (1+\epsilon)^2}\,+\,{y^2 \over
  (1-\epsilon)^2}, \;\;{\rm where}\;
\ \ \epsilon= {a-b \over a+b}.
\end{eqnarray}

Coupling these analytic forms with further assumptions about the fidelity with which mass traces light described in the next subsection, the SHMF is derived using strong lensing constraints from the HSTFF observations for these three clusters.

\subsection{Relating Mass to Light}

One of the key features and facilities of parametric modeling is the flexibility afforded in modeling the precise relationship between mass and light. Guided by empirically observed correlations between internal properties of individual, bright, early-type cluster galaxies we adopt those to couple the mass of the dark matter subhalo to the properties of the galaxy it hosts in our modeling scheme. In addition, we also assume that the ellipticity and the orientation of the dark matter subhaloes associated with early-type cluster members is aligned with that of the galaxies themselves. These simple assumptions are inputs while generating the best-fit lensing mass model for the cluster. The adopted set of physically motivated, empirically determined scaling laws for relating the dark matter distribution of the subhaloes to the light distribution of the cluster galaxies are:
\begin{eqnarray}
{\frac{\sigma_0}{\sigma_{0*}}} \,=\,\left({L \over L^*}\right)^{1 \over 4};\,\,
{\frac{r_0}{r_{0*}}}\,=\,\left({L \over L^*}\right)^{1 \over 2};\,\,
{\frac {r_t}{r_{t*}}}\,=\,\left({L \over L^*}\right)^{\alpha}.
\end{eqnarray}
These scalings lead to a set of models for cluster members where the total mass $M_{\rm ap}$ enclosed within an aperture $r_{t*}$ and the total mass-to-light ratio $M/L$ scale with the total luminosity as:
\begin{eqnarray}
M_{\rm ap}\,\propto\,{\sigma_{0*}^2}{r_{t*}}\,\left({L \over L^*}\right)^{{1 \over
2}+\alpha},\,\,{M_{\rm ap}/L}\,\propto\,
{\sigma_{0*}^2}\,{r_{t*}}\left( {L \over L^*} \right)^{\alpha-1/2},
\end{eqnarray}
where $\alpha$ determines the typical size scale of the galaxy halo. For a value of  $\alpha$ = 0.5, the model galaxy has constant mass to light ratio with luminosity though not as a function of radius within. Here we first explore $\alpha = 0.5$ as in previous work, as this leads to a scaling law that is empirically motivated by the Kormendy and the Faber-Jackson relations for early-type galaxies \citep{faber76,kormendy77,natarajan09,limousin07a}. The Kormendy relation relates the spatial scale to the luminosity, while the more general form of the Faber-Jackson relation is used to relate the velocity dispersion to the luminosity.  In practice, the constant mass to light ratio relation for $\alpha = 0.5 $ has proven to provide a good fit, and so far strong lensing data have not ruled out this hypothesis. With a choice of $\alpha=0.8$, we would have ended up with the fundamental plane relation $M/L \sim L^{0.3}$ \citep{halkola06, jorgenson96}. In recent work, however, modeling cluster galaxies similarly using data of the lensing cluster Abell 383 and combining with measured values for the central velocity dispersion for a handful of galaxies, \cite{monna15} report reasonable agreement with $\alpha = 0.5$ in the case of one galaxy. In more recent work, with velocity dispersion measurements for 5 galaxies near a strongly lensed arc in the cluster Abell 611, they report departure from scaling relations indicated by the galaxy to galaxy variation in the estimated truncation radii \citep{monna17}. The conclusion is that the efficiency for tidal stripping varies for galaxies, in fact, such a systematic difference between early and late-types was found for cluster galaxies in the merging lensing cluster Cl\,0024+14 by \cite{natarajan09}.  We explore $\alpha = 0.5$ as well as the scaling between mass and light that is found in the Illustris simulation with the assumed sub-grid models for modeling galaxy formation. In this paper, in addition to using the scalings implied by the Faber-Jackson \& Kormendy laws, we also derive the SHMF from observational data under the assumption that light traces mass as it does in the Illustris full physics run. Writing out the general scaling relations as: $\sigma \propto L^a$ and $r_t \propto L^b$ and $M/L \propto L^c$; the Faber-Jackson case corresponds to: $a = 0.25; b = 0.5; c = 1$ and in Illustris we find $a = 0.18; b = 0.16; c = 0.49$. Adopting this new set of scaling laws, we re-ran \textsc{Lenstool} to obtain the best-fit mass distribution for all three clusters and extracted the resultant SHMF.  In the results section of the paper, we plot the SHMFs derived for both these sets of assumed scaling laws. We are thus able to assess if and how this assumption of how light traces mass impacts our results. We note that the evidence thus far from other independent studies of the relationship between mass and light strongly support the fact that light traces mass effectively both on cluster scales \citep{kneib03} and on galaxy scales \citep{mandelbaum06,newman15}. The Bayesian evidence for the best-fit mass models for these two sets of scaling laws are virtually indistinguishable. This suggests that the SHMF we have derived even from the high quality HSTFF data  is not very sensitive to our detailed assumption of how light traces mass.

\section{Previous Lensing Substructure tests of LCDM Clusters}

In earlier work, we quantified substructure derived from WFPC-2 observations of lensing clusters and compared results with cosmological simulations. The results of the first attempt to do so were presented in \cite{pn04} and subsequently in \cite{natarajan07}. Results reported in both these papers used HST WFPC-2 imaging and a comparison with the {\it Millennium simulation}. In the first paper, results of the direct comparison of the lensing derived substructure mass function with that obtained from the simulated clusters using only dark matter particles  was performed. In the second paper, a semi-analytic model for galaxy formation was painted on to the dark matter only {\it Millennium Simulation} that enabled mimicking of the selection criteria adopted in the lensing analysis. That is, the dark matter halos hosting the brightest cluster members were extracted from the simulation after the semi-analytic model had been implemented to "form" realistic galaxies. 

In \cite{natarajan07}, we presented high resolution mass models for five HST cluster lenses, and performed a detailed comparison of the SHMF, the velocity dispersion and aperture radii function with an ensemble of cluster--sized haloes selected from the {\it Millennium Simulation} including an implementation of a semi-analytic model for the galaxy formation detailed in \cite{delucia06}. The construction of the mass models combining strong and weak lensing data for these massive clusters was performed using the same galaxy-galaxy lensing techniques outlined in the methodology section here. As described above, the goal was to quantify substructure under the assumption that bright early-type cluster galaxies are robust tracers. We derived the SHMF within a limited mass range $10^{11} - 10^{12.5}\, \Msun$ in the inner regions of these clusters. Upon detailed comparison with simulated {\it Millennium} clusters, remarkably we found consistency with the abundance of substructure given that they are completely independently determined. For the cases of clusters that were active mergers the match with simulations was less good.  

In both earlier works discussed above, we were sensitive only to substructures in a small mass range $\sim 10^{11}\,-\,10^{12.5}\,\Msun$. The mass spectrum of substructure over this mass range, and other sub halo properties retrieved from the lensing data were found to be consistent with the theoretical predictions of LCDM from the simulations. While this agreement suggested that there was no substructure "crisis" as claimed earlier in LCDM per se, to draw a more robust conclusion, a wider range of sub halo mass scales needed to be probed. And this is precisely what the HSTFF data affords us as we report below.  

\section{HST Frontier Fields cluster-lens mass models}

Using the extremely deep and high resolution HSTFF data for the massive lensing clusters Abell 2744, MACSJ\,0416 and MACSJ\,1149 that were made publicly available on the MAST Archive \footnote{URL: https://archive.stsci.edu/prepds/frontier/} at the Space Telescope Science Institute as part of the Frontier Fields Initiative, our collaboration CATS (Clusters As TelescopeS) identified all the multiple image systems and constructed comprehensive mass distributions for all three clusters including small scale clumps modeled as described above. Including the positions, brightnesses  and measured spectroscopic redshifts where available for the lensed images in the ACS field of view, all our constructed mass models have already been published \citep{jauzac14,jauzac15a,jauzac16}, and were built using the \textsc{Lenstool} software in its parametric mode \citep{jullo07}. Below, we provide a synopsis of the features of these mass models along with updates since their publication in order to illustrate the high data quality and resultant unprecedented precision of these models. The gain in precision of the overall mass model also enables more accurate characterization of the subhalo masses. 

\subsection{Mass distribution in Abell 2744}

We constructed a high-precision mass model of galaxy cluster Abell 2744 at $z = 0.308$, based on a strong- gravitational-lensing analysis of the entire HSTFF imaging data set, that includes both the ACS and WFC-3 observations. With the depth of this dataset in the visible and near-infrared, we identified 34 new multiply imaged background galaxy systems listed in \cite{jauzac15b}, bringing the total up to 61, leading to a final tally of 181 individual lensed images. While doing so, we corrected earlier erroneous identifications and inaccuracies in the positions of multiple systems in the northern part of the cluster core, namely the image System \#3). We then culled the multiple images that were less reliable after which running \textsc{Lenstool} with 54 multiply-imaged systems (154 total images) that were determined to be the most secure ones amongst the 61 listed in \cite{jauzac15b}. We modeled the cluster with two large-scale dark matter halos plus smaller galaxy-scale perturber halos associated with 733 individual cluster member galaxies. Our best-fit model, which only uses strong-lensing constraints, predicts image positions with an RMS error of 0.79$\arcsec$, that corresponds to an improvement of almost a factor of two over previous modeling attempts for this cluster. We find the total projected mass inside a 200 kpc aperture to be $2.162 \pm 0.005\,\times 10^{14}\Msun$. This gain in the accuracy of the mass modeling translates directly into an overall improvement of a factor 4 in the derived magnification map for the high-redshift lensed background galaxies that are brought into view by the cluster lens. Further details of this best-fit model can be found in \cite{jauzac15b}. We note that this model reconstruction extends radially out only to a fraction of the virial radius of Abell 2744 to $\sim\,{0.5\,R_{\rm vir}}.$ Table 1 below lists the details of the best-fit lensing mass model for Abell 2744.

\begin{table}
\begin{center}
\begin{tabular}[h!]{cccc}
\hline
\hline
\noalign{\smallskip}
Component  & \#1 & \#2 & L$^*$ elliptical galaxy \\
\hline
$\Delta$ \textsc{ra}  & $-4.8^{+0.2}_{-0.1}$  &  $-15.5^{+0.1}_{-0.2}$ & --  \\
$\Delta$ \textsc{dec} & $4.0^{+0.2}_{-0.1}$  & $-17.0^{+0.2}_{-0.1}$ & --  \\
$e$ & 0.30 $\pm$0.004 & 0.60 $\pm$ 0.01 & -- \\
$\theta$ & 64.2$^{+0.3}_{-0.2}$  & 40.5$^{+0.4}_{-0.5}$ & -- \\
r$_{\mathrm{core}}$ (\footnotesize{kpc}) & 205.0$^{+1.3}_{-1.5}$  & 39.6$^{+0.8}_{-0.6}$  & [0.15] \\
r$_{\mathrm{cut}}$ (\footnotesize{kpc}) & [1000] & [1000] & 18.0$^{+0.6}_{-1.0}$ \\
$\sigma$ (\footnotesize{km\,s$^{-1}$}) &  1296$^{+3}_{-5}$ & 564$^{+2}_{-2}$ & 154.3$\pm$ 1.8 \\
\noalign{\smallskip}
\hline
\hline
\end{tabular}
\caption{Abel 2744 best-fit PIEMD parameters for the two large-scale dark-matter halos, as well as for the L$^{*}$ elliptical galaxy. The model is built using 113 multiple images, includes 563 galaxy-scale perturbers. The best-fit RMS is 0.70 $\arcsec$.
Coordinates are quoted in arcseconds with respect to $\alpha=3.586259, \delta=-30.400174$.
Error bars correspond to the $1\sigma$ confidence level. Parameters in brackets are not optimised.
The reference magnitude for scaling relations is $mag_{\rm{F814W}} = 19.44$.
}
\label{table_a2744}
\end{center}
\end{table}

In September 2015, several selected independent lensing teams were asked to provide HFF mass models to the community through a mass mapping challenge. For this purpose, data were shared, including a large number of spectroscopic redshifts from \cite{wang15}, as well as new measurements from K. Sharon's team \citep{johnson14}. With this additional data, we revised our previously published mass model described above, after a group vote to select multiply imaged families, only keeping the most secure ones (voted as Gold, Silver and Bronze by all lensing teams) and more securely identified cluster members. While our overall mass model model did not change much from the one presented by \cite{jauzac15b}, we now include only 113 multiple images for the modeling. The best-fit mass model for the cluster comprises two cluster-scale halos as before, and 563 galaxy-scale perturbers. This final model predicts image positions with an even lower global RMS error of 0.70$\arcsec$, with similar mass estimation and precision on both mass and magnification as our initial HSTFF model. The viral radius for Abell 2744 lies at $\sim$ 2 Mpc and our mass model reliant on the HSTFF data extends out only to $\sim 0.5\,{\rm R_{\rm vir}}$ Substructure derived from this updated model is used here in our comparison with the {\it iCluster Zooms} as well as the Illustris $10^{14}$ {\it Illustris Haloes} and analytic predictions.

\begin{figure}
\includegraphics[width=1.0\columnwidth]{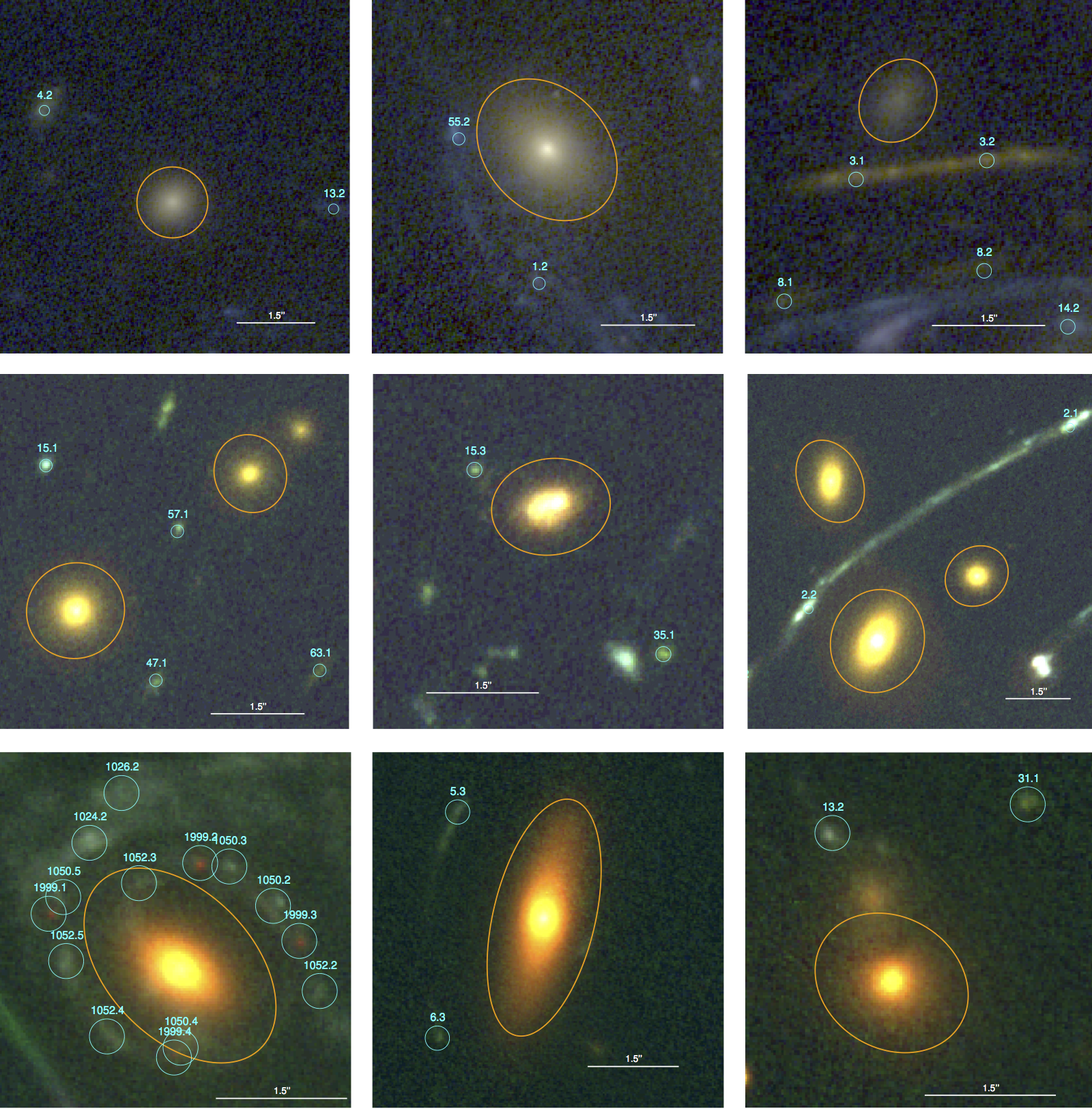}
\caption{Low luminosity cluster member galaxies that are included in the modeling as they are required to reproduce the observed multiple image configurations. Here we show 3 panels with 
examples from each of the 3 clusters lenses studied here Abell 2744, MACSJ\,0416 and MACSJ\,1149. The deeper HSTFF data reveal fainter cluster members as well as fainter lensed sources.}
\end{figure}

\subsection{Mass distribution in MACSJ\,0416}

MACSJ\,0416 (at $z=0.397$) was discovered as part of the MAssive Cluster Survey \citep[MACS;][]{ebeling01} and is classified as a merging system based on its double-peaked X-ray surface brightness distribution \citep{mann12}. Due to its exceedingly large Einstein radius, MACSJ\,0416 was selected as one of the five high-magnification clusters in the Cluster Lensing And Supernova survey with Hubble \citep[CLASH;][]{postman12}, thus providing HST imaging in 16 bands from the UV to the near-IR, with a typical depth of 1 orbit per passband. As expected for a highly elongated mass distribution quite typical of merging clusters, many multiple-image systems are produced and detected. The first detailed mass model of this complex merging cluster system was based on CLASH data, and was published by \cite{zitrin13}. This cluster was then selected as one of the six HSTFF targets. We constructed a high-precision mass model of the galaxy cluster MACSJ\,0416, based on a strong gravitational-lensing analysis of the HSTFF imaging data. Taking advantage of the unprecedented depth provided by HST/ACS observations in three passbands, we identified 51 new multiply imaged galaxies, quadrupling the previous census and bringing the grand total to 68, yielding a total of 194 individual lensed images. Having selected a subset of the 57 most securely identified multiply imaged systems, we obtain a best-fit mass model for the inner core of the cluster, consisting of two large-scale dark-matter halos and 98 accompanying galaxy-scale halos \citep{jauzac14}. This model predicts image positions with an RMS error of 0.68$\arcsec$, which constitutes an improvement of almost a factor of two over previous, pre-HFF mass models of this cluster. We find the total projected mass inside a 200 kpc aperture to be $(1.60 \pm0.01) \times 10^{14}\,\Msun$, a measurement that offers a three-fold improvement in precision from \cite{jauzac14}. The virial radius of MACSJ\,0416 extends out to $\sim$ 2.3 Mpc, here we note that the lens model reconstruction using HSTFF data extends radially out only to a fraction of the virial radius to $\sim\,{0.3\,R_{\rm vir}}$. Table 2 below lists the details of the best-fit lensing mass model for MACSJ\,0416.

\begin{table}
\begin{center}
\begin{tabular}[h!]{cccc}
\hline
\hline
\noalign{\smallskip}
Component  & \#1 & \#2 & L$^*$ elliptical galaxy \\
\hline
$\Delta$ \textsc{ra}  & $-5.9^{+0.4}_{-0.4}$  &  $23.6^{+0.3}_{-0.2}$ & --  \\
$\Delta$ \textsc{dec} & $3.5^{+0.3}_{-0.3}$  & $-43.4^{+0.2}_{-0.4}$ & --  \\
$e$ & 0.77 $\pm$0.01 & 0.64$\pm$ 0.01 & -- \\
$\theta$ & 147.3$^{+0.6}_{-0.7}$  & 126.8$^{+0.3}_{-0.3}$ & -- \\
r$_{\mathrm{core}}$ (\footnotesize{kpc}) & 72.8$^{+3.5}_{-2.5}$  & 95.8$^{+2.8}_{-2.5}$  & [0.15] \\
r$_{\mathrm{cut}}$ (\footnotesize{kpc}) & [1000] & [1000] & 27.5$^{+4.6}_{-4.1}$ \\
$\sigma$ (\footnotesize{km\,s$^{-1}$}) &  729$^{+16}_{-9}$ & 974$^{+13}_{-8}$ & 190.9$\pm$ 7.3 \\
\noalign{\smallskip}
\hline
\hline
\end{tabular}
\caption{MACSJ0416 best-fit PIEMD parameters for the two large-scale dark-matter halos, as well as for the L$^{*}$ elliptical galaxy. The model is built using 139 multiple images, includes 175 galaxy-scale perturbers. The best-fit RMS is 0.54 $\arcsec$.
Coordinates are quoted in arcseconds with respect to $\alpha=64.0381013, \delta=-24.0674860$.
Error bars correspond to the $1\sigma$ confidence level. Parameters in brackets are not optimised.
The reference magnitude for scaling relations is $mag_{\rm{F814W}} = 19.8$.
}
\label{table_m0416}
\end{center}
\end{table}

As for the case of Abell 2744, we revised our published strong-lensing mass model taking advantage of spectroscopic redshifts provided by the GLASS collaboration \citep[][Wang et al.\ in prep.]{hoag16} as well as considering the votes from all other lensing teams on the selection of secure multiple images. We also replaced our initial colour-magnitude selected cluster member catalogue with the \cite{Grillo15} catalogue as mentioned earlier. Our current mass model from which we derive the SHMF presented here now includes 139 multiple images. Our best-fit mass model comprises 2 cluster-scale halos, combined with 175 galaxy-scale perturbers, and predicts the image positions with an RMS error of 0.54$\arcsec$. In a recent preprint, \cite{caminha16} 
present an updated mass model that includes additional spectroscopic redshifts from archival Multi-User Spectroscopic Explorer (MUSE) data, and they report that the cluster galaxy catalog and the inferred sub-halo population are in good agreement with their earlier catalog that we have used here.

\subsection{Mass distribution in MACSJ\,1149}

For the cluster MACSJ\,1149 at $z{=}0.545$ our current model best-fit to the HSTFF lensing observations includes 12 new multiply imaged galaxies, bringing the total to only 22, comprising therefore a total of 65 individual lensed images. Unlike the first two HFF clusters, Abell 2744 and MACSJ\,0416, MACSJ\,1149 does not appear to be as powerful a lens \citep[see][for a more detailed discussion]{jauzac16}. As suggested in our pre-HFF models of the cluster (Richard et al.\ 2014), the inferred mass distribution here is exceedingly complex requiring 5 separate large-scale components whose spatial distribution and low masses make this cluster the least efficient and the least well constrained lens of the sample considered here. Our best-to-date model, which is due for significant improvements comprises of 5 large-scale clumps and 217 galaxy scale mass components. Our best-fit model predicts image positions with an RMS of 0.91$\arcsec$ which is larger than the RMS for our reconstructions of the mass in Abell 2744 and MACSJ\,0416. We estimate the total projected mass inside a 200 kpc aperture to be $(1.840 \pm 0.006) \times 10^{14}\,\Msun$. The integrated mass however reaches comparable precision with our models of the other two clusters. Off-set from the center, the supernova {\it SN Refsdal} was detected in this cluster. Models from several independent groups including ours predicted that six multiple images would be produced \citep{kelly15,rodney15,sharon15,Grillo15}. The mass model used to infer the SHMF here successfully predicted the appearance of the one of the multiple images seen in December 2015.  The virial radius of MACSJ\,1149 is $\sim$ 2 Mpc and we note here that our lensing model reconstruction extends radially out only to a fraction of the virial radius to $\sim\,{0.3\,R_{\rm vir}}$. Table 3 below lists the details of the best-fit lensing mass model for MACSJ\,1149.

\begin{table*}
\begin{center}
\begin{tabular}[h!]{cccccccc}
\hline
\hline
\noalign{\smallskip}
Clump  & $\Delta$ x  & $\Delta$ y & $e$ & $\theta$ & r$_{\mathrm{core}}$ (\footnotesize{kpc}) & r$_{\mathrm{cut}}$ (\footnotesize{kpc}) &$\sigma$ (\footnotesize{km\,s$^{-1}$})\\
\noalign{\smallskip}
\hline
\hline
\noalign{\smallskip}
\noalign{\smallskip}
\#1 &  $-1.95^{+0.10}_{-0.19}$  & 0.17 $^{+0.15}_{-0.22}$  &  0.58 $\pm$0.01  &  30.58$^{+0.35}_{-0.51}$  & 112.9$^{+3.6}_{-2.1}$  & [1000] & 1015$^{+7}_{-6}$ \\
\noalign{\smallskip}
\hline
\noalign{\smallskip}
\#2 &  -28.02$^{+0.26}_{-0.17}$ & -36.02$^{+0.27}_{-0.21}$ & 0.70$\pm$0.02 & 39.02$^{+2.23}_{-1.69}$ & 16.5$^{+2.7}_{-3.9}$  & [1000] & 331$^{+13}_{-9}$\\
\noalign{\smallskip}
\hline
\noalign{\smallskip}
\#3 & $-48.65^{+0.13}_{-0.49}$  & $-51.35^{+0.30}_{-0.22}$  &  0.35 $\pm$0.02  &  126.48$^{+7.11}_{-4.42}$  &  64.2$^{+6.8}_{-9.6}$  & [1000] & 286$^{+24}_{-16}$ \\
\noalign{\smallskip}
\hline
\noalign{\smallskip}
\#4 &  $17.62^{+0.28}_{-0.18}$  & 46.90 $^{+0.36}_{-0.28}$  &  0.15 $\pm$0.02  & 54.66$^{+3.51}_{-4.83}$  &  110.5$^{+1.2}_{-2.1}$  & [1000] & 688$^{+9}_{-17}$ \\
\noalign{\smallskip}
\hline
\noalign{\smallskip}
\#5 &  $-17.22^{+0.17}_{-0.18}$  & 101.85 $^{+0.08}_{-0.07}$  &  0.44 $\pm$0.05  &  62.29$^{+5.14}_{-4.61}$  &  2.1$^{+0.5}_{-0.1}$  & [1000] & 263$^{+8}_{-7}$ \\
\noalign{\smallskip}
\hline
\noalign{\smallskip}
\#6 &  [0.0]  & [0.0]  &  [0.2]  &  [34.0]  &  3.95$^{+0.57}_{-0.89}$  & 92.08$^{+6.50}_{-7.91}$ & 284$\pm$8 \\
\noalign{\smallskip}
\hline
\noalign{\smallskip}
\#7 &  [3.16]  & [-11.10]  &  0.22 $\pm$0.02  &  103.56$^{+7.09}_{-7.95}$  &  [0.15]  & 43.17$^{+1.34}_{-1.02}$ & 152$^{+2}_{-1}$ \\
\hline
\noalign{\smallskip}
L$^*$ elliptical galaxy & --  & --  & -- & -- & [0.15] & 52.48$^{+2.17}_{-0.89}$ & 148$^{+2}_{-3}$ \\
\noalign{\smallskip}
\hline
\hline
\end{tabular}
\caption{MACSJ1149 best-fit PIEMD parameters inferred for the five dark matter clumps considered in the optimization procedure. Clumps \#6 and \#7 are galaxy-scale halos that were modeled separately from scaling relations, to respectively model the BCG of the cluster as well as the cluster member responsible for the four multiple-images of \textit{SN Refsdal}.
The model is built using 65 multiple images, includes 217 galaxy-scale perturbers. The best-fit RMS is 0.91 $\arcsec$.
Coordinates are given in arcseconds with respect to
$\alpha=177.3987300, \delta=22.3985290$. Error bars correspond to the $1\sigma$ confidence level.
Parameters in brackets are not optimized. 
The reference magnitude for scaling relation is $mag_{\rm F814W} = 20.65$. 
}
\label{table_m1149}
\end{center}
\end{table*}

\section{Mapping substructure in Abell 2744, MACSJ\,0416 and MACSJ\,1149}

With the exquisite data from the HSTFF program for these clusters, and during the process of constructing the highest resolution mass models to date, we obtained constraints on subhalos associated with cluster member galaxies. Using the \textsc{Lenstool} software, the mass is partitioned into the large-scale clumps and galaxy-scale subhaloes as permitted by the input observed lensing constraints. \textsc{Lenstool} uses a Bayesian scheme with an MCMC algorithm to provide the best-fit suite of models given the priors while delineating the degeneracies amongst model parameters \citep{jullo07}. 

In Figure~2, we plot the luminosity distribution of the selected cluster members in all 3 clusters. We adopted a uniform luminosity cut and selected all cluster members with $L\,>0.01L_*$. This however yields different numbers for each cluster given their redshifts. With HSTFF data and our selected perturbing cluster members, we are able to now push down to two orders of magnitudes in mass below previous work in the determination of the SHMF. Here we present in Figure~3 the derived mass function of substructure over a mass range that spans $\sim 10^{9.5} - 10^{13}\,\Msun$ derived from the HSTFF data. This is remarkable as the $10^{9.5}\,\Msun$ clumps correspond to the dark matter halos associated with extremely low luminosity cluster members that are essentially dwarf galaxies. 

In Figure~1, we explicitly show thumbnails of low-luminosity cluster members that lie close to mulitple image systems in all three clusters. In many instances (of which only 3 are shown in Figure~1), these low luminosity cluster galaxies that lie in close proximity to multiply imaged systems are needed to accurately reproduce the geometry of some of the lensed images. Therefore, we include all low-luminosity cluster members down to $0.01\,L*$ in the lens modeling in our catalog of cluster members. And these clearly get folded in and contribute to output of our Bayesian analysis. This mass range of galaxies has not been accessible with any prior data-set of cluster lenses \citep{morishita16}.  We have also determined and overplot the errors arising from modeling in the derived SHMF for all three clusters. To compute these we used the standard deviation on the derived output values of the fiducial parameters ($r_{\rm t*}$ and $\sigma_{0*}$) for each cluster member from large number of models rather than just the best-fit model from the bayesian analysis. In Figure~2, we plot the luminosity distribution of the selected cluster members in all 3 clusters. We adopted a uniform luminosity cut and selected all cluster members with $L\,>0.01L_*$. This however yields different numbers for each cluster given their redshifts. With HSTFF data and our selected perturbing cluster members, we are able to now push down to two orders of magnitudes in mass below previous work in the determination of the SHMF. Here we present in Figure~3 the derived mass function of substructure over a mass range that spans $\sim 10^{9.5} - 10^{13}\,\Msun$ derived from the HSTFF data. This is remarkable as the $10^{9.5}\,\Msun$ clumps correspond to the dark matter halos associated with extremely low luminosity cluster members that are essentially dwarf galaxies. As is apparent from Figure~3, the HSTFF data do provide unique insights on the smallest galaxy halos that contribute to cluster lensing. We note that our completeness extends to $\sim 10^{10} \,\Msun$ in subhalo masses. While the mass function at lower masses is far from complete as is clear from the plots, we are for the first time obtaining an inventory, even if partial, on these scales within cluster lenses. These deep HSTFF data have offered a dramatic gain compared to earlier determinations of the small-scale substructure within clusters from WFPC-2 data.

\begin{figure}
\includegraphics[width=1.0\columnwidth]{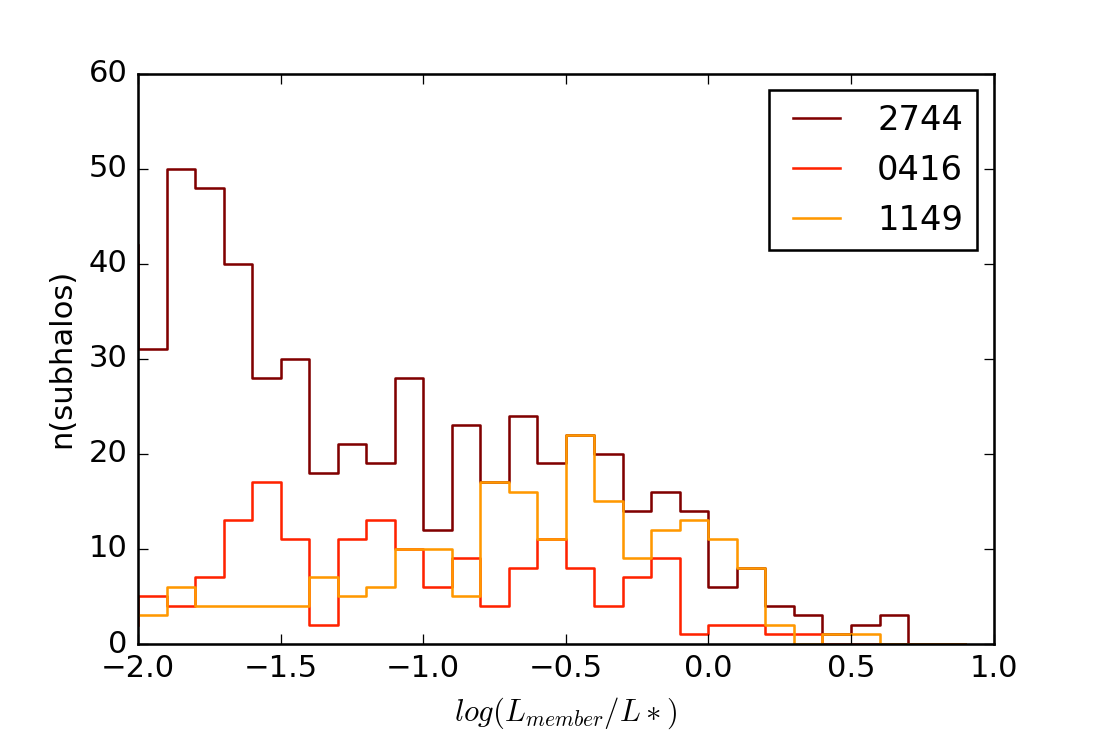}
\caption{The cluster galaxy luminosity selection for our analysis in Abell 2744, MACSJ\,0416 and MACSJ\,1149. As seen here clearly, the larger number of
identified strong lensing systems and their modeling plus the lower redshift of Abell 2744 allows us to probe more cluster galaxies down to $0.01\,L^*$; while the smaller number of strong lensing systems identified in MACSJ\,0416 and MACSJ\,1149 provide constraints on fewer cluster members down to the same limit luminosity cut. We use the selected number of 563 cluster galaxies in Abell 2744, 175 in MACSJ\,0416 and 217 in MACSJ\,1149 in all further analysis presented in this work.}
\end{figure}

\begin{figure}
\includegraphics[width=1.0\columnwidth]{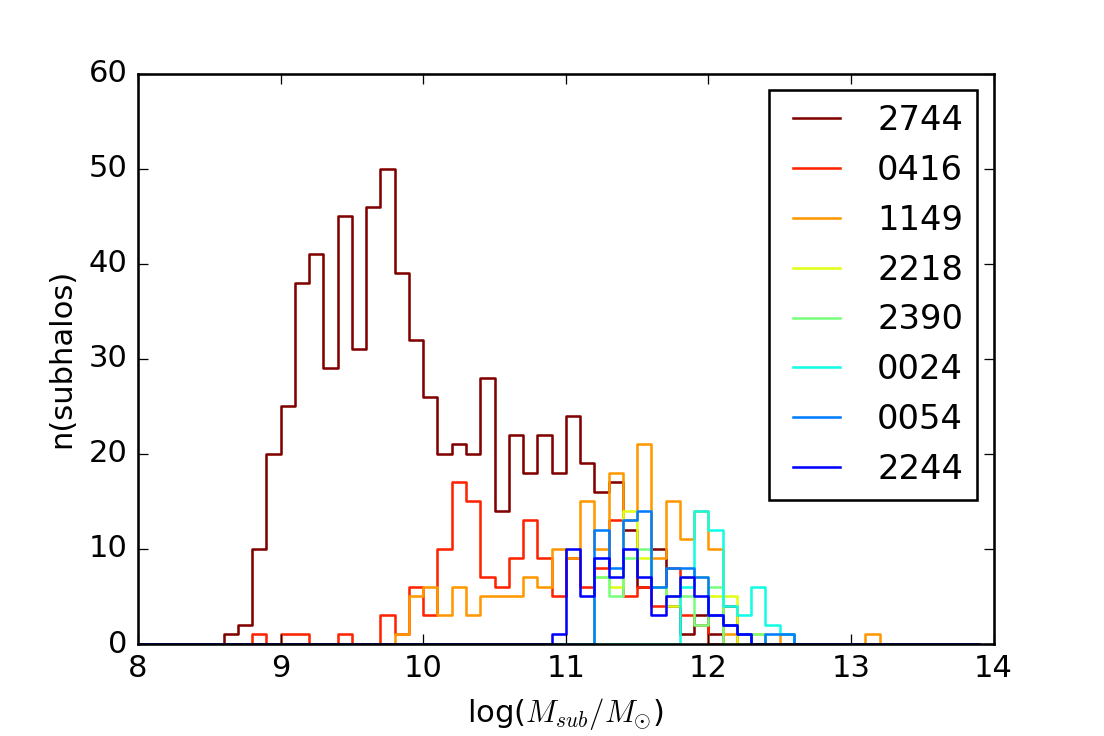}
\caption{The comparison of the SHMF derived from our current analysis of HSTFF data for Abell 2744, MACSJ\,0416 and MACSJ\,1149 with prior determinations from shallower HST data for the massive lensing clusters Abell 2218, CL0024, CL0054, Abell 2390 and CL2244 that are at a similar redshift range.}
\end{figure}

Note that there is a lot of cluster to cluster variation in the amount of substructure, and this is primarily due to the particulars of the geometry - the direction of elongation of the mass distribution, the differences in environment, redshift as well as dynamical state. Abell 2744 for instance, is a complex system with 3 other comparable mass sub-clusters actively undergoing a merger while MACSJ\,0416 consists of 2 large scale clumps that also appear to be merging. MACSJ\,1149 on the other hand has the least number of multiple images and the most complex mass distribution that at present is best fit with 5 merging subclusters, however these 5 components are all much less massive compared to the ones inferred to be merging in Abell 2744 and MACSJ\,0416.

\section{The Illustris simulations}

In this work, we compare the lensing data to a series of galaxy clusters simulated in a full cosmological context with the code AREPO \citep{springel10} that includes gravity, the hydrodynamics of gas and a series of of sub-grid prescriptions for star formation and feedback implementing the physics of galaxy formation. These constitute our full physics runs (abbreviated as FP). Here we focus on comparing our lensing derived SHMFs with the following sets of simulations: (i) a sample of cluster scale haloes with masses of about $10^{14}\,\Msun$ (in fact, we select halos with masses $\geq 10^{14}\,\Msun$ extracted from the Illustris Simulation \citep{vogelsberger14,Genel14,nelson15}; and (ii) ÒzoomÓ simulations of two massive cluster haloes with masses of $7\times 10^{14}\,\Msun$ and $2 \times 10^{15}\,\Msun$ respectively, chosen from the iCluster Simulation Suite (Popa et al. in prep; Pillepich et al. in prep). These will be referred to hereafter as $10^{14}$ {\it  Illustris Haloes} and {\it iCluster Zooms}. Both sets follow collisionless dark-matter and an equal initial number of baryons, and include an identical set of  physically motivated sub-grid models to implement galaxy formation \citep{vogelsberger14a,torrey14}. In both instances, the simulations have been setup and run with cosmological parameters consistent with WMAP9 results \citep{hinshaw13}, from an initial redshift of $z=126$ to $z=0$. In this work, we will also make use of the corresponding dark-matter only runs (DM), with identical initial conditions as the full physics runs described above but performed by taking only gravity into account. Illustris represents a state-of-the-art simulation of a $106.5\,{\rm Mpc^{3}}$ cosmological box and it is currently one of the highest resolution cosmological runs which simultaneously follows the evolution of haloes and galaxies all the way up to objects with total mass of about $2 \times 10^{14}\,\Msun$ (with a gravitational spatial resolution of about 1kpc and dark-matter mass resolution of about $6 \times 10^6\,\Msun$). From these runs 136 snapshots of output data are available, 36 of which are at redshift $z < 0.6$ and with an average time resolution of about $150-200\,{\rm Myrs}$. More information on the Illustris Simulation Suite can be found at http://www.illustris-project.org. 

The {\it iCluster Zooms} extend the massive range of the Ilustris box to haloes of about $10^{15}\,\Msun$. To compare with the HSTFF cluster lenses, here we use the $10^{15.3}\,\Msun$ [{\it iCluster Zoom 1}] and the $10^{14.8}\,\Msun$ [{\it iCluster Zoom 2}]. These iCluster Zooms that were selected from the {\it Millennium XXL} simulation box (with a size of 4.1 Gpc a side) and then re-simulated with AREPO and WMAP9-consistent cosmology with the so-called zoom technique utilizing the same Illustris galaxy formation model \citep[][see Popa et al in prep.; Pillepich et al. in prep. for more details]{angulo12}.  For these runs 256 output snapshots are available, with 73 snapshots below $z \sim 0.6$ with an average time spacing of about 70-80 Myrs. Cluster scale haloes in both the Illustris and iClusters Simulations are identified using the Friends-of-Friends algorithm (\cite{davis85}); bound sub-haloes within them are then identified using the SUBFIND algorithm \citep{springel01}, which in turn provides sub halo catalogs for the numerically determined SHMFs. The key motivation to choose the $10^{14}$ {\it  Illustris  Haloes} and {\it iCluster Zooms} for comparison to HSTFF data derives from the fact that they have been generated using a state-of-the-art baryonic and galaxy formation model which has been shown to reproduce fairly realistic populations of galaxies and which therefore allows us to straightforwardly mimic the luminosity selection criteria applied to the observational data. By contrast, in earlier such comparison work with cosmological simulations utilizing the {\it Millennium} run, we had to contend with post-hoc semi-analytic prescriptions for galaxy formation to replicate our selection criteria. Moreover, current availability of the full physics runs as well as the dark-matter only counterparts in the Illustris suite also offer us a unique handle to assess the effects of baryonic physics on the underlying SHMFs. 

The lensing derived enclosed mass within an aperture of 200 kpc is $\sim 2.1 \times 10^{14} \Msun$ for Abell 2744;  mass within an aperture of 200 kpc is $\sim 1.6 \times 10^{14} \Msun$ for MACSJ\,0416 and within 200 kpc is $\sim 1.8 \times 10^{14} \Msun$ for MACSJ\,1149. Only the $10^{15.3}$ {\it iCluster Zoom} run is truly comparable in terms of mass, and indeed in what follows the comparison between the FF data and simulations will be mostly focused on substructure derived from this halo. Yet, despite the overall mis-match in cluster masses of the Illustris box haloes, the $10^{14}${\it Illustris clusters} sample can provide a sense for the statistics as it comprises a total of 325 clusters across the entire redshift range $z = 0.2 - 0.6$, which we analyze and compare with the HSTFF sample to study the dependence of the SHMF on parent halo mass.

The {\it iCluster Zooms} employ a fixed co-moving softening length for the highest resolution dark matter particles $\epsilon_{\rm DM} = 2.84 {\rm kpc}$; a softening length for baryonic collisionless particles (stars and black holes) that is capped at a maximum value of $\epsilon_{\rm baryon} = 1.42 {\rm kpc}$. An adaptive softening scheme is adopted for the gas cells, wherein the softening length is proportional to the cell size. The mass resolution for the gas cells located in the high resolution region are successively refined to lie within a factor of 2 of the mass of baryonic particles (roughly $10^7\,\Msun$ while the masses of the high resolution dark matter particles are kept fixed at $\sim 5.8 \times 10^7\,\Msun$. An appropriately cascading scale is adopted for the medium and low resolution dark matter particles that get refined at subsequent levels. Details of the refinement scheme and the re-simulation methodology can be found in Popa et al. (in preparation) and Pillepich et al. (in preparation). 

Cataloging the abundance and mass spectrum of sub haloes bound to these selected cluster scale halos, we compare these LCDM predicted properties to those derived directly from the lensing data. For each selected simulated cluster from the $10^{14}\,{\it Illustris Haloes}$, the SHMF was computed and the mean and standard deviation computed for the entire sample of clusters. In order to estimate the variance from the {\it iCluster Zooms}, we computed the dispersion adopting the following method.  We compute the SHMF for each zoom cluster at $\approx$ 20 snapshots, corresponding to the $z_{cluster}\,\pm\,$0.1. Each SHMF is then scaled by the ratio of the halo mass at the cluster redshift snapshot to that at the given i-th snapshot, thus correcting for evolution with halo mass. The minimum and maximum SHMF thus obtained define the boundaries of the scatter that is plotted. Given that we have only two massive clusters that are truly comparable to the HSTFFs, this is the scheme we adopted to at arrive a rough estimate of the scatter.

From the $10^{14}$ {\it Illustris Haloes} and {\it iCluster Zooms}, we selected only the subhaloes that hosted luminous cluster members. We compare the data and simulations within the same projected area as spanned by the HSTFF lens models. he lens models extend out only to a fraction of the virial radius to only about $\sim\,30- 50\%$. Within this region we then count match by mimic-ing the selection of the same number of bright galaxies from the simulation and construct the mass function of their host dark matter sub haloes. TThe results of the comparisons of the count-matched sub haloes are detailed in the next section. In order to make an abundance matched comparison with our HSTFF datasets, we selected the subhaloes that hosted the 733 brightest galaxies in the {\it iCluster Zooms} and the $10^{14}$ {\it Illustris Haloes} for Abell 2744, 175 galaxies for comparison with MACSJ\,0416 and 217 for MACSJ\,1149. Ideally, we would have liked to adopt the same magnitude cut for cluster members in Illustris as done with the observational data in the K-band to select the equivalent  simulated cluster members. However, since the Illustris simulations are unable to match the observed luminosity function of cluster galaxies at these epochs we do not adopt this scheme. Work to improve the match with observed cluster galaxies is actively on-going within the Illustris collaboration (Pillepich et al., private communication). Instead, what we do is simply select the dark matter sub haloes that host the equivalent number of brightest cluster galaxies to compare with the lensing data. This in turn best mimics our observational selection. An important point has been made recently in the literature about the systematics introduced by the choice of halo-finder algorithm used in the determination of bound sub haloes from simulations and therefore the SHMF. \cite{onions12}, \cite{knebe13} and \cite{vandenbosch16} have shown that SHMFs determined by different halo-finders agree only to within $\sim$ 20\% at the low mass end. At the massive end of the SHMF, they report that sub halo finders that identify using density criteria in configuration space can under-predict by more than an order of magnitude. We need to be attentive to these systematic effects arising from different methods used to identify sub haloes while interpreting our results. It is known that there is evolution in the properties of bound subhaloes with parent halo mass \citep{vandenbosch05}. Therefore, to illustrate this dependence on cluster mass, we also examined the subhalo properties in Illustris clusters with lower masses in the range of $10^{13.5}-10^{14}\,\Msun$, even lower than those considered in the $10^{14}$ {\it Illustris Haloes}. In order to understand the effect of halo-to-halo variance on the high mass end of the SHMFs we we used analytic predictions for a cluster halo with mass $\sim 10^{15} \Msun$ within the equivalent spatial region that best reflects the FOV of HSTFF data. 

Finally, it has to be noted that highly efficient massive lenses like those selected in the HSTFF sample tend to have complicated mass distributions, enhanced surface mass densities due to interactions and on-going mergers that in fact make them desirable targets for study. The peculiar dynamics of these merging sub-clusters, with several components interacting, reflect rare geometries and phase-space configurations: that are not available in the entire Illustris box and amongst the {\it iCluster Zooms} even though the zoom runs have comparable masses to the HSTFF clusters studied here.

\section{Comparison with Illustris simulations}

To compare the results of the lensing analysis with Illustris simulations, we focus primarily on the {\it iCluster Zooms}, particularly on the more massive {\it iCluster Zoom 1}. A dark matter only run as well as one implementing the full physics was performed for the {\it iCluster Zooms}. We also study the $10^{14}$ {\it Illustris Haloes}  in the box, with $M\,>\,10^{14}\,\Msun$ within $\delta z \pm0.1$ in redshift of each of the HSTFF clusters $(z \sim 0.3 - 0.6)$. Our analysis is centered on comparison with the {\it iCluster Zooms}.

First, we compare the lensing derived SHMFs for each of the three HSTFF clusters studied here with that derived from {\it iCluster Zoom 1} over the same projected area as the data at the appropriate redshift. The ACS image footprint corresponds in radius to approximately one third to one half the viral radius for these clusters. We select sub haloes in {\it iCluster Zoom 1} and for the $10^{14}$ {\it Illustris Haloes} from within the corresponding projected radii corresponding to the ACS footprint for each cluster. We then proceed to make a member galaxy count-matched comparison after imposing this radial cut. This selection of sub haloes associated with the number of brightest cluster galaxies in each of these cluster lenses is referred to as the count-matched SHMF hereafter. We caution here that it is known that the luminosity function of real clusters is not appropriately reproduced by these simulations at the present time. 

Substructure is ubiquitous in the {\it iCluster Zooms} as well as in the $10^{14}$ {\it Illustris Haloes} clusters and the SHMF as predicted by LCDM is dominated by low mass halos in terms of their abundance.  The HSTFF data have helped us push the mass scale of detected subhaloes down by two orders of magnitude. However, we still have a resolution limit of $10^{9.5}\,\Msun$, even prior to which incompleteness starts to set in as seen in Figure~3. In simulated LCDM clusters, the substructure mass function extends well below this limit than we cannot probe in these cluster lenses even at this exquisite depth. Given this we make two kinds of selections within simulations: (i) an abundance/count matched version - wherein we select the dark matter subhaloes associated with the brightest 563, 175 and 217 cluster galaxies as in Abell 2744, MACSJ\,0416 and MACSJ\,1149 respectively;  and (ii) a selection that includes all subhaloes with masses $M\,>\,10^{9}\,\Msun$ that contain a luminous component. In the Illustris $10^{14}$ {\it Illustris Haloes} (run that includes the full sub-grid physics) we have 137 clusters at the redshift ($z \sim 0.3$) of Abell 2744; 117 clusters at the redshift ($z \sim 0.4$) of MACSJ\,0416 and 66 clusters at the redshift ($z \sim 0.5$) of MACSJ\,1149. We now proceed compare the SHMF derived from the HSTFF data with those derived from the appropriate redshift {\it iCluster Zoom 1} snapshot for each of the three clusters.

\begin{figure} 
\includegraphics[width=1.0\columnwidth]{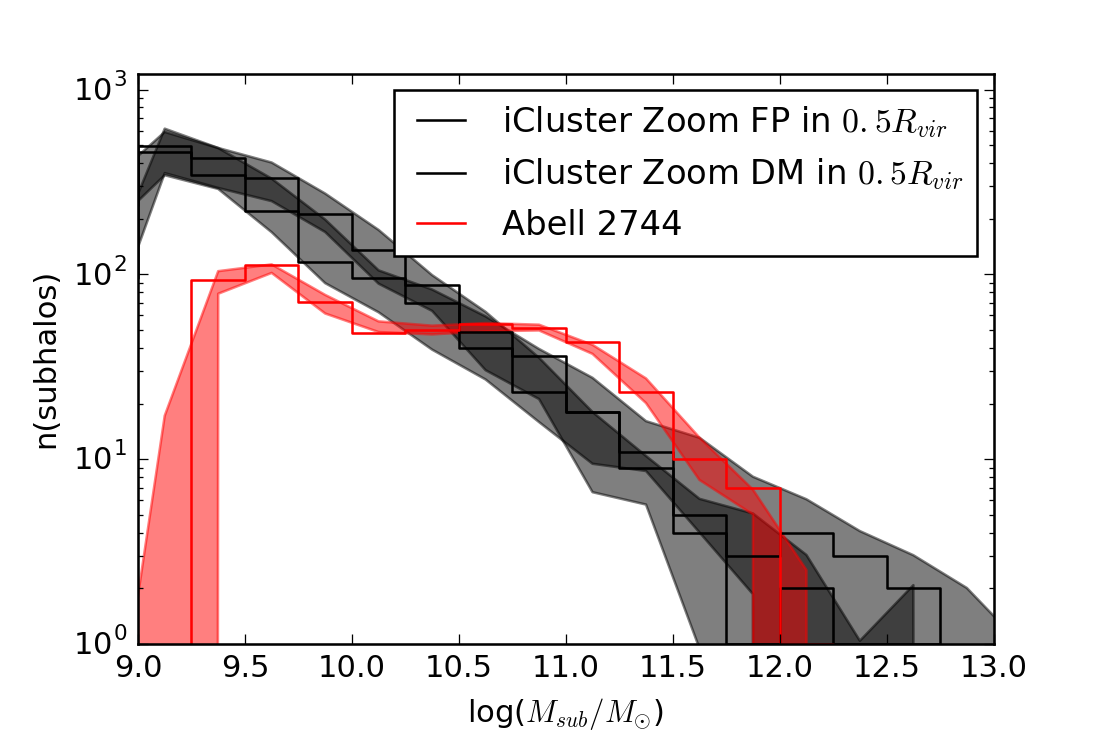}
\caption{Comparison of the SHMF derived for Abell 2744 (with overplotted modeling errors derived from the dispersion in N-Bayesian realizations for the key fiducial parameters $r_{\rm t*}$ and $\sigma_{0*}$) with Dark Matter only and Full Physics Illustris {\it iCluster Zooms}: The SHMF derived from the HST FF data (red histogram) for Abell 2744 is plotted along with that derived from the {\it iCluster Zooms}. The dark matter only run is plotted as a solid black histogram with the dispersion marked in the dark grey band and the full physics run (grey histogram) and corresponding dispersion shown as the light grey band. Sub haloes within $0.5\,R_{\rm vir}$ that corresponds to the FOV of ACS over which the lensing model has been reconstructed are extracted from {\it iCluster Zoom 1}. The full physics run of {\it iCluster Zoom 1} includes sub-grid models for the physical processes that are relevant to galaxy formation. From the full physics run we have selected only sub haloes that host a stellar component.}
\end{figure}

Firstly, we find that the SHMFs derived within $0.5\,{R_{\rm vir}}$ from the dark matter only {\it iCluster Zoom 1} snapshot and the full physics {\it iCluster Zoom 1} snapshot shown in Figure~4 are remarkably similar over the mass scales probed here. Both these SHMFs derived from the {\it iCluster Zoom 1} run are also in very good agreement with the lensing derived SHMF for Abell 2744 at the high-mass end and the diverge at the masses below $10^{10}\,\Msun$ due to the incompleteness in the lensing data. Further as seen in Figure~5, the lensing derived SHMFs with different assumptions for the fidelity of how light traces masses are also fairly similar. This suggests that given the current quality of data in the HSTFFs, the SHMF is fairly robust and cannot constrain the details of how galaxies populate dark matter halos for sub haloes more massive than $10^{10}\,\Msun$. Differences start to appear at lower sub halo masses. In Figure~6, we note that the count-matched SHMFs from Abell 2744 and {\it iCluster Zoom 1} are in excellent agreement both in amplitude and shape over 4 orders of magnitude in sub halo mass, from $10^{9-13}\,\Msun$, with a slight excess seen at $\sim 10^{11}\,\Msun$, which we show can be completely accounted for when  cosmic variance can be more accurately taken into account with the analytic calculation of the SHMF.
\begin{figure}
\includegraphics[width=1.0\columnwidth]{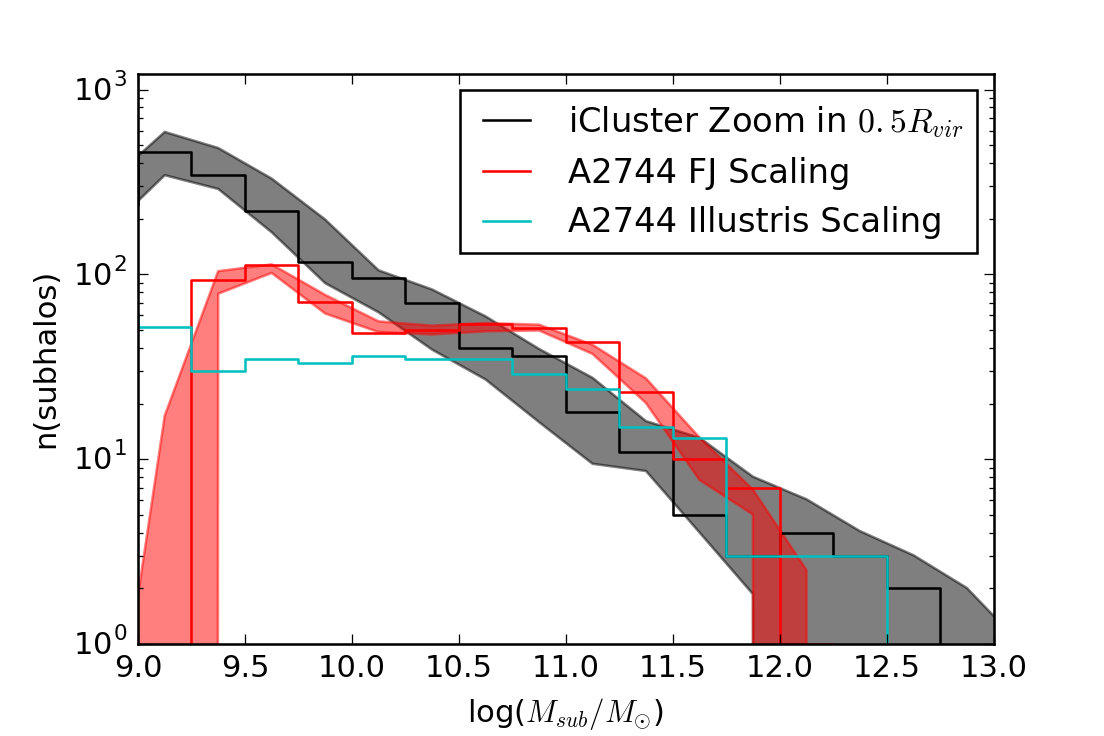}
\caption{Comparing lensing derived SHMFs (with estimated model errors) for different assumptions relating mass to light: here we plot the SHMFs derived from the best-fit lensing model for Abell 2744 using the Faber-Jackson and Kormendy luminosity scaling laws (FJ scaling), with that derived for the best-fit lens model using the scaling laws from the full physics run of the Illustris simulations (Illustris scaling).}
\end{figure}

In Figure~7, we plot the SHMF derived for the cluster lens Abell 2744 for the two scaling laws; that derived from {\it iCluster Zoom 1} as well as the analytically calculated SHMF for a $10^{15}\,\Msun$ cluster halo that now includes an estimate of the halo-to-halo scatter shown as the dull green band in Figure~7. We note the excellent agreement between the various independently determined SHMFs consistent with our estimate of cosmic variance.

We now examine the dependence of the SHMF on parent halo mass using the $10^{14}$ {\it Illustris Haloes} in Figure~8. As expected theoretically from the work of \cite{vandenbosch05}, the peak of the SHMF is sensitive to parent halo mass and tends to shift toward higher sub halo masses for more massive parent halos. The slope at the high mass end, however, appears to converge independent of the parent halo mass. The trends clearly show that the SHMF for Abell 2744 agrees best with that derived from the equivalent total mass cluster - the {\it iCluster Zoom 1} run. We investigate the role of various projections from the {\it iCluster Zoom 1} run to assess their contribution to the error budget in the derived SHMFs. For Abell 2744 as shown in Figure~9, we note that the SHMF derived from 3 independent projections from the {\it iCluster Zoom 1} snapshot converge for sub halo masses $> 10^{10.5}\,\Msun$ and the agreement in both slope and amplitude are excellent at the high mass end. Although the {\it iCluster Zoom 1} snapshot at $z \sim 0.3$ clearly offers the appropriate mass equivalent for the cluster lens Abell 2744, we find that there is considerable discrepancy when comparing the radial distribution of sub halos, shown in Figure~10. Simulated galaxies and consequently their host sub haloes appear to be much less concentrated in the inner regions compared to the real galaxies in Abell 2744. The mis-match in the radial distribution suggests that mass segregation is more efficient in observed cluster lenses while tidal stripping, tidal heating and dynamical friction might be over-efficient in simulations, leading to the dramatic reduction in the masses of in-falling sub haloes. Some of this disagreement could also arise from systematics introduced by algorithmic limitations of sub halo finders as noted earlier. What is clear though is that the HSTFF clusters represent transitory merging states of massive clusters that are not captured in simulation outputs. While the lensing signal itself is independent of the dynamical state of a cluster, the transient complex dynamics during an on-going merging event appears to alter the radial distribution of substructure significantly. In order to probe the role of the dynamical state, in future work, we intend to perform zoom-in runs while tracking the anatomy of the merger process by writing out output files more densely sampled in time, in particular just prior to and right after major sub-cluster mergers.

\begin{figure}
\includegraphics[width=1.0\columnwidth]{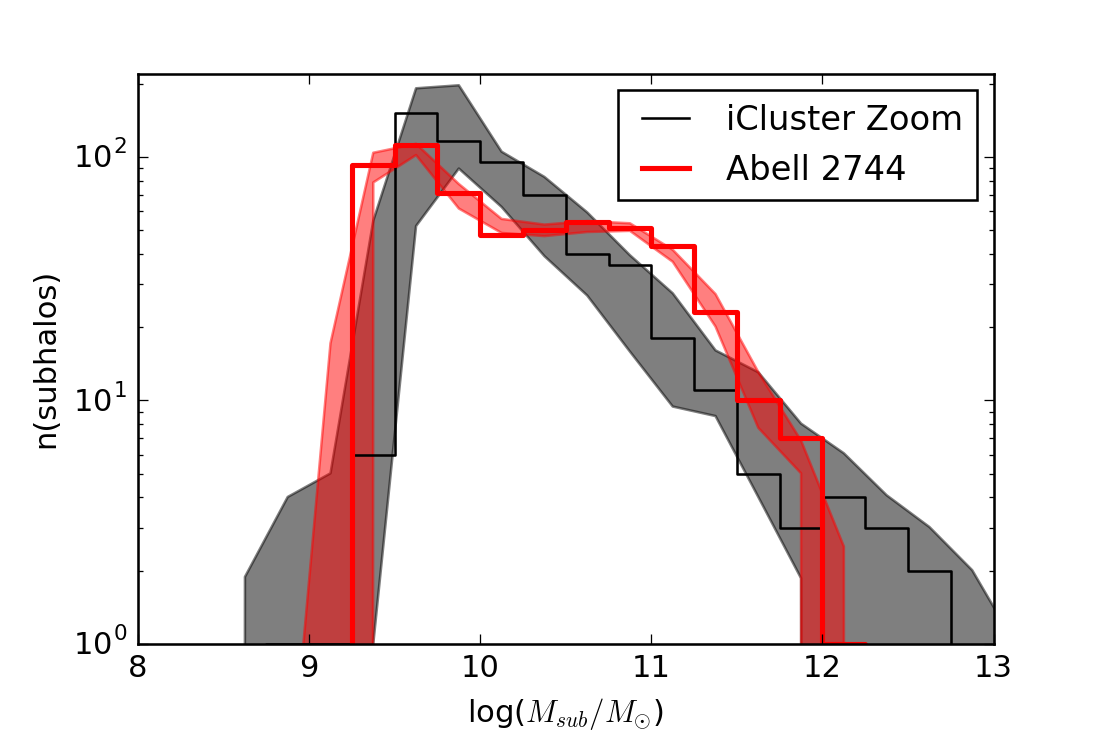}
\caption{Comparison of the count-matched SHMF derived for Abell 2744: the SHMF from the HST FF data (red histogram) and the count-matched SHMF from {\it iCluster Zoom 1} are plotted. Here mimicking the observational selection, only the dark matter subhaloes associated with the brightest 563 cluster galaxies that lie within $0.5\,R_{\rm vir}$ in the snapshot at $z \sim 0.3$ of {\it iCluster Zoom 1} are plotted.}
\end{figure}

\begin{figure} 
\includegraphics[width=1.0\columnwidth]{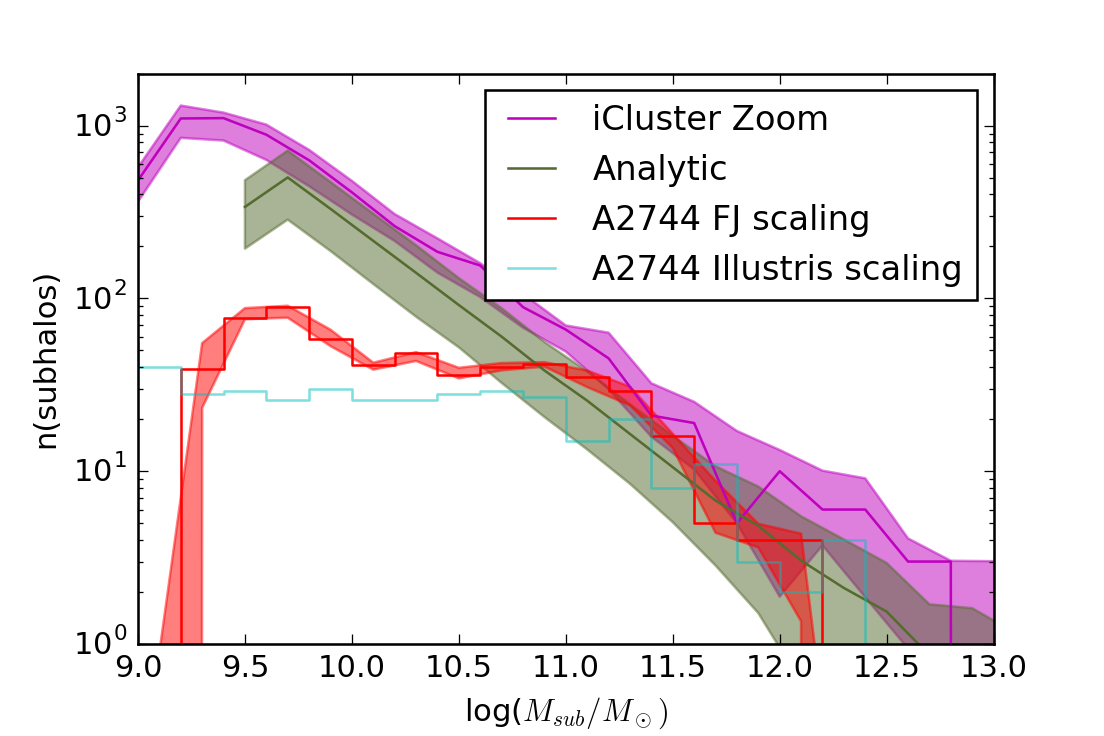}
\caption{Comparison with the analytically predicted SHMF: the lensing derived SHMF (with model errors) for the two independent scaling laws is overplotted along with the analytically calculated SHMF for a cluster halo with mass equivalent to that of Abell 2744. This analytic estimate includes the halo-to-halo scatter that is shown in the dull green band.}
\end{figure}

\begin{figure}
\includegraphics[width=1.0\columnwidth]{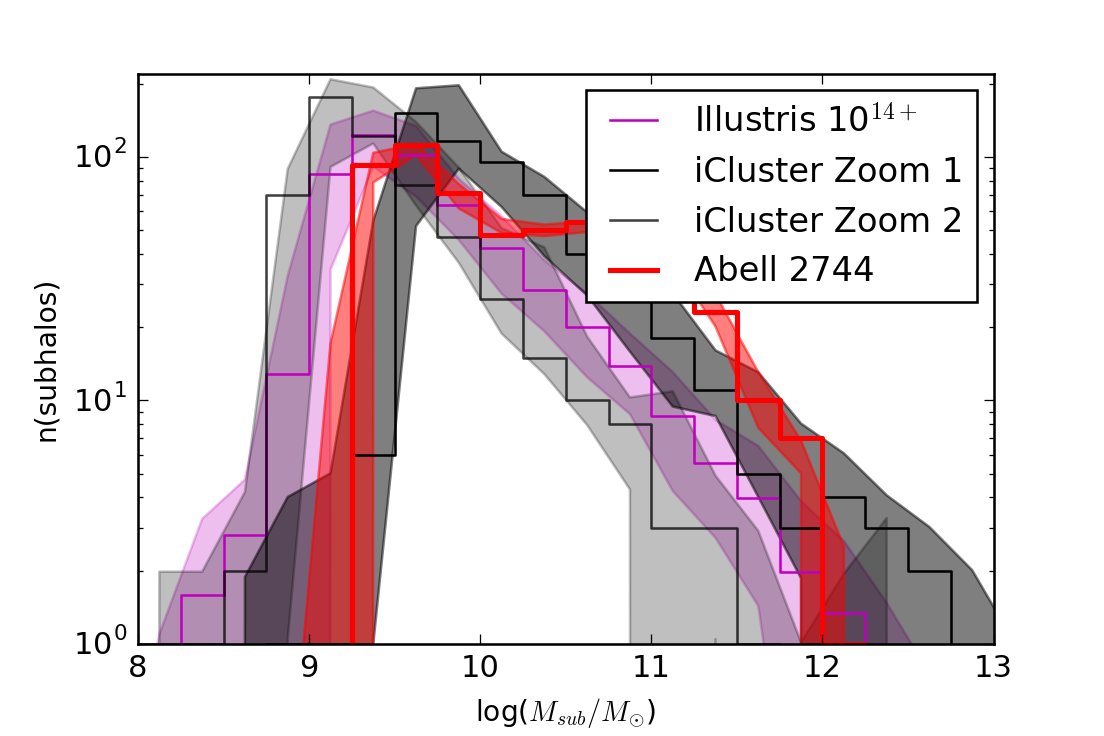}
\caption{Comparison of the SHMF derived for Abell 2744 (count-matched) with that of the SHMF derived from simulated massive clusters in Illustris to show dependence on parent halo mass. Here we plot the SHMF from the HST FF data (red histogram) and the count-matched SHMF from the two simulated zoomed in massive clusters with masses of $10^{15.3}\,\Msun$ {\it iCluster Zoom 1} and $10^{14.5}\,\Msun$ {\it iCluster Zoom 2} as well as from a larger sample of 137  $10^{14}\,\Msun$ {\it Illustris clusters}. Once again mimicking the observational selection, only the dark matter subhaloes associated with the brightest 563 cluster galaxies that lie within $0.5\,R_{\rm vir}$ - count matched to the HFF derived SHMF for Abell 2744 - are plotted.}
\end{figure}

\begin{figure}
\includegraphics[width=1.0\columnwidth]{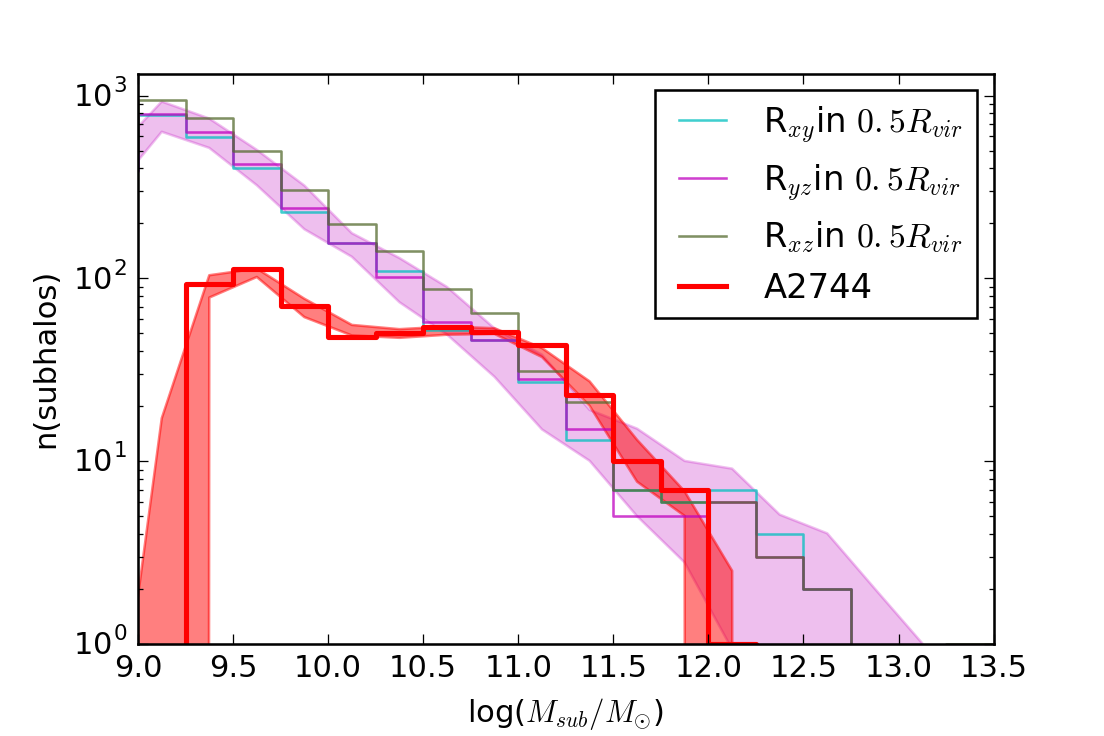}
\caption{Projection Effects on the derived SHMF from simulations: here we plot the SHMF derived from the zoomed in Illustris cluster {\it iCluster Zoom 1} derived by projecting along three distinct axes within $0.5\,R_{\rm vir}$. Note that the dispersion arising from projection effects is negligible.}
\end{figure}

\begin{figure}
\includegraphics[width=1.0\columnwidth]{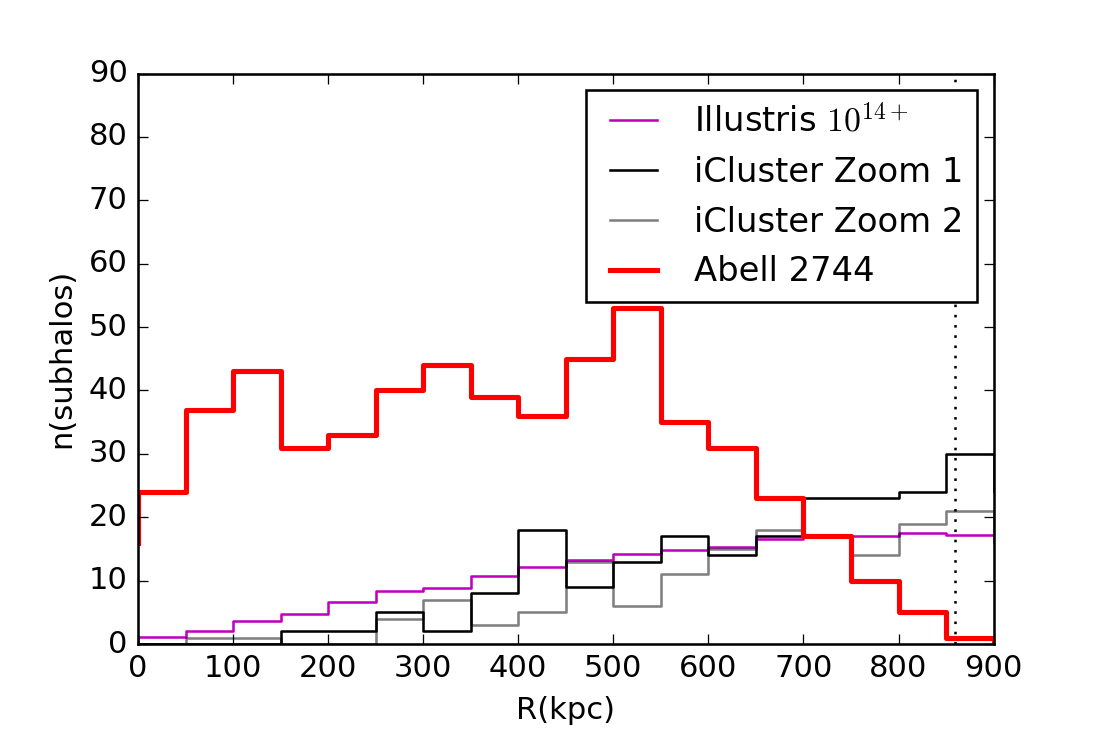}
\caption{Radial distribution of the SHMF derived from {\it iCluster Zoom 1} compared to that of the lensing derived SHMF for Abell 2744 from the HST FF data (red histogram). The snapshot was selected from the full physics run of {\it iCluster Zoom 1}. We clearly see that galaxies in {\it iCluster Zoom 1} are not as centrally concentrated as Abell 2744.}
\end{figure}

The mass distribution for MACSJ\,0416 is the best constrained of the 3 HSTFF clusters studied here since cluster membership has been largely spectroscopically determined. Abell 2744 has a larger number of identified cluster members (namely 563), though fewer of them are spectroscopically confirmed, while for MACSJ\,0416 despite having fewer selected cluster galaxies (numbering 175), they are all spectroscopically confirmed to be in the cluster.  Proceeding to compare the overall abundance and count matched version of the SHMF for MACSJ\,0416, once again we find excellent agreement with the appropriate redshift snapshot over the equivalent projected area corresponding to within $0.3\,R_{\rm vir}$ of {\it iCluster Zoom 1} grown in an LCDM cosmology. Once again at the high mass end of the SHMF, both the dark matter only snapshot and the full physics snapshot from {\it iCluster Zoom 1} agree rather well with the lensing derived SHMF (Figure~11). We note for MACSJ\,0416 as well, the weak dependence of the lensing derived SHMFs on the assumed scaling laws that are adopted to relate light and mass in the lensing inversion (Figure~12). There is once again, as for Abell 2744, strikingly good agreement between observations and simulations when we select dark matter sub haloes associated with the equivalent number of brightest cluster galaxies. In Figure~6, the SHMF derived from {\it iCluster Zoom 1} is number matched with the observational selection for Abell 2744 and includes only the DM haloes that host the brightest 563 cluster galaxies. The same is done with the appropriate redshift snapshot of {\it iCluster Zoom 1}, wherein only dark matter sub haloes hosting the brightest 175 galaxies within $0.3\,R_{\rm vir}$ are selected for comparison with the MACSJ\,0416 data, shown in Figure~13. The match is again excellent overall, and here too, we note a slight excess in the number of sub haloes derived from the HSTFF data compared to those selected from {\it iCluster Zoom 1} at $\sim 10^{11}\,\Msun$ as in the case for Abell 2744. However, upon overplotting the analytically determined SHMF that includes an estimate of the halo-to-halo scatter (Figure~14), the excess is entirely consistent with what is expected from cosmic variance. We find the same trends and dependence of the SHMF with parent halo mass for MACSJ\,0416 (Figure~15) as we did for Abell 2744. And once again projection effects do not scupper the robustness of the SHMF as a diagnostic of the underlying cosmological model (Figure~16). The radial distribution of sub haloes, that is essentially the radial distribution of early-type bright galaxies in MACSJ\,0416 is not reproduced by the simulated cluster {\it iCluster Zoom 1}. Once again, we find that sub haloes are more diffusely distributed in {\it iCluster Zooms} compared to the more concentration distribution seen in MACSJ\,0416 (Figure~17). 

In recently published work, \cite{Grillo15} independently determined substructure properties for MACSJ\,0416 employing the same methodology outlined here. The pre-HSTFF data mass model that they construct for this cluster using shallower CLASH survey data includes far fewer strong lensing constraints compared to our analysis here. However, for our analysis of the HSTFF data of MACSJ\,0416 presented here, we have used their cluster galaxy catalog as mentioned earlier. Adopting the same modeling methodology as us using \textsc{Lenstool}, their best-fit mass model also comprised of 2 large scale components and 175 galaxy scale components. Upon comparison of the derived mass function of dark matter substructure with a dark matter only cosmological simulation, they report a paucity of massive subhaloes in the inner regions of simulated LCDM clusters compared to the shallower CLASH data of MACSJ\,0416. However, as noted above, with the deeper HSTFF data for MACSJ\,0416 and performing a count-matched comparison with the full physics {\it iCluster Zoom 1} snapshot we find an excellent overall match with a slight excess that is fully accounted for by cosmic variance at sub halo masses of $\sim 10^{11}\,\Msun$ as well as systematics that derive from the choice of halo finding algorithm adopted for sub halo selection. In Figure~18 and Figure~19, we compare the radial distribution of sub haloes and their velocity dispersion distribution reported by Grillo et al. with our estimates derived from the {\it iCluster Zoom 1} run. We find very good agreement with the overall lens modeling of MACSJ\,0416 and consequently the inferred sub halo properties with their work. However, our better constrained mass model that incorporates many more observational constraints from the HSTFF data and comparison with {\it iCluster Zoom 1} that includes galaxy formation enables a more careful, detailed study of the SHMF. There are slight difference in the masses attributed to individual sub haloes by the best-fit models produced pre-HFF by Grillo et al. and our work here with HSTFF data for MACSJ\,0416, despite using the same lens inversion and analysis methodology. This is not unexpected, as the choice of priors adopted in the lens modeling does impact the final best-fit models that are derived. A comprehensive study comparing various mass modeling methodologies and their accuracy in the reconstruction of a simulated cluster where all components are known was undertaken recently, and the results of this extremely illuminating exercise are presented in \cite{Meneghettiff16}. In this study our model reconstruction of the simulated cluster {\it Ares} using the methodology described and adopted here was found to recover nearly unbiased substructure masses with good accuracy \citep[as shown in Figure~16 of][]{Meneghettiff16}.

\begin{figure}
\includegraphics[width=1.0\columnwidth]{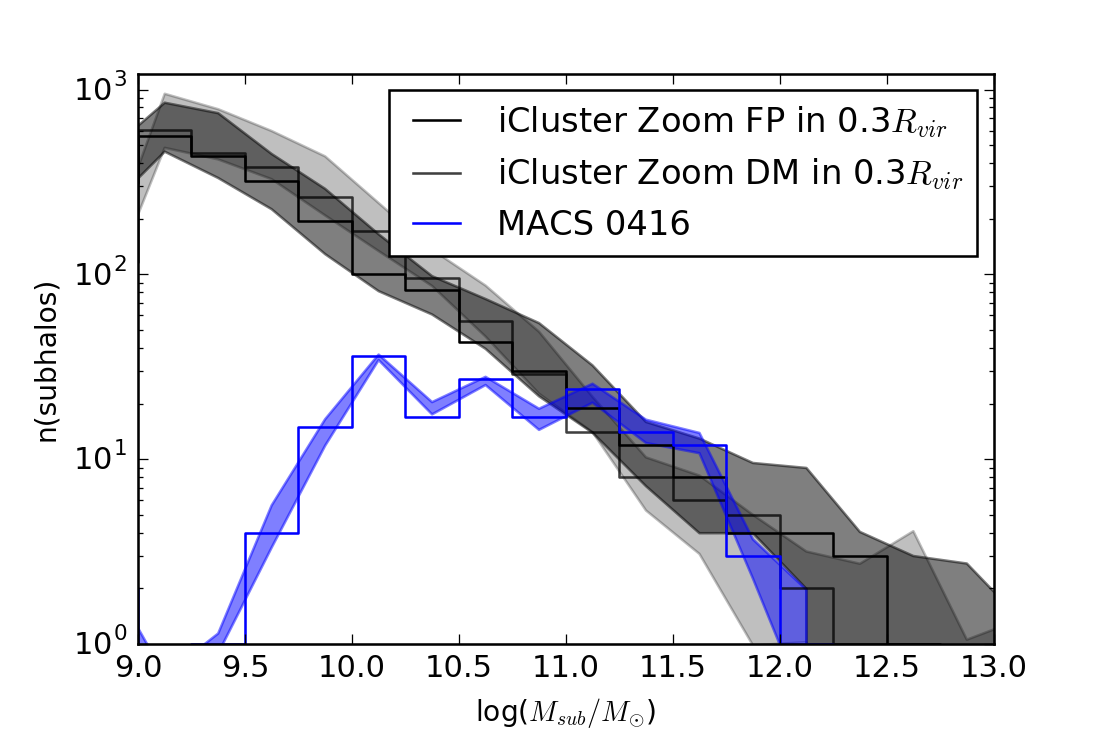}
\caption{Comparison of the SHMF derived for MACSJ\,0416 (with overplotted modeling errors derived from the dispersion in N-Bayesian realizations for the key fiducial parameters $r_{\rm t*}$ and $\sigma_{0*}$) from the HST FF data (blue histogram) with that derived from the dark matter only run of {\it iCluster Zoom 1} plotted as a solid black histogram with $\pm 1\sigma$ dispersion as the dark grey region. We also plot data from the SHMF derived from the full physics run of {\it iCluster Zoom 1} (dispersion around mean marked in lighter grey) within $0.3\,R_{\rm vir}$ that is equivalent to the ACS FOV over which the lensing model has been reconstructed. From the full physics run of {\it iCluster Zoom 1} we have selected all dark matter sub haloes that host a stellar component to derive the SHMF plotted here. The snapshots at $z \sim 0.4$ are plotted to compare with MACSJ\,0416.}
\end{figure}

\begin{figure}
\includegraphics[width=1.0\columnwidth]{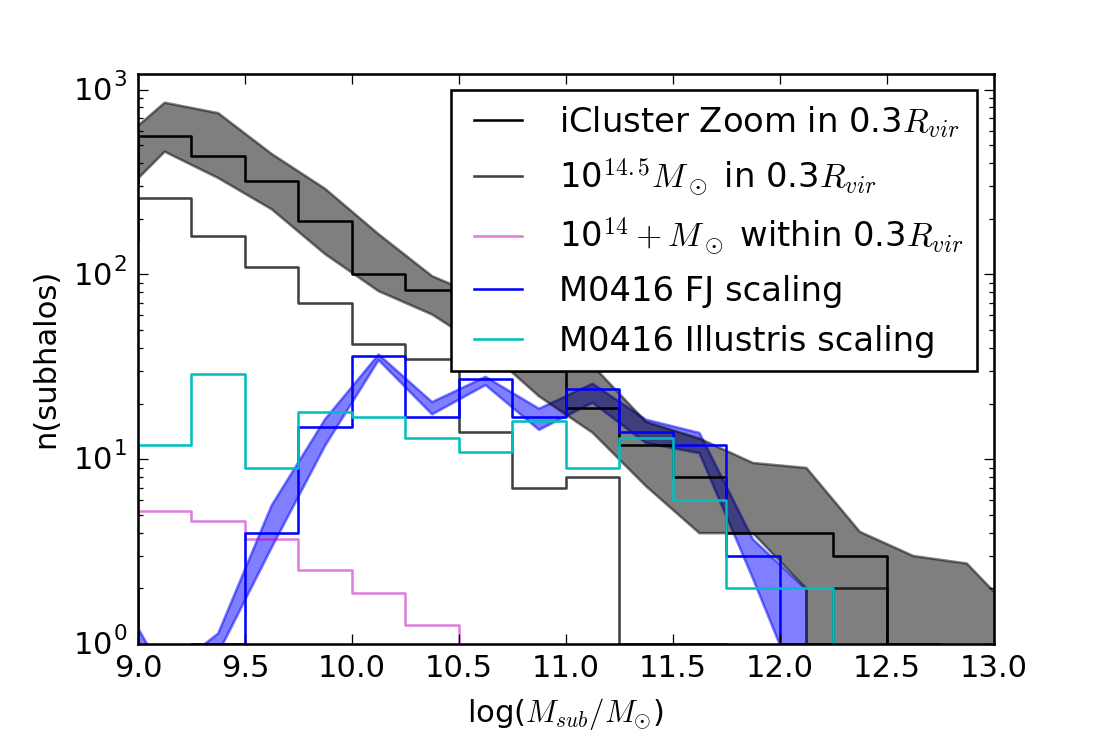}
\caption{Comparing lensing derived SHMFs (with model errors) for different assumptions relating mass to light: here we plot the SHMFs derived from the best-fit lensing model for MACSJ\,0416 using the Faber-Jackson and Kormendy luminosity scaling laws, with that derived for the best-fit lens model using the scaling laws from the full physics run of {\it iCluster Zoom 1}.}
\end{figure}

\begin{figure}
\includegraphics[width=1.0\columnwidth]{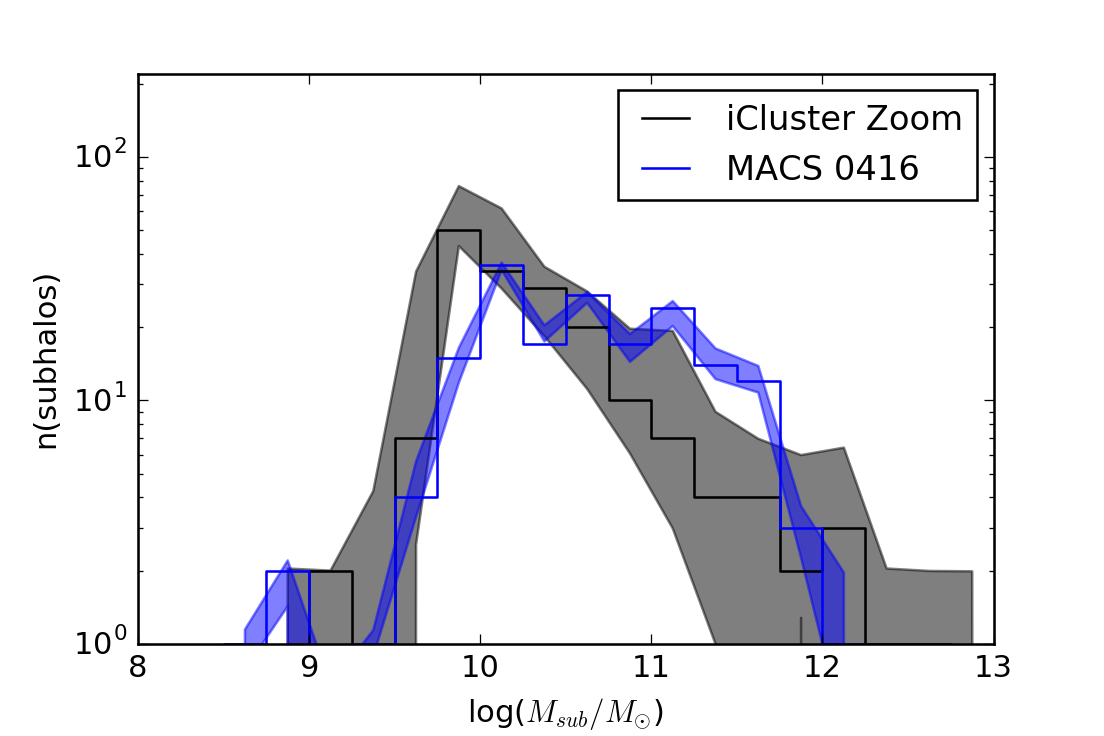}
\caption{Comparison of the count-matched SHMF derived for MACSJ\,0416 from the HST FF data (blue histogram) with that derived from {\it iCluster Zoom 1} (the $z \sim 0.4$ full physics run snapshot) now mimicking the observational selection.  Here only subhaloes that are associated with the brightest 175 cluster galaxies (number matched to the HSTFF data of MACSJ\,0416 are included within $0.3\,{\rm R_{vir}}$) in the solid black histogram. The $1\sigma$ cluster-to-cluster variation is once again plotted as the grey region.}
\end{figure}

\begin{figure} 
\includegraphics[width=1.0\columnwidth]{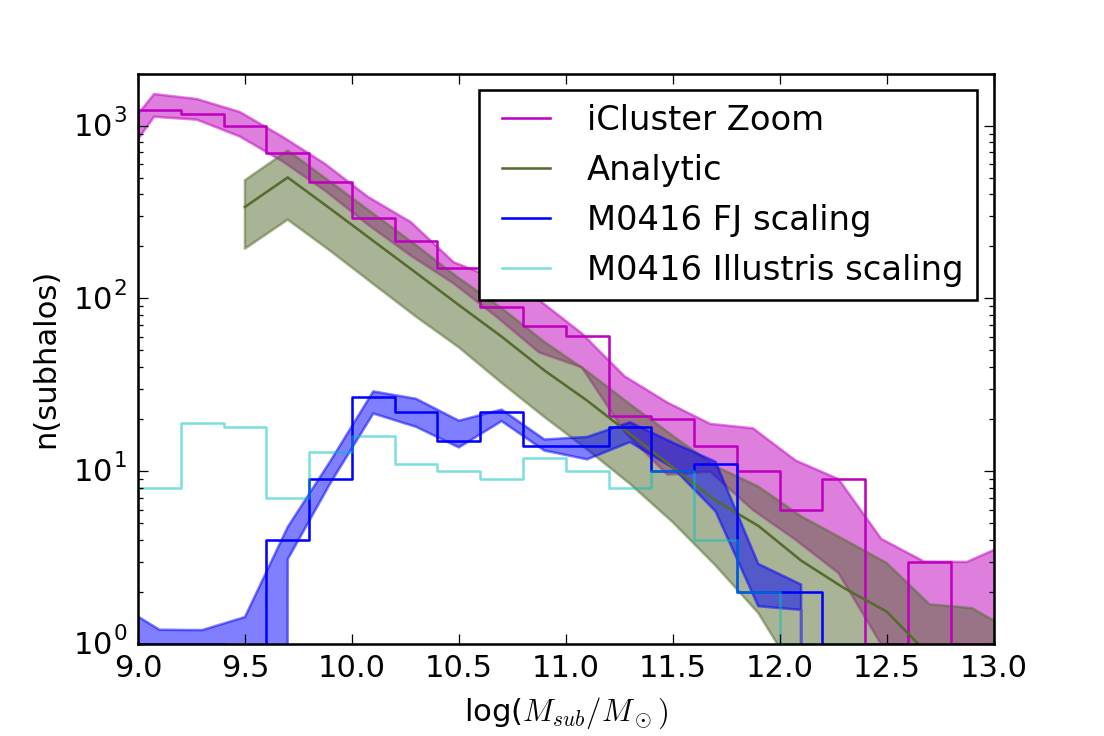}
\caption{Comparison with the analytically predicted SHMF: the lensing derived SHMF (with model errors)  for the two independent scaling laws is over-plotted along with the analytically calculated SHMF for a cluster halo with mass equivalent to that of MACSJ\,0416. This analytic estimate includes the halo-to-halo scatter that is shown in the dull green band.}
\end{figure}

\begin{figure}
\includegraphics[width=1.0\columnwidth]{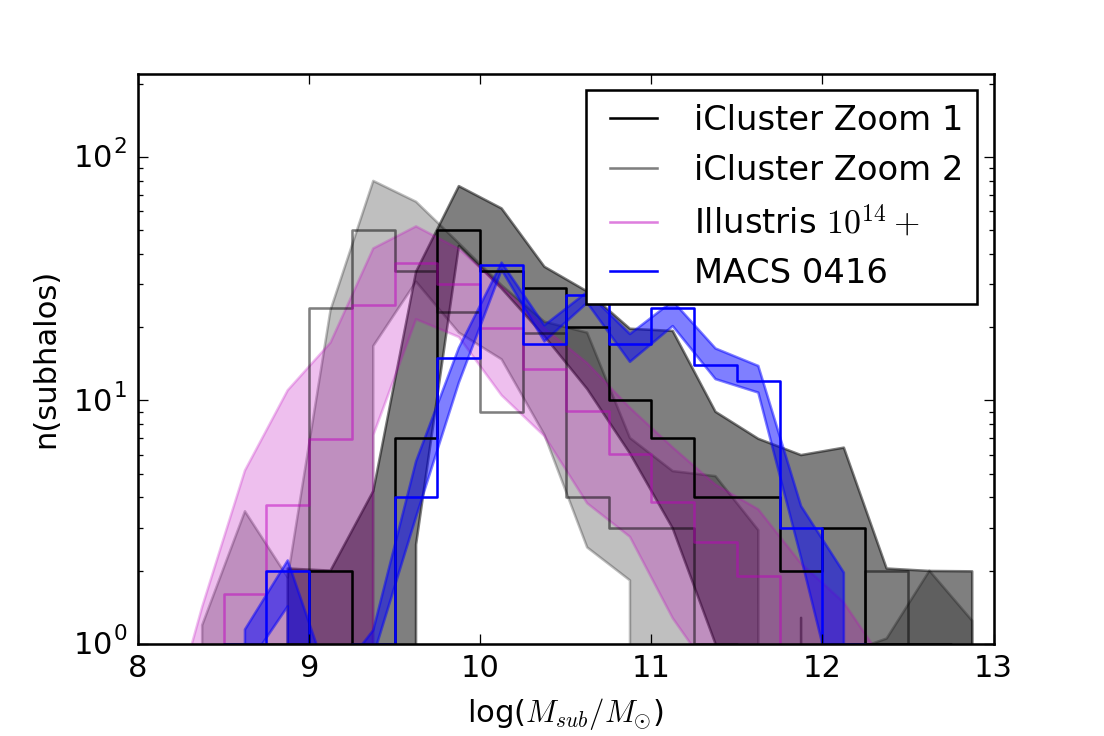}
\caption{Comparison of the SHMF derived for MACSJ\,0416 from the HST FF data (blue histogram) with that derived from the two massive zoomed-in simulated clusters {\it iCluster Zooms} (the $z \sim 0.4$ full physics run snapshots) and the $10^{14}$ {\it Illustris Haloes} of $M > 10^{14+}\,\Msun$ mimicking the observational selection in all of them.  Here only subhaloes that host the brightest 175 cluster galaxies (number matched to the HSTFF data of MACSJ\,0416 are included) selected from the simulations. The $1\sigma$ cluster-to-cluster variations are once again plotted as the bands.}
\end{figure}

\begin{figure}
\includegraphics[width=1.0\columnwidth]{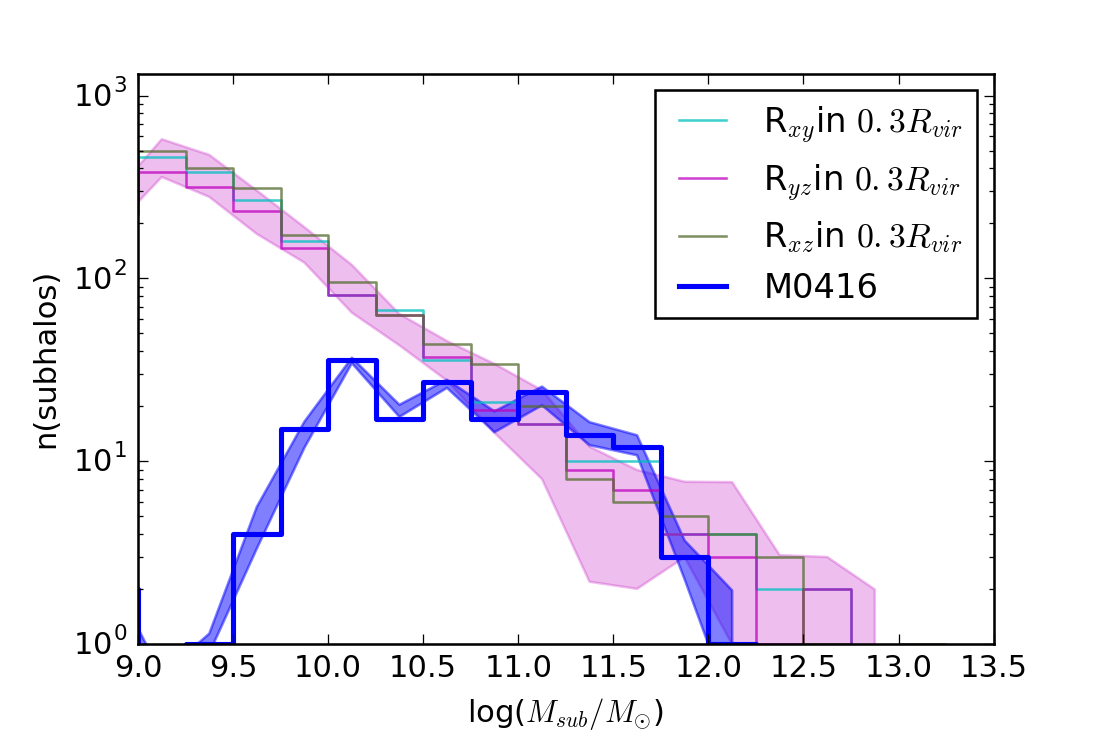}
\caption{Projection Effects on the derived SHMF: here we plot the SHMF derived from Illustris clusters derived by projecting along several distinct axes 
within $0.3\,R_{\rm vir}$. We note that projection effects cannot account for slight excess seen in the observationally determined SHMF in MACSJ\,0416 
(blue histogram) and the determinations from {\it iCluster Zoom 1}.}
\end{figure}

\begin{figure}
\includegraphics[width=1.0\columnwidth]{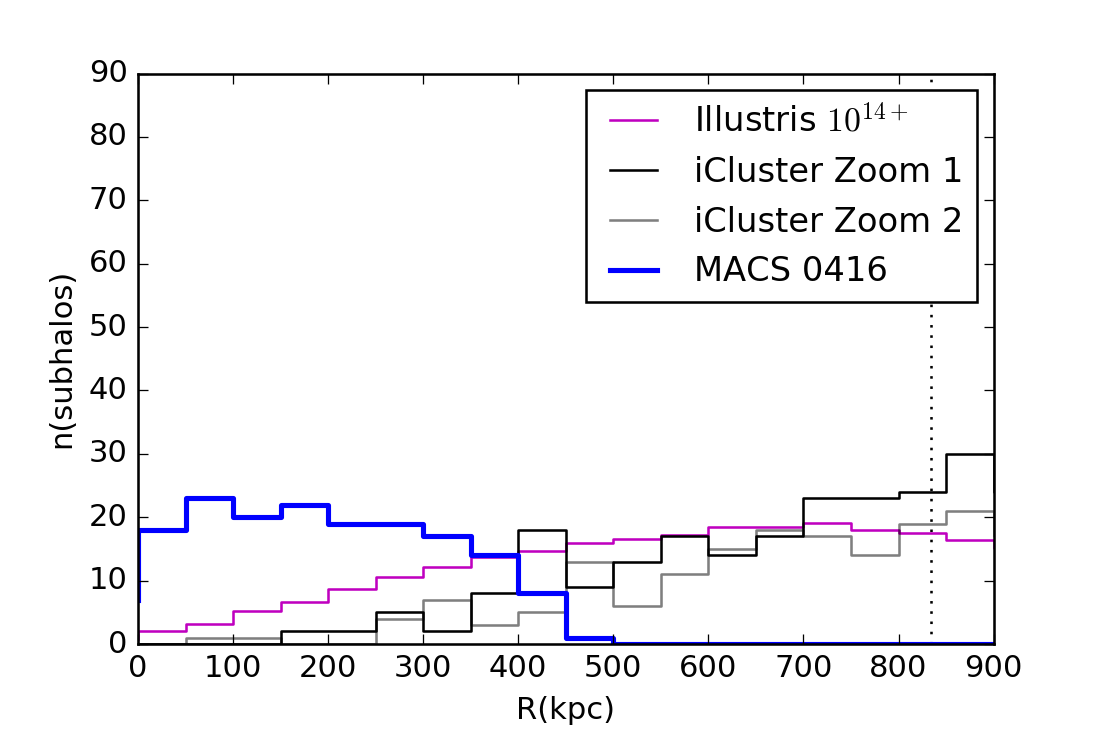}
\caption{Radial distribution of the SHMF derived from simulated Illustris clusters ($10^{14}$ {\it Illustris Haloes}) as well as the two zoomed-in runs {\it iCluster Zooms} compared to that of the lensing derived SHMF for MACSJ\,0416 from the HST FF data (blue histogram). Once again the most massive 117 simulated clusters selected at $z = 0.3-0.5$ from the $10^{14}$ {\it Illustris Haloes}  and the two massive zoomed-in cluster snapshots of the full physics run were selected. As clearly seen from the radial distribution of the count matched sub haloes that galaxies in Illustris are not as centrally concentrated as MACSJ\,0416. The radial distribution of sub haloes in Illustris are significantly more diffuse even when compared to the {\it iCluster Zooms}.}
\end{figure}

\begin{figure}
\includegraphics[width=1.0\columnwidth]{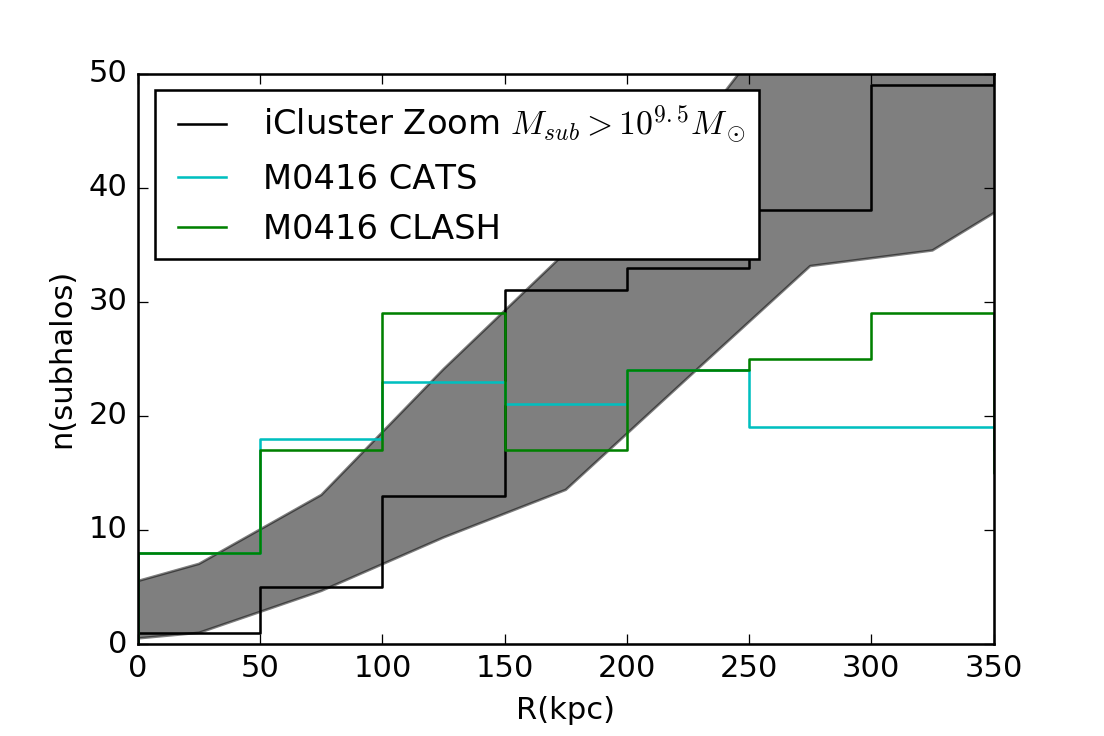}
\caption{Comparison of sub halo properties in MACSJ\,0416 inferred from lens modeling in this work and from the CLASH collaboration \citep{Grillo15} with that derived from the $10^{15}\,\Msun$ cluster from the Illustris suite - the radial distribution of subhaloes.}
\end{figure}

\begin{figure}
\includegraphics[width=1.0\columnwidth]{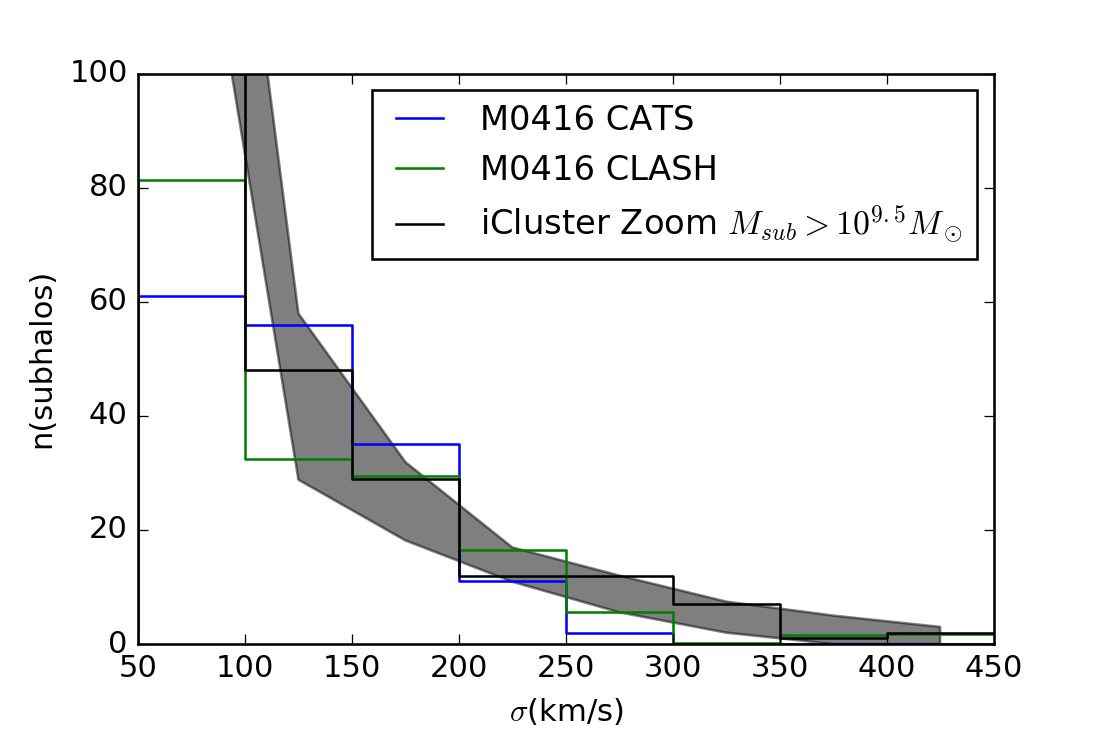}
\caption{Comparison of sub halo properties in MACSJ\,0416 inferred from lens modeling in this work and from the CLASH collaboration \citep{Grillo15} with that derived from the $10^{15}\,\Msun$ cluster from the Illustris suite - the velocity dispersion function of subhaloes. We find very good agreement between the 2 best-fit mass models and the {\it iCluster Zooms}}.
\end{figure}

\begin{figure}
\includegraphics[width=1.0\columnwidth]{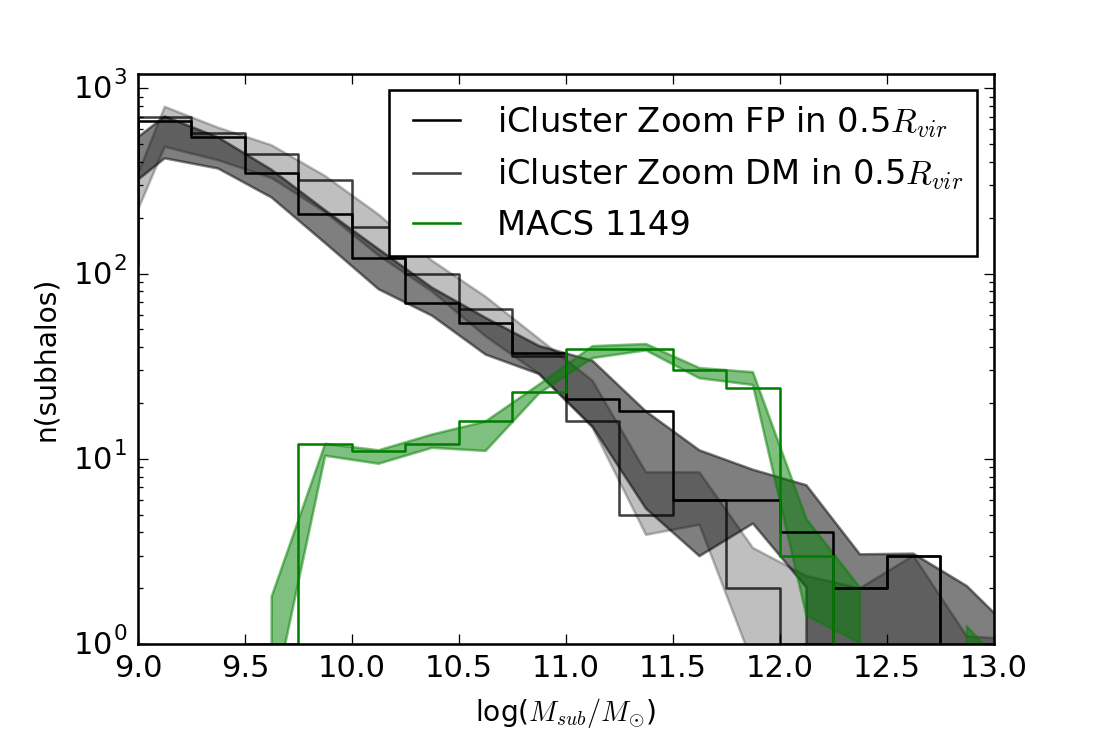}
\caption{Comparison of the SHMF derived for MACSJ\,1149 (with overplotted modeling errors derived from the dispersion in N-Bayesian realizations for the key fiducial parameters $r_{\rm t*}$ and $\sigma_{0*}$) from the HST FF data (green histogram) with that derived from the dark matter only snapshot of {\it iCluster Zoom 1} plotted as a solid black histogram with $\pm 1\sigma$ dispersion as the grey region. The $z \sim 0.55$ snapshot of {\it iCluster Zoom 1} was chosen to match the redshift of MACSJ\,1149. We also plot data of the SHMF derived from the full physics run of {\it iCluster Zoom 1} (dispersion around mean marked in lighter grey) within $0.5\,R_{\rm vir}$ that is equivalent to the ACS FOV over which the lensing model has been reconstructed.}
\end{figure}

\begin{figure}
\includegraphics[width=1.0\columnwidth]{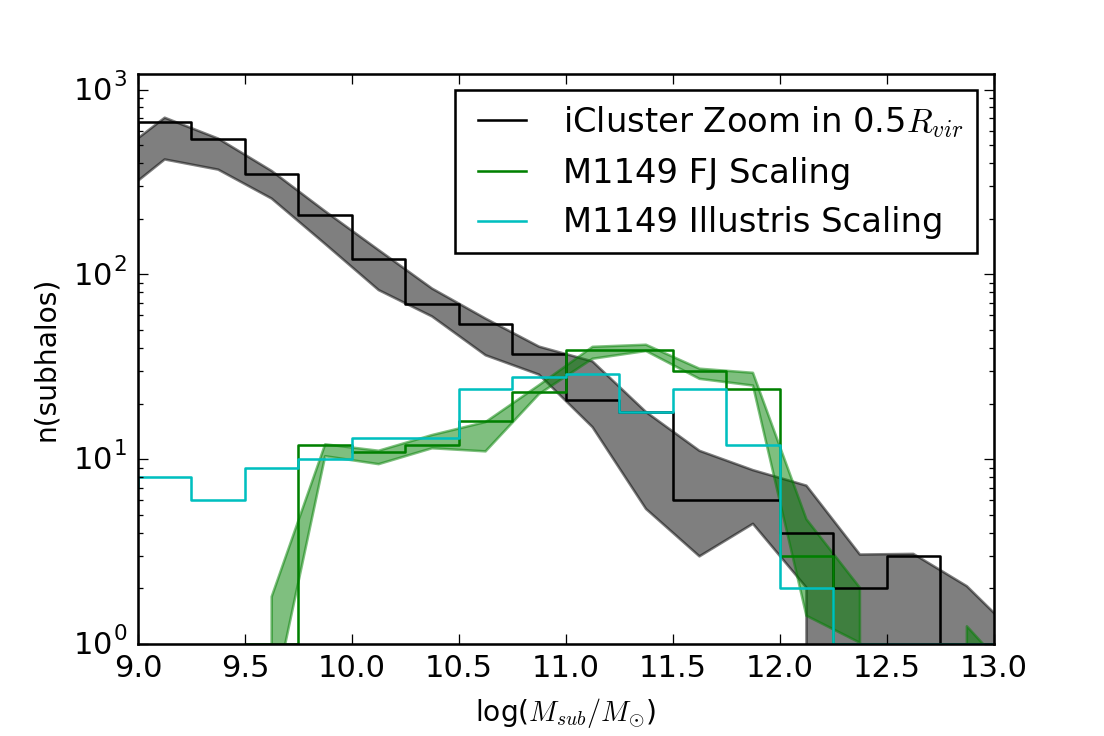}
\caption{Comparing lensing derived SHMFs for different assumptions relating mass to light: here we plot the SHMFs (with model errors) derived from the best-fit lensing model for MACSJ\,1149 using the Faber-Jackson and Kormendy luminosity scaling laws, with that derived for the best-fit lens model using the scaling laws from the full physics run of {\it iCluster Zoom 1}.}
\end{figure}

We present the results of similar comparisons now for MACSJ\,1149. As mentioned above, the mass distribution for the MACSJ\,1149 is least well constrained at the moment. This cluster is currently being followed-up with concerted spectroscopic campaigns led by several groups. Data sharing to obtain the best-to-date mass model and magnification map is expected to commence after the summer of 2016. Besides, in terms of gross properties although this cluster is the least massive, yet it appears to have an extremely complex spatial distribution. Its lower mass than Abell 2744 and MACSJ\,0416, makes it a significantly less efficient gravitational lens. MACSJ\,1149 appears to have an order of magnitude fewer multiply imaged sources identified at present and this of course directly translates into a less well determined mass distribution and therefore a less well constrained SHMF. Comparisons are once again drawn between the {\it iCluster Zoom 1} snapshot at $z \sim 0.55$ and equivalent projected area of the HSTFF data footprint which corresponds $\sim\,{0.5\,R_{\rm vir}}$. While the SHMFs derived from the dark matter only snapshot and the full physics run snapshot of {\it iCluster Zoom 1} are in good agreement with each other, there is considerable discrepancy with the lensing derived SHMF as seen in Figures~20 and 21. There is a notable excess in the abundance of lensing derived sub haloes in the mass bin ranging from $10^{11} - 10^{12}\,\Msun$ in the SHMF, and this remains prominent even in the count-count matched version plotted in Figure~22. In the case of MACSJ\,1149, this excess cannot be explained even when cosmic variance is taken into account with the analytic calculation (see Figure~23). The unparalleled complexity of MACSJ\,1149 and the fact that the {\it iCluster Zooms} are not dynamically equivalent can account for this discrepancy. The evolution of the SHMF with parent halo mass seen in Figure~24 is also more dramatic - this is of course not unexpected as this is the highest redshift cluster analyzed here and it is likely still very much in the process of assembly. The sub halo excess persists when comparing with various projections of {\it iCluster Zoom 1} (see Figure~25). Finally, we note that the radial distribution of sub haloes from the simulations is more concentrated in this case (Figure~26). Given that the MACSJ\,1149 has the least tightly constrained lensing mass model at present while having the most complex geometry of the three HSTFF clusters studied here, the slight disagreement between observations and simulations does not signal any tension with the LCDM paradigm. Once a more accurate mass model can be constructed, it would be instructive to re-do the above analysis, which we hope to tackle in future work once spectroscopic data is available to tighten model constraints.   

\begin{figure} 
\includegraphics[width=1.0\columnwidth]{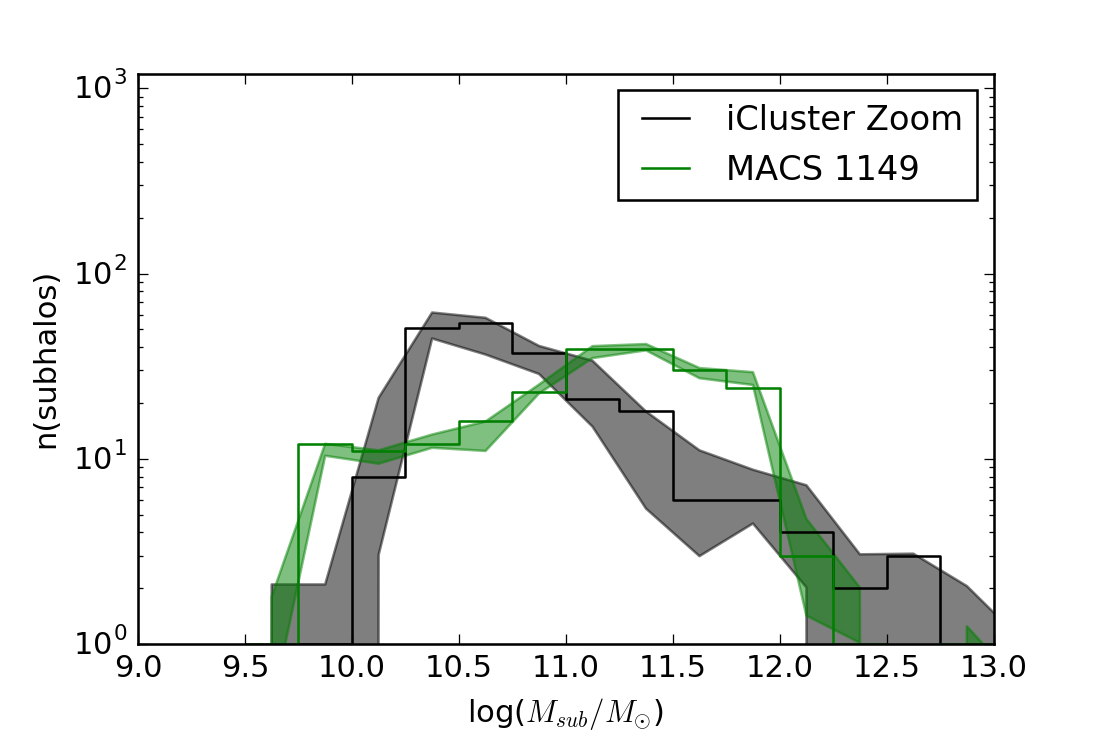}
\caption{Comparison of the count-matched SHMF derived for MACSJ\,1149 from the HST FF data (green histogram) with the {\it iCluster Zoom 1} snapshot. Once again the {\it iCluster Zoom 1} snapshot from the full physics run was selected, however here only the dark matter subhaloes associated with the brightest 217 cluster galaxies - number matched within ${\rm 0.5 R_{vir}}$ to the HFF derived SHMF are plotted.}
\end{figure}

\begin{figure} 
\includegraphics[width=1.0\columnwidth]{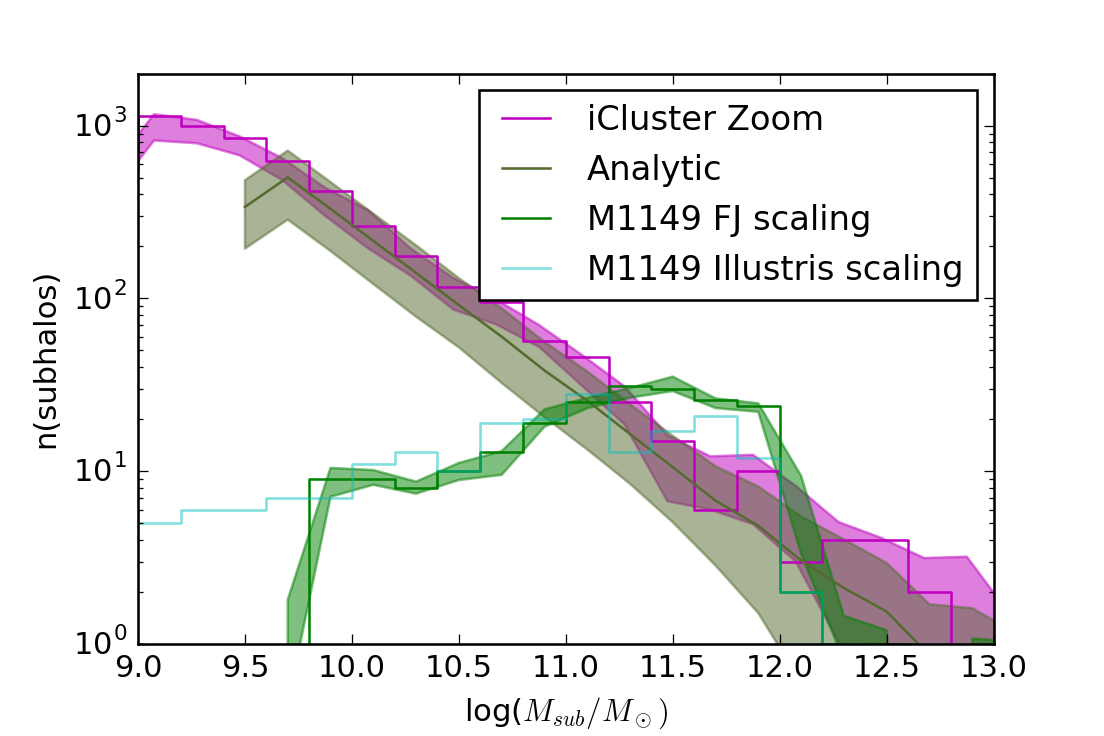}
\caption{Comparison with the analytically predicted SHMF: the lensing derived SHMF (with model errors) for the two independent scaling laws is overplotted along with the analytically calculated SHMF for a cluster halo with mass equivalent to that of MACSJ\,1149. This analytic estimate includes the halo-to-halo scatter that is shown in the dull green band. Note that the excess in substructure abundance seen in the mass range between $10^{11}\,\Msun - 10^{12}\,\Msun$ cannot be fully accounted for with cosmic variance alone.}
\end{figure}

\begin{figure} 
\includegraphics[width=1.0\columnwidth]{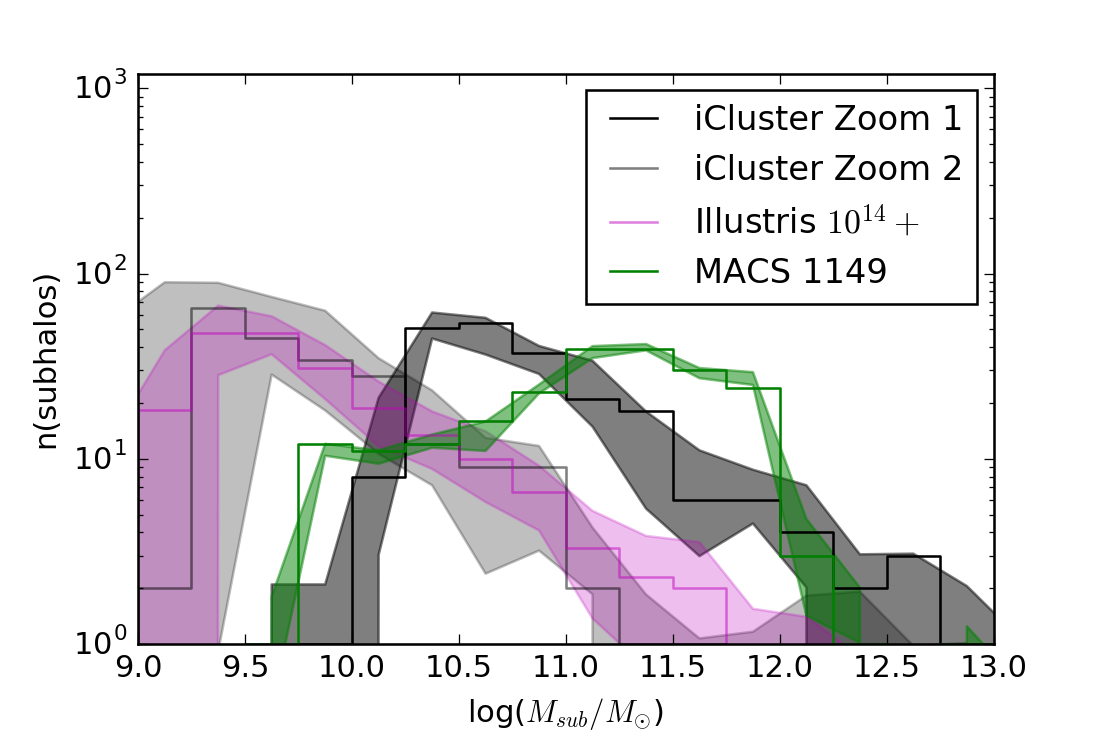}
\caption{Comparison of the SHMF derived for MACSJ\,1149 from the HST FF data (green histogram) with that derived from the two {\it iCluster Zooms} as well the larger sample $10^{14}$ {\it Illustris Haloes}  with $M > 10^{14+}\,\Msun$ clusters now mimicking the observational selection.  Here only subhaloes that host the brightest 217 cluster galaxies (number matched to the HSTFF data of MACSJ\,1149 are included) are selected from the simulations. The $1\sigma$ cluster-to-cluster variations are once again plotted as the bands.}
\end{figure}

\begin{figure}
\includegraphics[width=1.0\columnwidth]{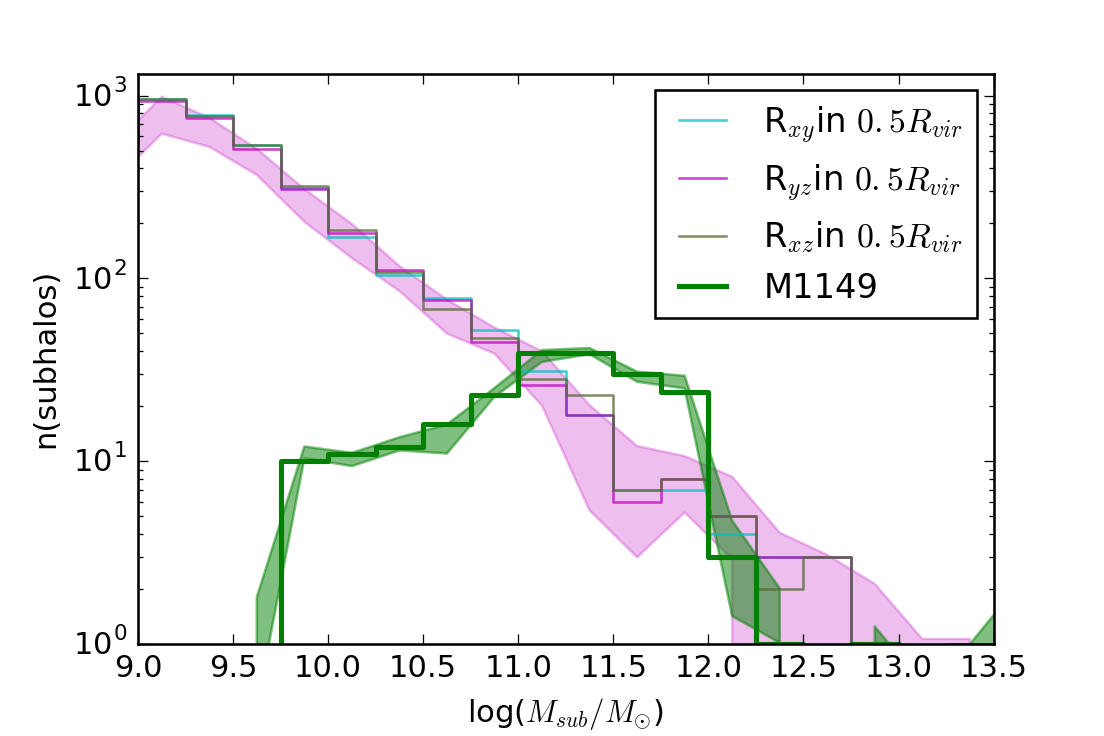}
\caption{Projection Effects on the derived SHMF: here we plot the SHMF derived from the {\it iCluster Zoom 1} snapshot derived by projecting along several distinct axes within $0.5\,R_{\rm vir}$ with the lensing derived SHMF for MACSJ\,1149.}
\end{figure}

\begin{figure}
\includegraphics[width=1.0\columnwidth]{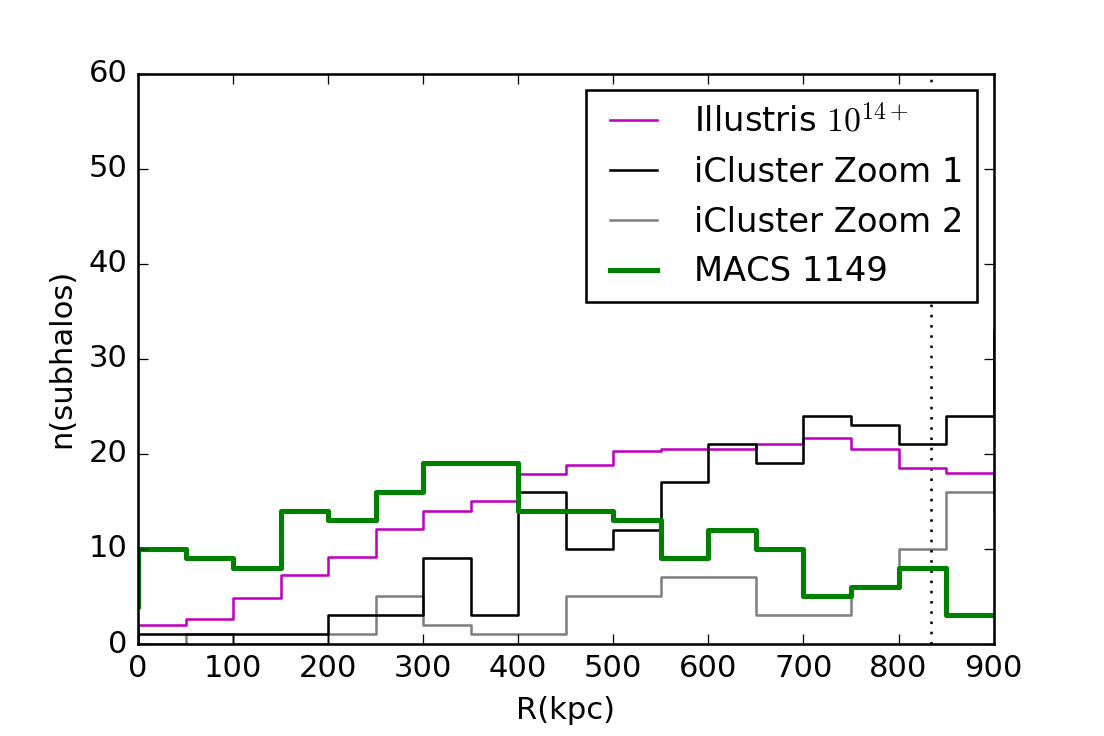}
\caption{Radial distribution of the sub halo mass function derived from simulated Illustris clusters and {\it iCluster Zoom 1} compared to that of the lensing derived SHMF for MACSJ\,1149 from the HST FF data (green histogram). Once again the most massive 66 simulated clusters with $M > 10^{14+}\,\Msun$ were selected at $z = 0.45-0.65$ from the Illustris simulation box comprising the $10^{14}$ {\it Illustris Haloes} were selected from the full physics run of Illustris. We also plot the radial distribution of the count matched sub haloes from two {\it iCluster Zooms}.  We clearly see in this case that surprisingly, galaxies in the simulations do better mimic the distribution in MACSJ\,1149 in the inner regions than for the other two HSTFF cluster lenses.}
\end{figure}

Finally, we present the 3-dimensional visualization of the substructure distribution derived from the HSTFF data for all three clusters in Figures~27-29.

\begin{figure}
\includegraphics[width=1.0\columnwidth]{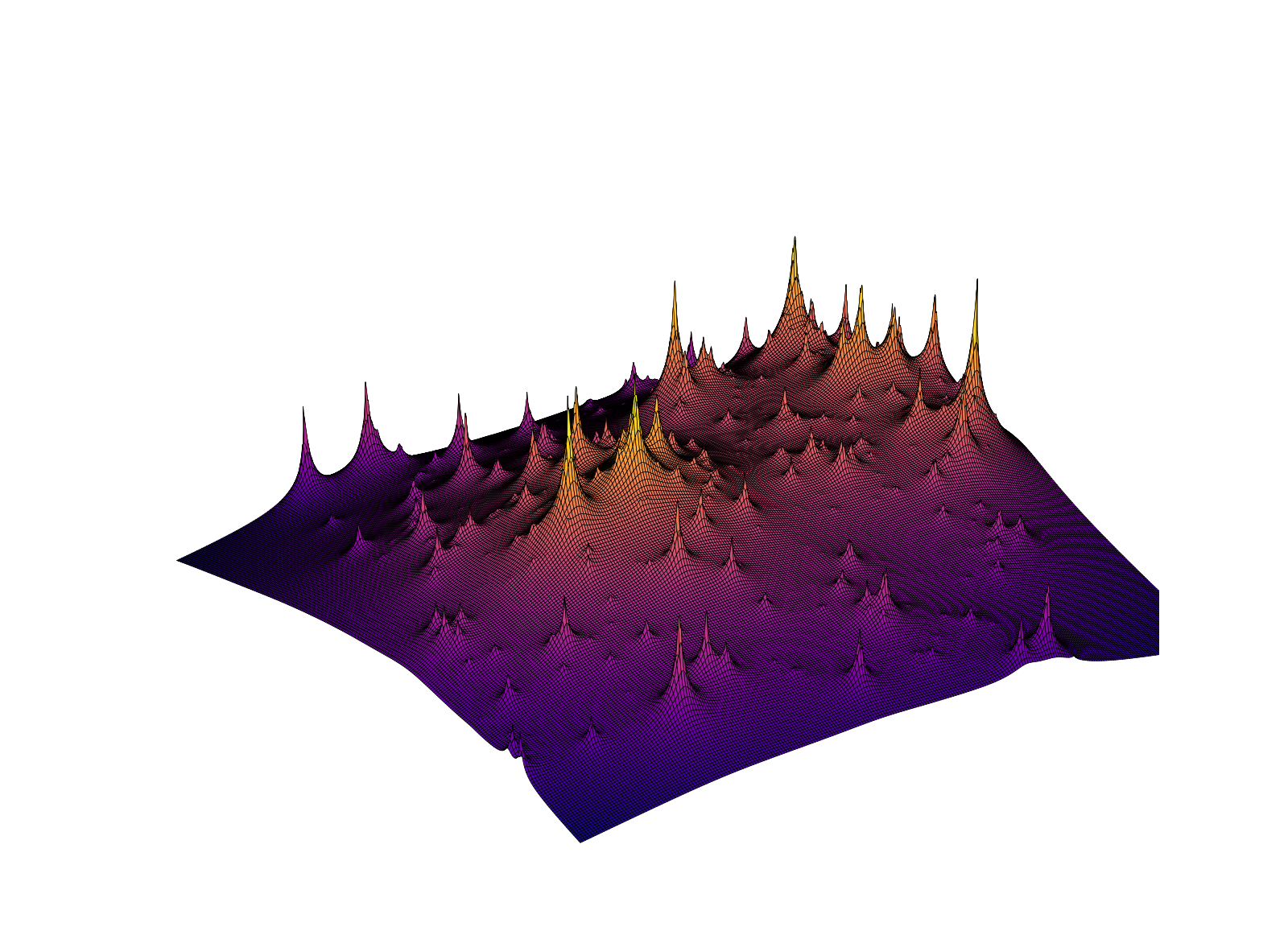}
\caption{3D visualization of the lensing derived substructure distribution for Abell 2744.}
\end{figure}

\begin{figure}
\includegraphics[width=1.0\columnwidth]{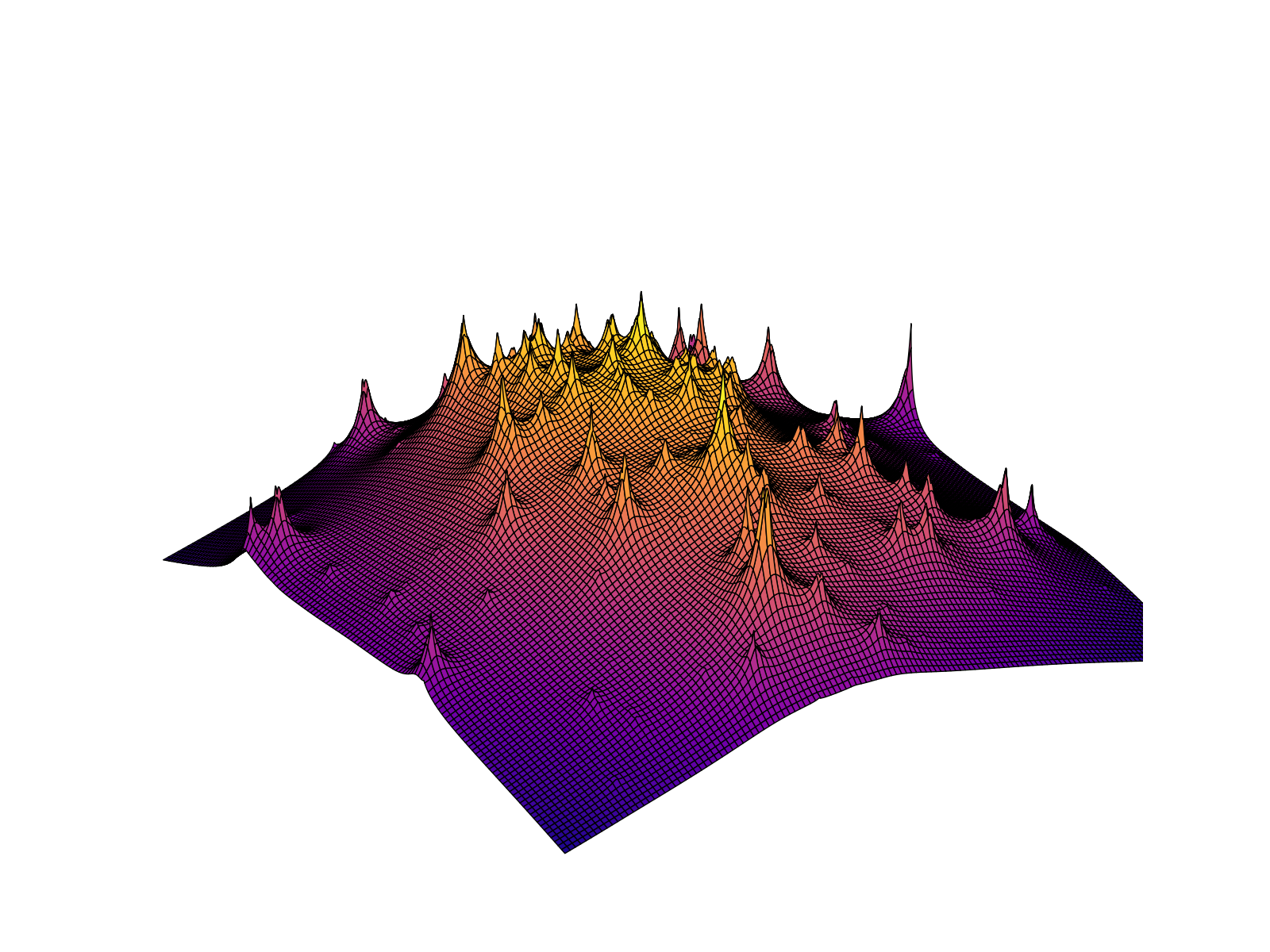}
\caption{3D visualization of the lensing derived substructure distribution for MACSJ\,0416.}
\end{figure}

\begin{figure} 
\includegraphics[width=1.0\columnwidth]{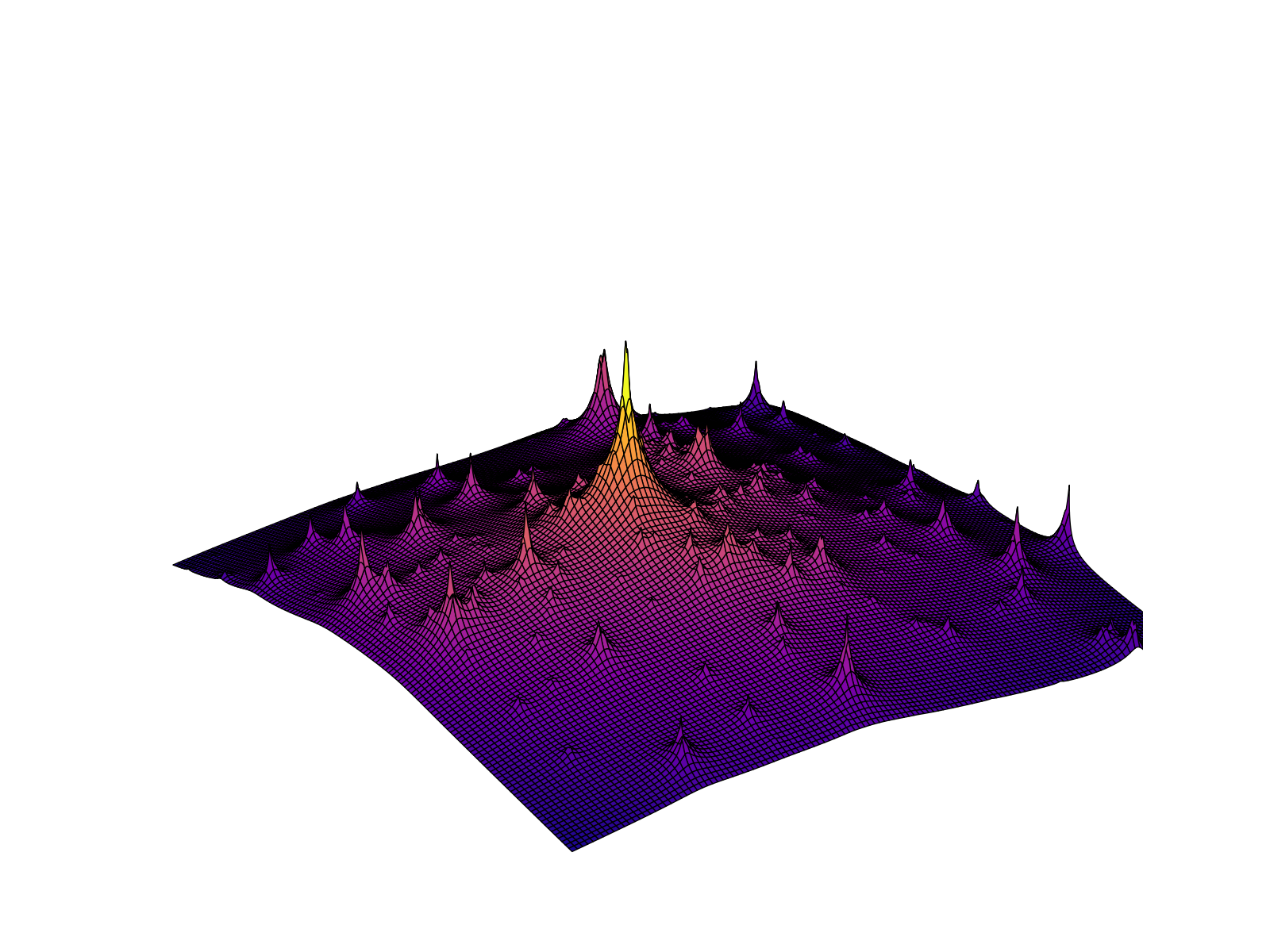}
\caption{3D visualization of the lensing derived substructure distribution for MACSJ\,1149.}
\end{figure}

\section{Discussion and Conclusions}

One of the key predictions of the standard LCDM model is the presence of abundant substructure within collapsed, virialized halos. The SHMF and other substructure properties like their radial distribution and velocity dispersion functions can be determined both analytically and derived from high resolution N-body simulations. Despite gaps in our understanding of the physics of galaxy formation, translating from observations of galaxies in clusters to the derived dark matter substructure using cluster lenses has offered an important test of the LCDM paradigm. Several criteria must be met by a large-scale numerical simulation of structure formation in order to permit a meaningful comparison with observations like the ones discussed by us here. Specifically, the comparison of a lensing-derived SHMF with dark-matter-only simulations (as performed in previous works) does not enable robust testing of the predictions of the LCDM paradigm. For the analysis presented in this paper, it was thus crucial to choose simulations that self-consistently include the effects of baryonic physics. We chose the Illustris suite of simulations which deploys the moving-mesh code AREPO and includes sub-grid models for galaxy formation and feedback. And importantly, we need to be able to mimic our observational selection when extracting information from the simulation. Zoom-in re-simulations of clusters whose mass is comparable to that of the HSTFF lenses ({\it iCluster Zooms}), extracted from the {\it Millennium} XXL simulation and re-run with AREPO by the Illustris collaboration, provide appropriate simulated counterparts to the HSTFF clusters as demonstrated here.

Reflecting the relatively small footprint of the HSTFF data, the lensing-derived mass models for all three cluster lenses studied here cover only the innermost regions. Within this area (and at redshifts matched to those of our HSTFF lenses) we find excellent agreement over four decades in mass (down to $M{\sim}10^{9.5-10}\,\Msun$) between the count-matched SHMFs for Abell 2744 and MACSJ\,0416 and those derived from the {\it iCluster Zooms} of comparable mass. 
Observed cluster lenses with extremely well calibrated, high resolution lensing mass reconstructions appear to have the same degree of substructure over 4 decades in mass compared to that found in a comparable mass simulated cluster evolving in a LCDM universe. We find that down to ${\sim}10^{10}\,\Msun$, mass and light appear to trace each other well. This might signal that light does effectively trace mass or that the current best-to-date SHMFs derived even from the high resolution HSTFF data cannot discriminate between disparate scaling laws since these encapsulate many complex processes that involve the interplay of baryons and dark matter. Even deeper data than the HSTFF it appears might be needed to provide new insights into the efficiency of galaxy formation. In order to explore and examine the detailed relationship between mass and light, we perform a variety of comparisons of the lensing-derived substructure mass function with that computed for the simulated cluster.  We note a small excess in the abundance of subhalos at masses $M{\sim}10^{11}\,\Msun$ for all three clusters studied here; in the case of Abell 2744 and MACSJ\,0416 this mild discrepancy can be entirely accounted for by cosmic variance, as evidenced in the comparison with the analytic prediction that estimates the halo-to-halo scatter and expected systematics arising from the choice of halo-finding algorithms. The agreement with simulations is poorer for the HSTFF target MACSJ\,1149, possibly because the mass distribution for this complex assembling structure is significantly less well constrained by the existing lensing data, which show fewer strong-lensing features, with less comprehensive spectroscopic confirmation than is available for the two other clusters studied here. Despite its scale-free nature, in LCDM the SHMF depends on the total mass of the parent halo, a trend demonstrated clearly by us over a range of redshifts from $z{\sim}0.3$ to 0.55. 

The observed radial distribution of subhaloes, however, is discrepant with predictions of simulations for all three of our target clusters. Possible reasons for this mismatch are limitations arising from approximations and assumptions made in sub-grid physics models, for example over-efficient AGN feedback that may preferentially suppress star formation in more massive halos and over-efficient tidal stripping of infalling galaxies. We stress again that selection for maximal lensing strength (as applied for the HSTFF) strongly favours dynamically complex and extremely massive systems. Our careful comparison of subhalo properties suggests that the single {\it iCluster Zoom} run used in the comparison fails to fully capture the dynamical complexity of disturbed, merging systems with rapidly evolving geometries. In fact we suspect that suitable counterparts (i.e., systems that match the HSTFF clusters in mass and explore the full range of dynamical states) do not exist in any available simulation volume. Most importantly, our analysis demonstrates that the HSTFF clusters are likely observed in short-lived, transient states that can be isolated in simulations only through deliberately selected zoom-in runs with extremely dense temporal sampling. Finally, from the discrepancies between observation and simulation in the radial distribution of subhaloes, it appears that tidal stripping and dynamical friction are over-efficient in current simulations. 

Clusters as massive as those deliberately selected for the HSTFF sample are extremely rare in the universe and hence unlikely to appear in present-day cosmological simulations. For example, despite its vast volume of (500 Mpc)$^{3}$, the {\it Millennium Simulation} contains no cluster analogous to the Bullet Cluster. In fact, the likelihood of finding a sub-cluster merger of total mass, relative velocity and merging geometry comparable to the Bullet Cluster was estimated at ${\sim}10^{-7}$ in LCDM  by \cite{springel07}. In a recent paper, comparing with the Millenium XXL simulation, \cite{schwinn17}, report not finding any cluster scale haloes with the equivalent number and radial distribution of massive substructures ($> 5 \times10^{13}\,\Msun$) similar to what is inferred from the HSTFF data of Abell 2744. Using extreme value statistics, they conclude that a simulation volume ten times larger than the Millennium XXL would be needed to find a cluster equivalent to Abell 2744.

In addition, cosmological simulations (including the Illustris suite) that employ AGN feedback mechanisms to regulate star formation in massive galaxies are known to inadequately reproduce the  observed luminosities, sizes and masses of cluster galaxies at the present time. Therefore, the disagreements that we find can be entirely explained by the paucity of simulated clusters that can be considered appropriate equivalents in both mass and complexity to the HSTFF clusters and our incomplete understanding of galaxy formation. In summary, we find that the concordance LCDM model provides an excellent description of the properties and abundance of substructure detected via strong gravitational lensing on cluster mass scales. 

\section*{Acknowledgements}

P.N. acknowledges many useful discussions with Dan Coe and Jennifer Lotz on the HSTFF clusters. She also thanks Angelo Ricarte for his help with the 3D rendering of the substructure distribution in these 3 HSTFF clusters. P.N. gratefully acknowledges support from an NSF theory program via the grant AST-1044455 that supported the early stages of this work.  MJ's work was supported by the Science and Technology Facilities Council [grant number ST/L00075X/1 \& ST/F001166/1] and used the DiRAC Data Centric system at Durham University, operated by the Institute for Computational Cosmology on behalf of the STFC DiRAC HPC Facility (www.dirac.ac.uk [www.dirac.ac.uk]). This equipment was funded by BIS National E-infrastructure capital grant ST/K00042X/1, STFC capital grant ST/H008519/1, and STFC DiRAC Operations grant ST/K003267/1 and Durham University. DiRAC is part of the National E-Infrastructure. JR acknowledges support from the ERC starting grant 336736 (CALENDS). EJ acknowledges the support of the OCEVU  Labex (ANR-11-LABX-0060)  and the A*Midex project (ANR-11- IDEX-0001-02). MM acknowledges support from the Italian Ministry of Foreign Affairs and International Cooperation, Directorate General for Country Promotion, from INAF via PRININAF2014 1.05.01.94.02, and from ASI via contract ASI/INAF/I/023/12/0. 

\bibliographystyle{mnras}
\bibliography{reference}

\begin{thebibliography}{}
\makeatletter
\relax
\def\mn@urlcharsother{\let\do\@makeother \do\$\do\&\do\#\do\^\do\_\do\%\do\~}
\def\mn@doi{\begingroup\mn@urlcharsother \@ifnextchar [ {\mn@doi@}
  {\mn@doi@[]}}
\def\mn@doi@[#1]#2{\def\@tempa{#1}\ifx\@tempa\@empty \href
  {http://dx.doi.org/#2} {doi:#2}\else \href {http://dx.doi.org/#2} {#1}\fi
  \endgroup}
\def\mn@eprint#1#2{\mn@eprint@#1:#2::\@nil}
\def\mn@eprint@arXiv#1{\href {http://arxiv.org/abs/#1} {{\tt arXiv:#1}}}
\def\mn@eprint@dblp#1{\href {http://dblp.uni-trier.de/rec/bibtex/#1.xml}
  {dblp:#1}}
\def\mn@eprint@#1:#2:#3:#4\@nil{\def\@tempa {#1}\def\@tempb {#2}\def\@tempc
  {#3}\ifx \@tempc \@empty \let \@tempc \@tempb \let \@tempb \@tempa \fi \ifx
  \@tempb \@empty \def\@tempb {arXiv}\fi \@ifundefined
  {mn@eprint@\@tempb}{\@tempb:\@tempc}{\expandafter \expandafter \csname
  mn@eprint@\@tempb\endcsname \expandafter{\@tempc}}}

\bibitem[\protect\citeauthoryear{{Angulo}, {Springel}, {White}, {Jenkins},
  {Baugh}  \& {Frenk}}{{Angulo} et~al.}{2012}]{angulo12}
{Angulo} R.~E.,  {Springel} V.,  {White} S.~D.~M.,  {Jenkins} A.,  {Baugh}
  C.~M.,   {Frenk} C.~S.,  2012, \mn@doi [\mnras]
  {10.1111/j.1365-2966.2012.21830.x}, \href
  {http://adsabs.harvard.edu/abs/2012MNRAS.426.2046A} {426, 2046}

\bibitem[\protect\citeauthoryear{{Atek} et~al.,}{{Atek} et~al.}{2014}]{atek14a}
{Atek} H.,  et~al., 2014, \mn@doi [\apj] {10.1088/0004-637X/786/1/60}, \href
  {http://adsabs.harvard.edu/abs/2014ApJ...786...60A} {786, 60}

\bibitem[\protect\citeauthoryear{{Balestra} et~al.,}{{Balestra}
  et~al.}{2016}]{Balestra16}
{Balestra} I.,  et~al., 2016, \mn@doi [\apjs] {10.3847/0067-0049/224/2/33},
  \href {http://adsabs.harvard.edu/abs/2016ApJS..224...33B} {224, 33}

\bibitem[\protect\citeauthoryear{{Bertin} \& {Arnouts}}{{Bertin} \&
  {Arnouts}}{1996}]{BA96}
{Bertin} E.,  {Arnouts} S.,  1996, \aap, 117, 393

\bibitem[\protect\citeauthoryear{{Bouwens} et~al.,}{{Bouwens}
  et~al.}{2014}]{Bouwens15}
{Bouwens} R.~J.,  et~al., 2014, \mn@doi [\apj] {10.1088/0004-637X/795/2/126},
  \href {http://adsabs.harvard.edu/abs/2014ApJ...795..126B} {795, 126}

\bibitem[\protect\citeauthoryear{{Brada{\v c}}, {Schneider}, {Lombardi}  \&
  {Erben}}{{Brada{\v c}} et~al.}{2005}]{bradac05}
{Brada{\v c}} M.,  {Schneider} P.,  {Lombardi} M.,   {Erben} T.,  2005, \mn@doi
  [\aap] {10.1051/0004-6361:20042233}, \href
  {http://adsabs.harvard.edu/abs/2005A%26A...437...39B} {437, 39}

\bibitem[\protect\citeauthoryear{{Brada{\v c}} et~al.,}{{Brada{\v c}}
  et~al.}{2014}]{Bradac14}
{Brada{\v c}} M.,  et~al., 2014, \mn@doi [\apj] {10.1088/0004-637X/785/2/108},
  \href {http://adsabs.harvard.edu/abs/2014ApJ...785..108B} {785, 108}

\bibitem[\protect\citeauthoryear{{Caminha} et~al.,}{{Caminha}
  et~al.}{2015}]{Caminha15}
{Caminha} G.~B.,  et~al., 2015, preprint, \href
  {http://adsabs.harvard.edu/abs/2015arXiv151204555C} {} (\mn@eprint {arXiv}
  {1512.04555})

\bibitem[\protect\citeauthoryear{{Caminha} et~al.,}{{Caminha}
  et~al.}{2016}]{caminha16}
{Caminha} G.~B.,  et~al., 2016, preprint, \href
  {http://adsabs.harvard.edu/abs/2016arXiv160703462C} {} (\mn@eprint {arXiv}
  {1607.03462})

\bibitem[\protect\citeauthoryear{{Clowe}, {De Lucia}  \& {King}}{{Clowe}
  et~al.}{2004}]{clowe04}
{Clowe} D.,  {De Lucia} G.,   {King} L.,  2004, \mn@doi [\mnras]
  {10.1111/j.1365-2966.2004.07723.x}, \href
  {http://adsabs.harvard.edu/cgi-bin/nph-bib_query?bibcode=2004MNRAS.350.1038C&db_key=AST}
  {350, 1038}

\bibitem[\protect\citeauthoryear{{Coe}, {Bradley}  \& {Zitrin}}{{Coe}
  et~al.}{2015}]{Coe15}
{Coe} D.,  {Bradley} L.,   {Zitrin} A.,  2015, \mn@doi [\apj]
  {10.1088/0004-637X/800/2/84}, \href
  {http://adsabs.harvard.edu/abs/2015ApJ...800...84C} {800, 84}

\bibitem[\protect\citeauthoryear{{D'Aloisio} \& {Natarajan}}{{D'Aloisio} \&
  {Natarajan}}{2011}]{D'Aloisio11}
{D'Aloisio} A.,  {Natarajan} P.,  2011, \mn@doi [\mnras]
  {10.1111/j.1365-2966.2010.17795.x}, \href
  {http://adsabs.harvard.edu/abs/2011MNRAS.411.1628D} {411, 1628}

\bibitem[\protect\citeauthoryear{{Davis}, {Efstathiou}, {Frenk}  \&
  {White}}{{Davis} et~al.}{1985}]{davis85}
{Davis} M.,  {Efstathiou} G.,  {Frenk} C.~S.,   {White} S.~D.~M.,  1985,
  \mn@doi [\apj] {10.1086/163168}, \href
  {http://adsabs.harvard.edu/abs/1985ApJ...292..371D} {292, 371}

\bibitem[\protect\citeauthoryear{{De Lucia}, {Springel}, {White}, {Croton}  \&
  {Kauffmann}}{{De Lucia} et~al.}{2006}]{delucia06}
{De Lucia} G.,  {Springel} V.,  {White} S.~D.~M.,  {Croton} D.,   {Kauffmann}
  G.,  2006, \mn@doi [\mnras] {10.1111/j.1365-2966.2005.09879.x}, \href
  {http://adsabs.harvard.edu/abs/2006MNRAS.366..499D} {366, 499}

\bibitem[\protect\citeauthoryear{{Deason} et~al.,}{{Deason}
  et~al.}{2014}]{deason14}
{Deason} A.~J.,  et~al., 2014, \mn@doi [\mnras] {10.1093/mnras/stu1764}, \href
  {http://adsabs.harvard.edu/abs/2014MNRAS.444.3975D} {444, 3975}

\bibitem[\protect\citeauthoryear{{Di Cintio} \& {Lelli}}{{Di Cintio} \&
  {Lelli}}{2016}]{dicintio16}
{Di Cintio} A.,  {Lelli} F.,  2016, \mn@doi [\mnras] {10.1093/mnrasl/slv185},
  \href {http://adsabs.harvard.edu/abs/2016MNRAS.456L.127D} {456, L127}

\bibitem[\protect\citeauthoryear{{Diego}, {Broadhurst}, {Molnar}, {Lam}  \&
  {Lim}}{{Diego} et~al.}{2015}]{diego15}
{Diego} J.~M.,  {Broadhurst} T.,  {Molnar} S.~M.,  {Lam} D.,   {Lim} J.,  2015,
  \mn@doi [\mnras] {10.1093/mnras/stu2660}, \href
  {http://adsabs.harvard.edu/abs/2015MNRAS.447.3130D} {447, 3130}

\bibitem[\protect\citeauthoryear{{Ebeling}, {Edge}  \& {Henry}}{{Ebeling}
  et~al.}{2001}]{ebeling01}
{Ebeling} H.,  {Edge} A.~C.,   {Henry} J.~P.,  2001, \mn@doi [\apj]
  {10.1086/320958}, \href {http://adsabs.harvard.edu/abs/2001ApJ...553..668E}
  {553, 668}

\bibitem[\protect\citeauthoryear{{Ebeling}, {Edge}, {Mantz}, {Barrett},
  {Henry}, {Ma}  \& {van Speybroeck}}{{Ebeling} et~al.}{2010}]{ebeling10}
{Ebeling} H.,  {Edge} A.~C.,  {Mantz} A.,  {Barrett} E.,  {Henry} J.~P.,  {Ma}
  C.~J.,   {van Speybroeck} L.,  2010, \mn@doi [\mnras]
  {10.1111/j.1365-2966.2010.16920.x}, \href
  {http://adsabs.harvard.edu/abs/2010MNRAS.407...83E} {407, 83}

\bibitem[\protect\citeauthoryear{{Ebeling}, {Ma}  \& {Barrett}}{{Ebeling}
  et~al.}{2014}]{ebeling14}
{Ebeling} H.,  {Ma} C.-J.,   {Barrett} E.,  2014, \mn@doi [\apjs]
  {10.1088/0067-0049/211/2/21}, \href
  {http://adsabs.harvard.edu/abs/2014ApJS..211...21E} {211, 21}

\bibitem[\protect\citeauthoryear{{Eckert} et~al.,}{{Eckert}
  et~al.}{2015}]{eckert15}
{Eckert} D.,  et~al., 2015, \mn@doi [\nat] {10.1038/nature16058}, \href
  {http://adsabs.harvard.edu/abs/2015Natur.528..105E} {528, 105}

\bibitem[\protect\citeauthoryear{{Faber} \& {Jackson}}{{Faber} \&
  {Jackson}}{1976}]{faber76}
{Faber} S.~M.,  {Jackson} R.~E.,  1976, \mn@doi [\apj] {10.1086/154215}, \href
  {http://adsabs.harvard.edu/abs/1976ApJ...204..668F} {204, 668}

\bibitem[\protect\citeauthoryear{{Finkelstein}}{{Finkelstein}}{2015}]{Finkelstein15}
{Finkelstein} S.~L.,  2015, preprint, \href
  {http://adsabs.harvard.edu/abs/2015arXiv151105558F} {} (\mn@eprint {arXiv}
  {1511.05558})

\bibitem[\protect\citeauthoryear{{Gao}, {White}, {Jenkins}, {Stoehr}  \&
  {Springel}}{{Gao} et~al.}{2004}]{gao04}
{Gao} L.,  {White} S.~D.~M.,  {Jenkins} A.,  {Stoehr} F.,   {Springel} V.,
  2004, \mn@doi [\mnras] {10.1111/j.1365-2966.2004.08360.x}, \href
  {http://adsabs.harvard.edu/abs/2004MNRAS.355..819G} {355, 819}

\bibitem[\protect\citeauthoryear{{Genel} et~al.,}{{Genel}
  et~al.}{2014}]{Genel14}
{Genel} S.,  et~al., 2014, \mn@doi [\mnras] {10.1093/mnras/stu1654}, \href
  {http://adsabs.harvard.edu/abs/2014MNRAS.445..175G} {445, 175}

\bibitem[\protect\citeauthoryear{{Grillo} et~al.,}{{Grillo}
  et~al.}{2015}]{Grillo15}
{Grillo} C.,  et~al., 2015, \mn@doi [\apj] {10.1088/0004-637X/800/1/38}, \href
  {http://adsabs.harvard.edu/abs/2015ApJ...800...38G} {800, 38}

\bibitem[\protect\citeauthoryear{{Halkola}, {Seitz}  \& {Pannella}}{{Halkola}
  et~al.}{2006}]{halkola06}
{Halkola} A.,  {Seitz} S.,   {Pannella} M.,  2006, \mn@doi [\mnras]
  {10.1111/j.1365-2966.2006.10948.x}, \href
  {http://adsabs.harvard.edu/abs/2006MNRAS.372.1425H} {372, 1425}

\bibitem[\protect\citeauthoryear{{Helmi}, {Cooper}, {White}, {Cole}, {Frenk}
  \& {Navarro}}{{Helmi} et~al.}{2011}]{helmi11}
{Helmi} A.,  {Cooper} A.~P.,  {White} S.~D.~M.,  {Cole} S.,  {Frenk} C.~S.,
  {Navarro} J.~F.,  2011, \mn@doi [\apjl] {10.1088/2041-8205/733/1/L7}, \href
  {http://adsabs.harvard.edu/abs/2011ApJ...733L...7H} {733, L7}

\bibitem[\protect\citeauthoryear{{Hinshaw} et~al.,}{{Hinshaw}
  et~al.}{2013}]{hinshaw13}
{Hinshaw} G.,  et~al., 2013, \mn@doi [\apjs] {10.1088/0067-0049/208/2/19},
  \href {http://adsabs.harvard.edu/abs/2013ApJS..208...19H} {208, 19}

\bibitem[\protect\citeauthoryear{{Hoag} et~al.,}{{Hoag} et~al.}{2016}]{hoag16}
{Hoag} A.,  et~al., 2016, preprint, \href
  {http://adsabs.harvard.edu/abs/2016arXiv160300505H} {} (\mn@eprint {arXiv}
  {1603.00505})

\bibitem[\protect\citeauthoryear{{Jauzac} et~al.,}{{Jauzac}
  et~al.}{2014}]{jauzac14}
{Jauzac} M.,  et~al., 2014, \mn@doi [\mnras] {10.1093/mnras/stu1355}, \href
  {http://adsabs.harvard.edu/abs/2014MNRAS.443.1549J} {443, 1549}

\bibitem[\protect\citeauthoryear{{Jauzac} et~al.,}{{Jauzac}
  et~al.}{2015a}]{jauzac15a}
{Jauzac} M.,  et~al., 2015a, \mn@doi [\mnras] {10.1093/mnras/stu2425}, \href
  {http://adsabs.harvard.edu/abs/2015MNRAS.446.4132J} {446, 4132}

\bibitem[\protect\citeauthoryear{{Jauzac} et~al.,}{{Jauzac}
  et~al.}{2015b}]{jauzac15b}
{Jauzac} M.,  et~al., 2015b, \mn@doi [\mnras] {10.1093/mnras/stv1402}, \href
  {http://adsabs.harvard.edu/abs/2015MNRAS.452.1437J} {452, 1437}

\bibitem[\protect\citeauthoryear{{Jauzac} et~al.,}{{Jauzac}
  et~al.}{2016}]{jauzac16}
{Jauzac} M.,  et~al., 2016, \mn@doi [\mnras] {10.1093/mnras/stw069}, \href
  {http://adsabs.harvard.edu/abs/2016MNRAS.457.2029J} {457, 2029}

\bibitem[\protect\citeauthoryear{{Johnson}, {Sharon}, {Bayliss}, {Gladders},
  {Coe}  \& {Ebeling}}{{Johnson} et~al.}{2014}]{johnson14}
{Johnson} T.~L.,  {Sharon} K.,  {Bayliss} M.~B.,  {Gladders} M.~D.,  {Coe} D.,
   {Ebeling} H.,  2014, \mn@doi [\apj] {10.1088/0004-637X/797/1/48}, \href
  {http://adsabs.harvard.edu/abs/2014ApJ...797...48J} {797, 48}

\bibitem[\protect\citeauthoryear{{Jorgensen}, {Franx}  \&
  {Kjaergaard}}{{Jorgensen} et~al.}{1996}]{jorgenson96}
{Jorgensen} I.,  {Franx} M.,   {Kjaergaard} P.,  1996, \mn@doi [\mnras]
  {10.1093/mnras/280.1.167}, \href
  {http://adsabs.harvard.edu/abs/1996MNRAS.280..167J} {280, 167}

\bibitem[\protect\citeauthoryear{{Jullo}, {Kneib}, {Limousin},
  {El{\'{\i}}asd{\'o}ttir}, {Marshall}  \& {Verdugo}}{{Jullo}
  et~al.}{2007}]{jullo07}
{Jullo} E.,  {Kneib} J.-P.,  {Limousin} M.,  {El{\'{\i}}asd{\'o}ttir} {\'A}.,
  {Marshall} P.~J.,   {Verdugo} T.,  2007, \mn@doi [New Journal of Physics]
  {10.1088/1367-2630/9/12/447}, \href
  {http://adsabs.harvard.edu/abs/2007NJPh....9..447J} {9, 447}

\bibitem[\protect\citeauthoryear{{Jullo}, {Natarajan}, {Kneib}, {D'Aloisio},
  {Limousin}, {Richard}  \& {Schimd}}{{Jullo} et~al.}{2010}]{Jullo10}
{Jullo} E.,  {Natarajan} P.,  {Kneib} J.-P.,  {D'Aloisio} A.,  {Limousin} M.,
  {Richard} J.,   {Schimd} C.,  2010, \mn@doi [Science]
  {10.1126/science.1185759}, \href
  {http://adsabs.harvard.edu/abs/2010Sci...329..924J} {329, 924}

\bibitem[\protect\citeauthoryear{{Kelly} et~al.,}{{Kelly}
  et~al.}{2015}]{kelly15}
{Kelly} P.~L.,  et~al., 2015, \mn@doi [Science] {10.1126/science.aaa3350},
  \href {http://adsabs.harvard.edu/abs/2015Sci...347.1123K} {347, 1123}

\bibitem[\protect\citeauthoryear{{Klypin}, {Kravtsov}, {Valenzuela}  \&
  {Prada}}{{Klypin} et~al.}{1999}]{klypin99}
{Klypin} A.,  {Kravtsov} A.~V.,  {Valenzuela} O.,   {Prada} F.,  1999, \mn@doi
  [\apj] {10.1086/307643}, \href
  {http://adsabs.harvard.edu/abs/1999ApJ...522...82K} {522, 82}

\bibitem[\protect\citeauthoryear{{Knebe} et~al.,}{{Knebe}
  et~al.}{2013}]{knebe13}
{Knebe} A.,  et~al., 2013, \mn@doi [\mnras] {10.1093/mnras/sts173}, \href
  {http://adsabs.harvard.edu/abs/2013MNRAS.428.2039K} {428, 2039}

\bibitem[\protect\citeauthoryear{{Kneib} \& {Natarajan}}{{Kneib} \&
  {Natarajan}}{2011}]{KN11}
{Kneib} J.-P.,  {Natarajan} P.,  2011, \mn@doi [\aapr]
  {10.1007/s00159-011-0047-3}, \href
  {http://adsabs.harvard.edu/abs/2011A%26ARv..19...47K} {19, 47}

\bibitem[\protect\citeauthoryear{{Kneib} et~al.,}{{Kneib}
  et~al.}{2003}]{kneib03}
{Kneib} J.,  et~al., 2003, \apj, 598, 804

\bibitem[\protect\citeauthoryear{{Kormendy}}{{Kormendy}}{1977}]{kormendy77}
{Kormendy} J.,  1977, \mn@doi [\apj] {10.1086/155687}, \href
  {http://adsabs.harvard.edu/abs/1977ApJ...218..333K} {218, 333}

\bibitem[\protect\citeauthoryear{{Lam}, {Broadhurst}, {Diego}, {Lim}, {Coe},
  {Ford}  \& {Zheng}}{{Lam} et~al.}{2014}]{lam14}
{Lam} D.,  {Broadhurst} T.,  {Diego} J.~M.,  {Lim} J.,  {Coe} D.,  {Ford}
  H.~C.,   {Zheng} W.,  2014, \mn@doi [\apj] {10.1088/0004-637X/797/2/98},
  \href {http://adsabs.harvard.edu/abs/2014ApJ...797...98L} {797, 98}

\bibitem[\protect\citeauthoryear{{Laporte} et~al.,}{{Laporte}
  et~al.}{2015}]{Laporte15}
{Laporte} N.,  et~al., 2015, \mn@doi [\aap] {10.1051/0004-6361/201425040},
  \href {http://adsabs.harvard.edu/abs/2015A%26A...575A..92L} {575, A92}

\bibitem[\protect\citeauthoryear{{Limousin}, {Kneib}, {Bardeau}, {Natarajan},
  {Czoske}, {Smail}, {Ebeling}  \& {Smith}}{{Limousin}
  et~al.}{2007a}]{limousin07a}
{Limousin} M.,  {Kneib} J.~P.,  {Bardeau} S.,  {Natarajan} P.,  {Czoske} O.,
  {Smail} I.,  {Ebeling} H.,   {Smith} G.~P.,  2007a, \mn@doi [\aap]
  {10.1051/0004-6361:20065543}, \href
  {http://adsabs.harvard.edu/cgi-bin/nph-bib_query?bibcode=2007A%26A...461..881L&db_key=AST}
  {461, 881}

\bibitem[\protect\citeauthoryear{{Limousin} et~al.,}{{Limousin}
  et~al.}{2007b}]{limousin07b}
{Limousin} M.,  et~al., 2007b, \mn@doi [\apj] {10.1186/383259}, \href
  {http://adsabs.harvard.edu/abs/2006astro.ph.12165L} {668, 643}

\bibitem[\protect\citeauthoryear{{Lotz}}{{Lotz}}{2015}]{lotz15}
{Lotz} J.,  2015, IAU General Assembly, \href
  {http://adsabs.harvard.edu/abs/2015IAUGA..2255460L} {22, 2255460}

\bibitem[\protect\citeauthoryear{{Lovell}, {Frenk}, {Eke}, {Jenkins}, {Gao}  \&
  {Theuns}}{{Lovell} et~al.}{2014}]{lovell14}
{Lovell} M.~R.,  {Frenk} C.~S.,  {Eke} V.~R.,  {Jenkins} A.,  {Gao} L.,
  {Theuns} T.,  2014, \mn@doi [\mnras] {10.1093/mnras/stt2431}, \href
  {http://adsabs.harvard.edu/abs/2014MNRAS.439..300L} {439, 300}

\bibitem[\protect\citeauthoryear{{Madau}, {Diemand}  \& {Kuhlen}}{{Madau}
  et~al.}{2008}]{kuhlen08}
{Madau} P.,  {Diemand} J.,   {Kuhlen} M.,  2008, \mn@doi [\apj]
  {10.1086/587545}, \href {http://adsabs.harvard.edu/abs/2008ApJ...679.1260M}
  {679, 1260}

\bibitem[\protect\citeauthoryear{{Mandelbaum}, {Seljak}, {Kauffmann}, {Hirata}
  \& {Brinkmann}}{{Mandelbaum} et~al.}{2006}]{mandelbaum06}
{Mandelbaum} R.,  {Seljak} U.,  {Kauffmann} G.,  {Hirata} C.~M.,   {Brinkmann}
  J.,  2006, \mn@doi [\mnras] {10.1111/j.1365-2966.2006.10156.x}, \href
  {http://adsabs.harvard.edu/abs/2006MNRAS.368..715M} {368, 715}

\bibitem[\protect\citeauthoryear{{Mann} \& {Ebeling}}{{Mann} \&
  {Ebeling}}{2012}]{mann12}
{Mann} A.~W.,  {Ebeling} H.,  2012, \mn@doi [\mnras]
  {10.1111/j.1365-2966.2011.20170.x}, \href
  {http://adsabs.harvard.edu/abs/2012MNRAS.420.2120M} {420, 2120}

\bibitem[\protect\citeauthoryear{{McLeod}, {McLure}  \& {Dunlop}}{{McLeod}
  et~al.}{2016}]{McLeod16}
{McLeod} D.~J.,  {McLure} R.~J.,   {Dunlop} J.~S.,  2016, preprint, \href
  {http://adsabs.harvard.edu/abs/2016arXiv160205199M} {} (\mn@eprint {arXiv}
  {1602.05199})

\bibitem[\protect\citeauthoryear{{Medezinski}, {Umetsu}, {Okabe}, {Nonino},
  {Molnar}, {Massey}, {Dupke}  \& {Merten}}{{Medezinski}
  et~al.}{2016}]{medezinski16}
{Medezinski} E.,  {Umetsu} K.,  {Okabe} N.,  {Nonino} M.,  {Molnar} S.,
  {Massey} R.,  {Dupke} R.,   {Merten} J.,  2016, \mn@doi [\apj]
  {10.3847/0004-637X/817/1/24}, \href
  {http://adsabs.harvard.edu/abs/2016ApJ...817...24M} {817, 24}

\bibitem[\protect\citeauthoryear{{Meneghetti} et~al.,}{{Meneghetti}
  et~al.}{2016}]{Meneghettiff16}
{Meneghetti} M.,  et~al., 2016, preprint, \href
  {http://adsabs.harvard.edu/abs/2016arXiv160604548M} {} (\mn@eprint {arXiv}
  {1606.04548})

\bibitem[\protect\citeauthoryear{{Merten}, {Cacciato}, {Meneghetti}, {Mignone}
  \& {Bartelmann}}{{Merten} et~al.}{2009}]{Merten09}
{Merten} J.,  {Cacciato} M.,  {Meneghetti} M.,  {Mignone} C.,   {Bartelmann}
  M.,  2009, \mn@doi [\aap] {10.1051/0004-6361/200810372}, \href
  {http://adsabs.harvard.edu/abs/2009A%26A...500..681M} {500, 681}

\bibitem[\protect\citeauthoryear{{Merten} et~al.,}{{Merten}
  et~al.}{2011}]{merten11}
{Merten} J.,  et~al., 2011, \mn@doi [\mnras]
  {10.1111/j.1365-2966.2011.19266.x}, \href
  {http://adsabs.harvard.edu/abs/2011MNRAS.417..333M} {417, 333}

\bibitem[\protect\citeauthoryear{{Monna} et~al.,}{{Monna}
  et~al.}{2015}]{monna15}
{Monna} A.,  et~al., 2015, \mn@doi [\mnras] {10.1093/mnras/stu2534}, \href
  {http://adsabs.harvard.edu/abs/2015MNRAS.447.1224M} {447, 1224}

\bibitem[\protect\citeauthoryear{{Monna} et~al.,}{{Monna}
  et~al.}{2017}]{monna17}
{Monna} A.,  et~al., 2017, \mn@doi [\mnras] {10.1093/mnras/stw3048}, \href
  {http://adsabs.harvard.edu/abs/2017MNRAS.465.4589M} {465, 4589}

\bibitem[\protect\citeauthoryear{{Moore}, {Ghigna}, {Governato}, {Lake},
  {Quinn}, {Stadel}  \& {Tozzi}}{{Moore} et~al.}{1999}]{moore99}
{Moore} B.,  {Ghigna} S.,  {Governato} F.,  {Lake} G.,  {Quinn} T.,  {Stadel}
  J.,   {Tozzi} P.,  1999, \mn@doi [\apjl] {10.1086/312287}, \href
  {http://adsabs.harvard.edu/abs/1999ApJ...524L..19M} {524, L19}

\bibitem[\protect\citeauthoryear{{Morishita} et~al.,}{{Morishita}
  et~al.}{2016}]{morishita16}
{Morishita} T.,  et~al., 2016, preprint, \href
  {http://adsabs.harvard.edu/abs/2016arXiv160700384M} {} (\mn@eprint {arXiv}
  {1607.00384})

\bibitem[\protect\citeauthoryear{{Natarajan} \& {Kneib}}{{Natarajan} \&
  {Kneib}}{1997}]{natarajan97}
{Natarajan} P.,  {Kneib} J.-P.,  1997, \mnras, \href
  {http://adsabs.harvard.edu/cgi-bin/nph-bib_query?bibcode=1997MNRAS.287..833N&db_key=AST}
  {287, 833}

\bibitem[\protect\citeauthoryear{{Natarajan} \& {Springel}}{{Natarajan} \&
  {Springel}}{2004}]{pn04}
{Natarajan} P.,  {Springel} V.,  2004, \mn@doi [\apjl] {10.1086/427079}, \href
  {http://adsabs.harvard.edu/abs/2004ApJ...617L..13N} {617, L13}

\bibitem[\protect\citeauthoryear{{Natarajan}, {Kneib}, {Smail}  \&
  {Ellis}}{{Natarajan} et~al.}{1998}]{natarajan98}
{Natarajan} P.,  {Kneib} J.-P.,  {Smail} I.,   {Ellis} R.~S.,  1998, \apj,
  \href {http://adsabs.harvard.edu/abs/1998ApJ...499..600N} {499, 600}

\bibitem[\protect\citeauthoryear{{Natarajan}, {De Lucia}  \&
  {Springel}}{{Natarajan} et~al.}{2007}]{natarajan07}
{Natarajan} P.,  {De Lucia} G.,   {Springel} V.,  2007, \mn@doi [\mnras]
  {10.1111/j.1365-2966.2007.11399.x}, \href
  {http://adsabs.harvard.edu/abs/2007MNRAS.376..180N} {376, 180}

\bibitem[\protect\citeauthoryear{{Natarajan}, {Kneib}, {Smail}, {Treu},
  {Ellis}, {Moran}, {Limousin}  \& {Czoske}}{{Natarajan}
  et~al.}{2009}]{natarajan09}
{Natarajan} P.,  {Kneib} J.-P.,  {Smail} I.,  {Treu} T.,  {Ellis} R.,  {Moran}
  S.,  {Limousin} M.,   {Czoske} O.,  2009, \mn@doi [\apj]
  {10.1088/0004-637X/693/1/970}, \href
  {http://adsabs.harvard.edu/abs/2009ApJ...693..970N} {693, 970}

\bibitem[\protect\citeauthoryear{{Nelson} et~al.,}{{Nelson}
  et~al.}{2015}]{nelson15}
{Nelson} D.,  et~al., 2015, \mn@doi [Astronomy and Computing]
  {10.1016/j.ascom.2015.09.003}, \href
  {http://adsabs.harvard.edu/abs/2015A%26C....13...12N} {13, 12}

\bibitem[\protect\citeauthoryear{{Newman}, {Ellis}  \& {Treu}}{{Newman}
  et~al.}{2015}]{newman15}
{Newman} A.~B.,  {Ellis} R.~S.,   {Treu} T.,  2015, \mn@doi [\apj]
  {10.1088/0004-637X/814/1/26}, \href
  {http://adsabs.harvard.edu/abs/2015ApJ...814...26N} {814, 26}

\bibitem[\protect\citeauthoryear{{Onions} et~al.,}{{Onions}
  et~al.}{2012}]{onions12}
{Onions} J.,  et~al., 2012, \mn@doi [\mnras]
  {10.1111/j.1365-2966.2012.20947.x}, \href
  {http://adsabs.harvard.edu/abs/2012MNRAS.423.1200O} {423, 1200}

\bibitem[\protect\citeauthoryear{{Owers}, {Randall}, {Nulsen}, {Couch}, {David}
   \& {Kempner}}{{Owers} et~al.}{2011}]{Owers11}
{Owers} M.~S.,  {Randall} S.~W.,  {Nulsen} P.~E.~J.,  {Couch} W.~J.,  {David}
  L.~P.,   {Kempner} J.~C.,  2011, \mn@doi [\apj] {10.1088/0004-637X/728/1/27},
  \href {http://adsabs.harvard.edu/abs/2011ApJ...728...27O} {728, 27}

\bibitem[\protect\citeauthoryear{{Pontzen} \& {Governato}}{{Pontzen} \&
  {Governato}}{2014}]{pontzen14}
{Pontzen} A.,  {Governato} F.,  2014, \mn@doi [\nat] {10.1038/nature12953},
  \href {http://adsabs.harvard.edu/abs/2014Natur.506..171P} {506, 171}

\bibitem[\protect\citeauthoryear{{Postman} et~al.,}{{Postman}
  et~al.}{2012}]{postman12}
{Postman} M.,  et~al., 2012, \mn@doi [\apjs] {10.1088/0067-0049/199/2/25},
  \href {http://adsabs.harvard.edu/abs/2012ApJS..199...25P} {199, 25}

\bibitem[\protect\citeauthoryear{{Read} \& {Gilmore}}{{Read} \&
  {Gilmore}}{2005}]{read05}
{Read} J.~I.,  {Gilmore} G.,  2005, \mn@doi [\mnras]
  {10.1111/j.1365-2966.2004.08424.x}, \href
  {http://adsabs.harvard.edu/abs/2005MNRAS.356..107R} {356, 107}

\bibitem[\protect\citeauthoryear{{Richard} et~al.,}{{Richard}
  et~al.}{2014}]{richard14}
{Richard} J.,  et~al., 2014, \mn@doi [\mnras] {10.1093/mnras/stu1395}, \href
  {http://adsabs.harvard.edu/abs/2014MNRAS.444..268R} {444, 268}

\bibitem[\protect\citeauthoryear{{Rocha}, {Peter}, {Bullock}, {Kaplinghat},
  {Garrison-Kimmel}, {O{\~n}orbe}  \& {Moustakas}}{{Rocha}
  et~al.}{2013}]{rocha13}
{Rocha} M.,  {Peter} A.~H.~G.,  {Bullock} J.~S.,  {Kaplinghat} M.,
  {Garrison-Kimmel} S.,  {O{\~n}orbe} J.,   {Moustakas} L.~A.,  2013, \mn@doi
  [\mnras] {10.1093/mnras/sts514}, \href
  {http://adsabs.harvard.edu/abs/2013MNRAS.430...81R} {430, 81}

\bibitem[\protect\citeauthoryear{{Rodney} et~al.,}{{Rodney}
  et~al.}{2015}]{rodney15}
{Rodney} S.~A.,  et~al., 2015, preprint, \href
  {http://adsabs.harvard.edu/abs/2015arXiv150506211R} {} (\mn@eprint {arXiv}
  {1505.06211})

\bibitem[\protect\citeauthoryear{{Rosati} et~al.,}{{Rosati}
  et~al.}{2014}]{Rosati14}
{Rosati} P.,  et~al., 2014, The Messenger, \href
  {http://adsabs.harvard.edu/abs/2014Msngr.158...48R} {158, 48}

\bibitem[\protect\citeauthoryear{{Schmidt} et~al.,}{{Schmidt}
  et~al.}{2014}]{schmidt14}
{Schmidt} K.~B.,  et~al., 2014, \mn@doi [\apjl] {10.1088/2041-8205/782/2/L36},
  \href {http://adsabs.harvard.edu/abs/2014ApJ...782L..36S} {782, L36}

\bibitem[\protect\citeauthoryear{{Schwinn}, {Jauzac}, {Baugh}, {Bartelmann},
  {Eckert}, {Harvey}, {Natarajan}  \& {Massey}}{{Schwinn}
  et~al.}{2016}]{schwinn17}
{Schwinn} J.,  {Jauzac} M.,  {Baugh} C.~M.,  {Bartelmann} M.,  {Eckert} D.,
  {Harvey} D.,  {Natarajan} P.,   {Massey} R.,  2016, preprint, \href
  {http://adsabs.harvard.edu/abs/2016arXiv161102790S} {} (\mn@eprint {arXiv}
  {1611.02790})

\bibitem[\protect\citeauthoryear{{Sharon} \& {Johnson}}{{Sharon} \&
  {Johnson}}{2015}]{sharon15}
{Sharon} K.,  {Johnson} T.~L.,  2015, \mn@doi [\apjl]
  {10.1088/2041-8205/800/2/L26}, \href
  {http://adsabs.harvard.edu/abs/2015ApJ...800L..26S} {800, L26}

\bibitem[\protect\citeauthoryear{{Springel}}{{Springel}}{2010}]{springel10}
{Springel} V.,  2010, \mn@doi [\mnras] {10.1111/j.1365-2966.2009.15715.x},
  \href {http://adsabs.harvard.edu/abs/2010MNRAS.401..791S} {401, 791}

\bibitem[\protect\citeauthoryear{{Springel} \& {Farrar}}{{Springel} \&
  {Farrar}}{2007}]{springel07}
{Springel} V.,  {Farrar} G.~R.,  2007, \mn@doi [\mnras]
  {10.1111/j.1365-2966.2007.12159.x}, \href
  {http://adsabs.harvard.edu/abs/2007MNRAS.380..911S} {380, 911}

\bibitem[\protect\citeauthoryear{{Springel}, {Yoshida}  \& {White}}{{Springel}
  et~al.}{2001}]{springel01}
{Springel} V.,  {Yoshida} N.,   {White} S.~D.~M.,  2001, \mn@doi [\na]
  {10.1016/S1384-1076(01)00042-2}, \href
  {http://adsabs.harvard.edu/abs/2001NewA....6...79S} {6, 79}

\bibitem[\protect\citeauthoryear{{Torrealba}, {Koposov}, {Belokurov}  \&
  {Irwin}}{{Torrealba} et~al.}{2016}]{torrealba16}
{Torrealba} G.,  {Koposov} S.~E.,  {Belokurov} V.,   {Irwin} M.,  2016,
  preprint, \href {http://adsabs.harvard.edu/abs/2016arXiv160107178T} {}
  (\mn@eprint {arXiv} {1601.07178})

\bibitem[\protect\citeauthoryear{{Torrey}, {Vogelsberger}, {Genel}, {Sijacki},
  {Springel}  \& {Hernquist}}{{Torrey} et~al.}{2014}]{torrey14}
{Torrey} P.,  {Vogelsberger} M.,  {Genel} S.,  {Sijacki} D.,  {Springel} V.,
  {Hernquist} L.,  2014, \mn@doi [\mnras] {10.1093/mnras/stt2295}, \href
  {http://adsabs.harvard.edu/abs/2014MNRAS.438.1985T} {438, 1985}

\bibitem[\protect\citeauthoryear{{Torri}, {Meneghetti}, {Bartelmann},
  {Moscardini}, {Rasia}  \& {Tormen}}{{Torri} et~al.}{2004}]{torri04}
{Torri} E.,  {Meneghetti} M.,  {Bartelmann} M.,  {Moscardini} L.,  {Rasia} E.,
   {Tormen} G.,  2004, \mn@doi [\mnras] {10.1111/j.1365-2966.2004.07508.x},
  \href
  {http://adsabs.harvard.edu/cgi-bin/nph-bib_query?bibcode=2004MNRAS.349..476T&db_key=AST}
  {349, 476}

\bibitem[\protect\citeauthoryear{{Treu} et~al.,}{{Treu} et~al.}{2016}]{treu16}
{Treu} T.,  et~al., 2016, \mn@doi [\apj] {10.3847/0004-637X/817/1/60}, \href
  {http://adsabs.harvard.edu/abs/2016ApJ...817...60T} {817, 60}

\bibitem[\protect\citeauthoryear{{Umetsu}, {Zitrin}, {Gruen}, {Merten},
  {Donahue}  \& {Postman}}{{Umetsu} et~al.}{2016}]{umetsu16}
{Umetsu} K.,  {Zitrin} A.,  {Gruen} D.,  {Merten} J.,  {Donahue} M.,
  {Postman} M.,  2016, \mn@doi [\apj] {10.3847/0004-637X/821/2/116}, \href
  {http://adsabs.harvard.edu/abs/2016ApJ...821..116U} {821, 116}

\bibitem[\protect\citeauthoryear{{Vogelsberger} et~al.,}{{Vogelsberger}
  et~al.}{2014a}]{vogelsberger14}
{Vogelsberger} M.,  et~al., 2014a, \mn@doi [\mnras] {10.1093/mnras/stu1536},
  \href {http://adsabs.harvard.edu/abs/2014MNRAS.444.1518V} {444, 1518}

\bibitem[\protect\citeauthoryear{{Vogelsberger} et~al.,}{{Vogelsberger}
  et~al.}{2014b}]{vogelsberger14a}
{Vogelsberger} M.,  et~al., 2014b, \mn@doi [\nat] {10.1038/nature13316}, \href
  {http://adsabs.harvard.edu/abs/2014Natur.509..177V} {509, 177}

\bibitem[\protect\citeauthoryear{{Wang} et~al.,}{{Wang} et~al.}{2015}]{wang15}
{Wang} X.,  et~al., 2015, preprint, \href
  {http://adsabs.harvard.edu/abs/2015arXiv150402405W} {} (\mn@eprint {arXiv}
  {1504.02405})

\bibitem[\protect\citeauthoryear{{Wetzel}, {Hopkins}, {Kim}, {Faucher-Giguere},
  {Keres}  \& {Quataert}}{{Wetzel} et~al.}{2016}]{wetzel16}
{Wetzel} A.~R.,  {Hopkins} P.~F.,  {Kim} J.-h.,  {Faucher-Giguere} C.-A.,
  {Keres} D.,   {Quataert} E.,  2016, preprint, \href
  {http://adsabs.harvard.edu/abs/2016arXiv160205957W} {} (\mn@eprint {arXiv}
  {1602.05957})

\bibitem[\protect\citeauthoryear{{Wong}, {Ammons}, {Keeton}  \&
  {Zabludoff}}{{Wong} et~al.}{2012}]{wong12}
{Wong} K.~C.,  {Ammons} S.~M.,  {Keeton} C.~R.,   {Zabludoff} A.~I.,  2012,
  \mn@doi [\apj] {10.1088/0004-637X/752/2/104}, \href
  {http://adsabs.harvard.edu/abs/2012ApJ...752..104W} {752, 104}

\bibitem[\protect\citeauthoryear{{Wong}, {Zabludoff}, {Ammons}, {Keeton},
  {Hogg}  \& {Gonzalez}}{{Wong} et~al.}{2013}]{wong13}
{Wong} K.~C.,  {Zabludoff} A.~I.,  {Ammons} S.~M.,  {Keeton} C.~R.,  {Hogg}
  D.~W.,   {Gonzalez} A.~H.,  2013, \mn@doi [\apj]
  {10.1088/0004-637X/769/1/52}, \href
  {http://adsabs.harvard.edu/abs/2013ApJ...769...52W} {769, 52}

\bibitem[\protect\citeauthoryear{{Zitrin} et~al.,}{{Zitrin}
  et~al.}{2013}]{zitrin13}
{Zitrin} A.,  et~al., 2013, \mn@doi [\apjl] {10.1088/2041-8205/762/2/L30},
  \href {http://adsabs.harvard.edu/abs/2013ApJ...762L..30Z} {762, L30}

\bibitem[\protect\citeauthoryear{{van den Bosch} \& {Jiang}}{{van den Bosch} \&
  {Jiang}}{2016}]{vandenbosch16}
{van den Bosch} F.~C.,  {Jiang} F.,  2016, \mn@doi [\mnras]
  {10.1093/mnras/stw440}, \href
  {http://adsabs.harvard.edu/abs/2016MNRAS.458.2870V} {458, 2870}

\bibitem[\protect\citeauthoryear{{van den Bosch}, {Tormen}  \& {Giocoli}}{{van
  den Bosch} et~al.}{2005}]{vandenbosch05}
{van den Bosch} F.~C.,  {Tormen} G.,   {Giocoli} C.,  2005, \mn@doi [\mnras]
  {10.1111/j.1365-2966.2005.08964.x}, \href
  {http://adsabs.harvard.edu/abs/2005MNRAS.359.1029V} {359, 1029}

\makeatother
\end{thebibliography}

\end{document}